\documentstyle[12pt,epsfig]{article}

\def\tstrut{\vrule height2.5ex depth0pt width0pt} % used in tables
\def\er#1#2{\relax\ifmmode{}^{+#1}_{-#2}\else$^{+#1}_{-#2}$\fi}
\newcommand{\be}{\begin{equation}}
\newcommand{\bea}{\begin{eqnarray}}
\newcommand{\ee}{\end{equation}}
\newcommand{\eea}{\end{eqnarray}}

\begin{document}
%
%\thispagestyle{empty}
%\begin{flushright}
%UG-DFM-1/99 \\
%nucl-th/9906437
%\end{flushright}
%  \vspace*{5mm}

\begin{center}
{\Large \bf The Inverse Amplitude Method in  $\pi\pi$
Scattering in Chiral Perturbation Theory to Two
Loops}

  J. Nieves\footnote{e-mail: jmnieves@ugr.es}, M. Pav\'on Valderrama
and E. Ruiz Arriola\footnote{e-mail:earriola@ugr.es} \\ [2em]
Departamento de F\'{\i}sica Moderna, Universidad de Granada, E-18071
Granada, Spain.\\

\end{center}

%\vskip.2cm
\begin{abstract}
The inverse amplitude method is used to unitarize the two loop
$\pi\pi$ scattering amplitudes of SU(2) Chiral Perturbation Theory in
the $I=0,J=0$, $I=1,J=1$ and $I=2,J=0$ channels. An error analysis in
terms of the low energy one-loop parameters $\bar l_{1,2,3,4,}$ and
existing experimental data is undertaken. A comparison to standard
resonance saturation values for the two loop coefficients $\bar
b_{1,2,3,4,5,6} $ is also carried out. Crossing violations are
quantified and the convergence of the expansion is discussed. 
\vskip.3cm 
\noindent
\centerline{\it 
}

\centerline{\it Keywords: Unitarization, Inverse Amplitude Method,}
\centerline{\it Chiral Perturbation Theory,$\pi\pi$-Scattering, Error
analysis. }
\end{abstract}

%\date{\today} 

\section{Introduction} 

The $\pi\pi$ reaction is one of the theoretically cleanest processes
in hadronic physics. This is due to the fact that crossing, unitarity
and analyticity impose severe constraints to the scattering amplitude
\cite{Ro71}. Thus, a lot of attention has been paid to the study of
this process and the determination of partial wave phase shifts
\cite{pa73}-\cite{fp77}. The current theoretical setup for such an approach
is Chiral Perturbation Theory (ChPT) which is an Effective Field
Theory embodying all these constraints, and leads to a perturbative
expansion of the scattering partial wave amplitude in the $(I,J)$
isospin-spin channel\footnote{We use the normalization such that the
partial wave cross section is $ \sigma_{IJ} = (2J+1) ( 4 \pi / s)
|t_{IJ}(s)|^2 $. See also Eq.~(\ref{eq:fIJ}) below.}
\begin{eqnarray}
t_{IJ} (s) =  t^{(2)}_{IJ} (s) +  t^{(4)}_{IJ} (s) +
t^{(6)}_{IJ} (s) + \cdots
\label{eq:chpt} 
\end{eqnarray} 
Here the expansion parameter turns out to be $\lambda^2 \equiv m_\pi^2
/ (4\pi f_\pi)^2 \sim 0.01 $ with $m_\pi = 139.57 {\rm MeV} $ the
physical pion mass and $f_\pi = 92.3 {\rm MeV} $ the weak pion decay
constant. Hence $t_{IJ}^{(n)}$ turns out to be proportional to
$\lambda^{2n}$. In the $\pi\pi$ scattering case, the chiral expansion
generates a hierarchy of corrections which depend on an increasing
number of dimensionless and renormalization scale independent
parameters.  To lowest non-trivial order (LO,tree level) current
algebra unique predictions for scattering amplitudes,
$t_{IJ}^{(2)}(s)$ in terms of $f_\pi $ and $m_\pi$ are generated
\cite{We66}. At next-to-leading order (NLO,one loop) four low energy
parameters $\bar l_{1,2,3,4}$ determine the amplitude
$t_{IJ}^{(4)}(s)$ \cite{GL84,gl85}. At next-to-next-to-leading order
(NNLO, two loops) the amplitude $t_{IJ}^{(6)}(s)$ can be expressed in
terms of six parameters $\bar b_{1,2,3,4,5,6} $
\cite{mksf95,bc97}. Unlike the one loop parameters $\bar l_{1,2,3,4}$
which can be fixed from ChPT calculations confronted to experimental
data from several sources, the two loop coefficients $\bar
b_{1,2,3,4,5,6} $ are a bit more difficult to fix directly from
experiment since the amount of data close enough to threshold is
scarce. Because of this problem and motivated by the success of the
resonance saturation hypothesis at the one-loop level and at a
renormalization scale $\mu = 750 \pm 250 {\rm MeV}$ \cite{egpr89},
values for the two-loop $\bar b$'s have been suggested on the basis of
this hypothesis at that scale \cite{bc97}. This makes the, in
principle, renormalization scheme independent $\bar b$'s to have a
spurious scale dependence.  Since it is not really known what
uncertainty should be assigned to this hypothesis, it has been
suggested to ascribe a $100\%$ error on the contributions to the low
energy parameters determined by resonances \cite{gir97}.

\begin{table}[t]
%\vspace{-0.3cm}
\begin{center}
\begin{tabular}{c|c|c|c|c|c}  
 & $ \bar l_1 $ & $\bar l_2 $ & $\bar l_3 $ & $\bar l_4 $ & $r(\bar l_1
 , \bar l_2 )$ \\
\hline\tstrut 
Set {\bf Ic \,} ($K_{l4}$ decays) & $0.3 \pm 1.2 $ &
$4.77 \pm {0.45} $ & $ 2.9 \pm {2.4} $ & $4.4 \pm{0.3} $ & 
$-0.69 $ \\
Set {\bf II \,} ($D-$waves)& $-0.8 \pm 4.8 $ &
$4.5 \pm {1.1} $ & $ 2.9 \pm {2.4} $ & $4.4 \pm{0.3} $ & 
$-0.75 $ \\
Set {\bf III \,}(Roy sum rules)& $-0.9 \pm 1.2 $ &
$4.34 \pm {0.25} $ & $ 2.9 \pm {2.4} $ & $4.4 \pm{0.3} $ & 
$-0.22 $ 
\end{tabular}
\end{center}
\caption{\footnotesize One loop $\bar l_{1,2,3,4} $ low energy
parameters in ChPT for the parameter sets {\bf Ic}, {\bf II} and {\bf
III} of Ref.~\cite{EJ00a} as well as the correlation coefficient
between $\bar l_1$ and $\bar l_2$. In the present paper we use only
Set {\bf Ic} and Set {\bf III}.}
\label{tab:lbar} 
\end{table}

The numerical consideration of errors from ChPT requires taking into
account consistent sets of low energy parameters, both $\bar
l_{1,2,3,4}$ and $\bar b_{1,2,3,4,5,6}$, deduced from several sources
\cite{bcg94,ABT00a,ABT00b,bct98}. Moreover, there seems to be strong
anti-correlations between $\bar l_1$ and $\bar l_2$ on the light of
two loop $K_{l4}$ analysis \cite{ABT00a,ABT00b}. This point has been
thoroughly discussed in a previous work by two of us \cite{EJ00a} and
refer to it for further details. There, error propagation has been
undertaken \`a la Monte Carlo, instead of using parametric statistics,
by generating a synthetic set of primary data. The basic assumption is
that primary quantities, i.e., those obtained either directly from
experiment or from an acceptable $\chi^2$ fit, i.e., with a $\chi^2 /
{\rm d.o.f}. \sim 1 $, are Gauss distributed although perhaps with
correlations.  In practice, the distributions are represented by a
population of finite but sufficiently large number ($N=10^4 $) of
samples.  The discussion in that work amounts to have three compatible
sets of one loop $\bar l$ and two loop $\bar b$ low energy parameter
distributions deduced from several sources. They are sumarized in
Tables ~\ref{tab:lbar} and \ref{tab:bbar}.  For consistency, we
keep the same notation of our previous work. Set {\bf Ic} in
Ref.~\cite{EJ00a} corresponds to $K_{l4}$ decays, following the fits
of Ref.~\cite{ABT00a} and some statistical modeling designed to
reproduce the fragmentary information given in the $K_{l4}$ analysis
of Ref.~\cite{ABT00a}. Set {\bf II} corresponds to use $D-$waves as
proposed in the two loop $\pi\pi$ scattering calculation in standard
ChPT \cite{bc97}. Finally, Set {\bf III} denotes the values of the low
energy parameters obtained through Roy sum rules following the lines
of Ref.~\cite{gir97}. We have found in Ref.~\cite{EJ00a} that the
anti-correlations between $\bar l_1$ and $\bar l_2$ persist in Set
{\bf III}, although they are not that strong.  In the present work we
discard Sets {\bf Ia} and {\bf Ib} for being somewhat unrealistic. We
also disregard Set {\bf II} because it produces too large errors as
compared to Sets {\bf Ic} and {\bf III}.

Despite of its great success, ChPT does not incorporate exact
unitarity to a given order of the expansion and hence cannot account
for resonances, in particular for the $\rho$ and $\sigma$ resonances
which appear in the $I=1,J=1$ and $I=0,J=0$ channels respectively. To
deal with this problem, several unitarization methods have been
devised in the past in the context of $\pi\pi$
scattering~\cite{Tr88}-\cite{EJ00b} (for a recent and short review see
e.g. Ref.~\cite{RGNP00}). In those methods, unitarity is restored in
the different partial wave amplitudes, while crossing is violated
\cite{BP97,Ha99,CB01}. In the complex energy plane this corresponds to
exactly take into account the unitarity right hand elastic cut ( $ s >
4 m_\pi^2 $) but to approximate the left hand cut ( $ s < 0 $
). Detailed quantitative studies reveal that the approximation used to
take into account the left hand cut does not become critical to
describe phase shifts in the scattering region, but it may
significantly influence the violation of crossing and the values of
the low energy parameters. Thus, there is some confidence that
unitarization methods can indeed be used to enlarge the domain of
applicability of ChPT to the study of intermediate energy hadronic
reactions. Among these unitarization approaches the Inverse Amplitude
Method (IAM) has successfully been applied to the description of
meson-meson scattering incorporating up to one loop perturbative
constraints. Original applications of IAM involved dispersion relation
arguments \cite{DP93,DP97a}, which became rather cumbersome when
incorporating coupled channels such as $K \bar K$ in $\pi\pi$
scattering. An algebraic derivation was soon found \cite{OOP98} to
provide an almost trivial generalization to the coupled channel case
and a complete one loop analysis for all meson-meson channels has been
carried out very recently in Ref.~\cite{GP01}. In addition to its very
simple implementation from the standard chiral expansion, what makes
this method particularly attractive is the fact that no new constants
arise besides those already present in standard ChPT. Furthermore, the
IAM method offers the possibility of systematic improvement according
to the chiral expansion. Given the great success of this unitarization
method in several meson-meson reactions including up to one loop
corrections, there seems almost obvious to extend the calculation to
the, in principle, more accurate description up to two loops. As we
will show below, such an extension is not as trivial as one might
think. Actually, there have been a previous calculation \cite{Ha97}
where an analysis of the IAM on the light of ChPT to two loops has
been undertaken. The conclusion was that the IAM is a well converging
scheme. It is fair to say that no effort was done to assign
uncertainties in the low energy parameters, making somewhat hard to
decide not only on the convergence itself but also on the
compatibility with standard ChPT.

Recently, theoretical restrictions for the $s-$wave scattering lengths
have been obtained from an analysis of Roy equations \cite{AC00}.
Unprecedented accuracy is obtained if in addition to the relativistic,
crossing and unitarity demands from local quantum field theory, chiral
symmetry constraints and the corresponding chiral expansion are
implemented.  The recent work \cite{CGL01} on matching the Roy
equation analysis \cite{AC00} to the two loop ChPT expansion
\cite{mksf95,bc97} has produced, using parametric statistics, the
smallest error estimates for the low energy parameters, so far.  Roy
equations provide an extremely elegant framework to incorporate both
crossing, analyticity and unitarity constraints on $\pi\pi$ scattering
amplitudes.  The set of non-linear inhomogeneous integral equations
are not autonomous but require some high energy tails obtained
from experiment as input. In the low energy regime these so called
driving terms can be described as polynomials which coefficients can
be mapped to the low energy parameters of ChPT in the common region of
validity of Roy equations and ChPT.  As a theoretical tool, Roy
equations cannot be bitten by unitarization methods since the latter
violate crossing to some extent. On the other hand, Roy equations have
not been generalized yet to other processes different from $\pi\pi$
scattering and require a knowledge of high energy data which may not
always be available or accurate enough
\footnote{See however the recent work on $\pi N$
scattering\cite{BL01}}.  Taking this fact into account and the time
elapsed since the original work \cite{Ro71} and the recent update
\cite{AC00}, it would be desirable but it seems unlikely that Roy
equation techniques will become overnight a daily tool in hadronic
physics.  In contrast, unitarization methods based on ChPT require in
principle no more work than ChPT itself which works well in the
threshold region, but are able to describe, in addition, resonance
physics and have been successfully applied to a variety of
problems. Because of this the unitarization of $\pi\pi$ scattering
amplitudes {\it \`a la} IAM provides a model case where we can learn
about the virtues and drawbacks of the method and also on its
convergence properties.

In the present work we study the IAM of unitarization of the two loop
ChPT amplitudes including a detailed error analysis based on the
presently available information on the low energy parameters obtained
from ChPT. By pursuing such a calculation we want to answer the
question of whether or not low energy information plus unitarization
reproduces the data beyond the domain of applicability of standard
ChPT. In common with other unitarized calculations it is not clear how
to avoid the unavoidable and prejudiced choice of a particular
unitarization method. Moreover, given the unitarization method it is
hard to estimate uncertainties due to higher orders in the
expansion. Our only hint so far, is to compare successive orders in
the scattering phase shifts with their corresponding error-bars and
determine whether or not practical convergence requirements are
met. We do this analysis using Set {\bf Ic} and Set {\bf III} of one-
and two loop low energy parameters $\bar l_{1,2,3,4}$ and $\bar
b_{1,2,3,4,5,6} $ respectively of our previous work \cite{EJ00a}.

The paper is structured as follows. In Sect.~\ref{sec:chpt} we provide
some basic definitions in order to fix notation. We also estimate
unitarity violations and the failure of a perturbative definition of
phase-shifts to describe the data in the region above threshold. In
Sect.~\ref{sec:iam} we analyze the IAM phase-shift predictions as well
as the corresponding threshold parameters, together with estimates on
the amount of crossing violations in terms of Roskies sum rules
\cite{Ro70,CP71} inherent to any unitarization method. Motivated by
previous experiences such crossing violations can be amended by a
suitable generalization of the IAM. This point is analyzed in
Sect.~\ref{sec:giam}. Although, there has been some work on one loop
IAM of unitarization, we present in Sect.~\ref{sec:conv} an updated
analysis from the point of view of the convergence of the expansion
for the unitarized phase shifts. Finally, we draw our main conclusions
in Sect.~\ref{sec:concl}. In the Appendix~\ref{sec:app1} we provide
some information not presented in our previous paper \cite{EJ00a} and
also relevant for the present work such as correlation matrices of
both low energy constants and threshold parameters.

\section{ChPT to two loops and unitarity violations} \label{sec:chpt}

\subsection{Basic definitions} 

Let $t_{IJ} (s)$ be the partial wave scattering
amplitude for the reaction $\pi\pi \to \pi \pi $ at the Center of Mass
(CM) energy $\sqrt{s} $ in the $IJ$ isospin-spin channel 
\begin{eqnarray} 
t_{IJ}(s) = { e^{2 i \delta_{IJ} (s)} -1  \over 2 i \sigma (s) } 
\label{eq:fIJ} 
\end{eqnarray} 
with $\sigma (s) = \sqrt{1-4 m_\pi^2/s} $ the CM momentum and
$\delta_{IJ}(s) $ the corresponding phase-shifts. Two particle
unitarity corresponds to real $ \delta_{IJ}(s) $ and can be written as
a non-linear relation on the amplitude
\begin{eqnarray}
{\rm Im} t_{IJ} (s) = \sigma (s) | t_{IJ} (s)|^2   
\end{eqnarray} 
or, equivalently, as a linear relation on the inverse amplitude
\begin{eqnarray}
{\rm Im} t_{IJ}^{-1} (s) = -  \sigma (s) 
\label{eq:unitarity}   
\end{eqnarray} 
For a $\pi\pi$ scattering amplitude calculated in the chiral expansion
sketched in Eq.~(\ref{eq:chpt}), the lowest order amplitude
$t_{IJ}^{(2)} (s) $ is a real function for $S-$ and $P-$waves, and
vanishes for $D-$ and higher ones.  The NLO amplitude $t^{(4)}_{IJ}
(s)$ develops an imaginary part for $S-$ and $P-$ waves but becomes
real for $D-$waves, and so on. The exact unitarity relation of
Eq.~(\ref{eq:unitarity}) requires, at a perturbative level, the set of
relations
\begin{eqnarray}
{\rm Im} t^{(2)}_{IJ} (s) &=& 0 \\ {\rm Im} t^{(4)}_{IJ} (s) &=&
\sigma (s) |t^{(2)}_{IJ} (s)|^2 \\ {\rm Im} t^{(6)}_{IJ} (s) &=& 2
\sigma(s) t^{(2)}_{IJ} (s) {\rm Re} t^{(4)}_{IJ} (s)
\label{eq:pert-uni}
\end{eqnarray} 
Standard ChPT fulfills exact crossing symmetry at any order of the
expansion, but violates unitarity. Two aspects are related to this
violation. In the first place, a necessary condition for unitarity is
the fulfillment of the inequality
\begin{eqnarray}
 \sigma(s)  \left| t_{IJ} (s) \right| = | \sin \delta_{IJ} (s) | \le 1   
\end{eqnarray} 
This unitarity limit yields values of $s$ a bit too high\footnote{For
instance, for the isoscalar $S-$ wave one gets at LO the inequality
fulfilled in the range  $s < 4 \sqrt{\pi} f_\pi \sim 670 {\rm MeV} $. We will
show below that unitarity violations take place at significantly lower
energies.}. A better way to quantify the {\it unitarity violation} is
by defining the quantity
\begin{eqnarray}
{\rm U}_{IJ} (s) = \left| 1+ 2 {\rm i} \sigma(s) t_{IJ}(s) \right|
\label{eq:uni_viol} 
\end{eqnarray} 
Below the two pion production threshold $\sqrt{s}= 4 m_\pi \sim 560 {\rm
MeV} $ the elastic unitarity condition requires $U_{IJ} (s)=|e^{ 2{\rm i}
\delta_{IJ}(s)}|= 1 $. Strictly speaking, if unitarity is violated,
$U_{IJ}(s) \neq 1 $, there is no way besides perturbation theory for a
real phase shift to fulfill simultaneously Eq.~(\ref{eq:chpt}) and
Eq.~(\ref{eq:fIJ}).  Expanding Eq.~(\ref{eq:fIJ}) according to
Eq.~(\ref{eq:chpt}) the standard ChPT phase shift may be computed
yielding
\begin{eqnarray}
\delta_{IJ}^{\rm ChPT} (s) &=& {1\over 2i} \ln \left[ 1 + 2 i
\sigma(s) t_{IJ} (s) \right] = \sigma(s) t_{IJ}^{(2)} (s) + \sigma(s)
\left[ t_{IJ}^{(4)}(s) - i \sigma(s) t_{IJ}^{(2)}(s)^2 \right]
\nonumber \\ &+& \sigma(s) \left[ t_{IJ}^{(6)}(s) - 2 i \sigma(s)
t_{IJ}^{(2)}(s) t_{IJ}^{(4)}(s) -\frac43 \sigma(s)^2 t_{IJ}^{(2)}(s)^3
\right] + \dots
\label{eq:delta_chpt}
\end{eqnarray}
The elastic unitarity condition corresponds to $\delta_{IJ}^{\rm ChPT}
(s)$ being real, which is automatically fulfilled if the perturbative
unitarity relations, Eq.~(\ref{eq:pert-uni}), are used, and one
effectively gets
\begin{eqnarray}
\delta_{IJ}^{\rm ChPT} (s) &=& \sigma(s) t_{IJ}^{(2)} (s) + \sigma(s)
{\rm Re} t_{IJ}^{(4)}(s) + \sigma(s) \left[ {\rm Re} t_{IJ}^{(6)}(s) +
\frac23 \sigma(s)^2 t_{IJ}^{(2)}(s)^3 \right] + \dots
\end{eqnarray}

Close to threshold, the scattering amplitude can be written in terms of
the threshold parameters, scattering lengths, $a_{IJ}$, and slopes,
$b_{IJ}$, defined by 
\begin{eqnarray}
t_{IJ} (s) = 2 m_\pi \left(s/4-m_\pi^2\right)^J \left[ a_{IJ} + b_{IJ}
(s/4- m_\pi^2) + \cdots \right]
\label{eq:threshold} 
\end{eqnarray} 
The scattering amplitudes $t_{IJ} (s)$ present kinematical zeros of
order $J$ at $s=4m_\pi^2$. Chiral symmetry implies the existence of
dynamical zeros for the $S-$waves, named chiral or Adler zeros.  In
ChPT, Adler zeros may be determined perturbatively, i.e.
\begin{eqnarray}
t_{IJ} ( s_A ) = t_{IJ}^{(2)} (s_A ) + t_{IJ}^{(4)} (s_A ) +
t_{IJ}^{(6)} (s_A ) + \cdots 
\end{eqnarray}
Expanding the solution $ s_A = s_A^{(2)} + s_A^{(4)} + s_A^{(6)} +
\cdots $, we get\footnote{For a general function there would be a term
involving the second derivative of $t_{IJ}^{(2)}$. This term
disappears from the formulas since $t_{IJ}^{(2)}$ is a linear function
of $s$.}
\begin{eqnarray} 
t_{IJ}^{(2)} (s_A^{(2)} ) &=& 0 \\ s_A^{(4)} &=& - t_{IJ}^{(4)}
(s_A^{(2)} ) / t_{IJ}^{(2)} (s_A^{(2)})' \\ s_A^{(6)} &=& -
t_{IJ}^{(6)} (s_A^{(2)} ) / t_{IJ}^{(2)} (s_A^{(2)})' +
\frac{t^{(4)}_{IJ} (s_A^{(2)})' t^{(4)}_{IJ}
(s_A^{(2)})}{[t^{(2)}_{IJ} (s_A^{(2)})']^2 }
\label{eq:adler-zeros}
\end{eqnarray} 
At lowest order, non-kinematical zeros are located at 
\begin{eqnarray}
s_A^{(2)} &=& \frac12 m_\pi^2 \qquad I=0 \quad J=0 \\ s_A^{(2)} &=& 2
m_\pi^2 \qquad I=2 \quad J=0
\label{eq:adler0}
\end{eqnarray}

\begin{figure}[t]
\begin{center}
\epsfig{figure=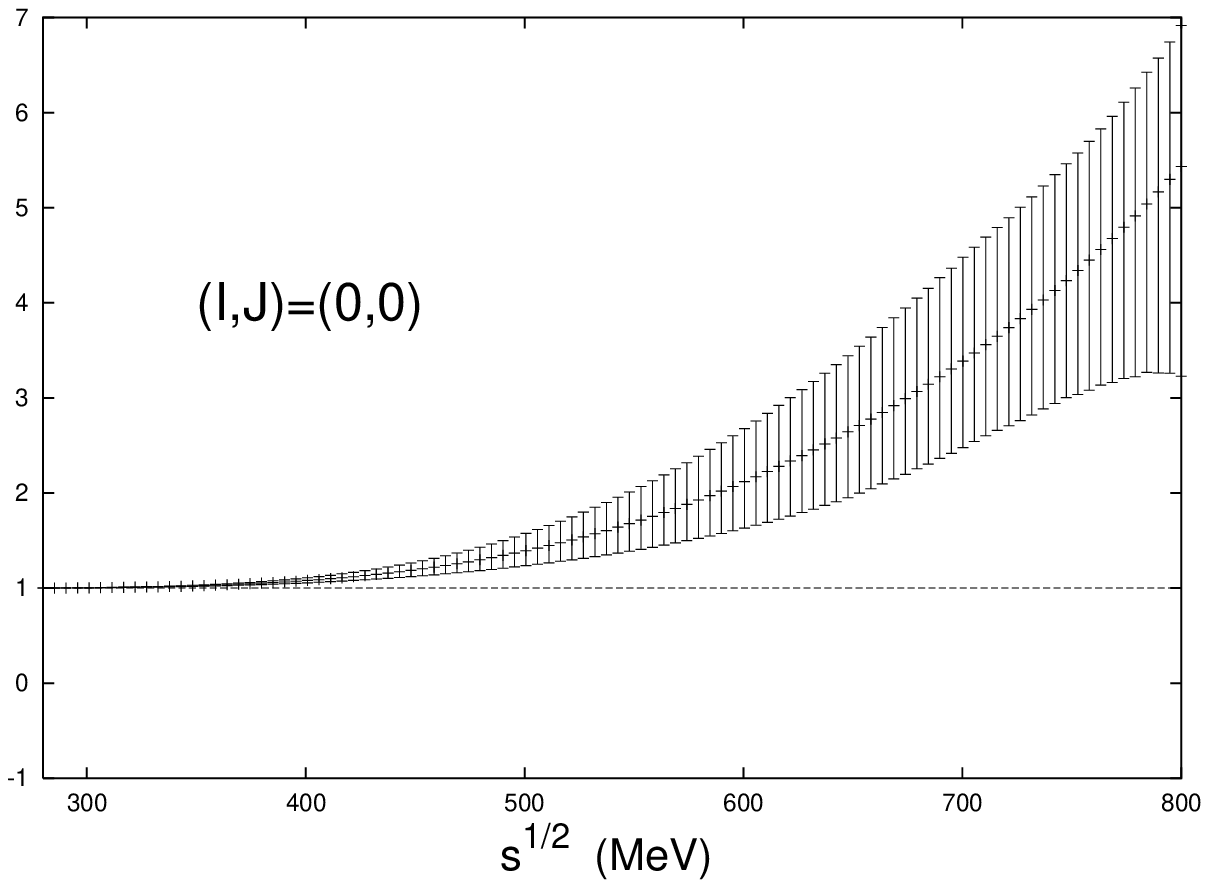,height=5.5cm,width=5.5cm}\epsfig{figure=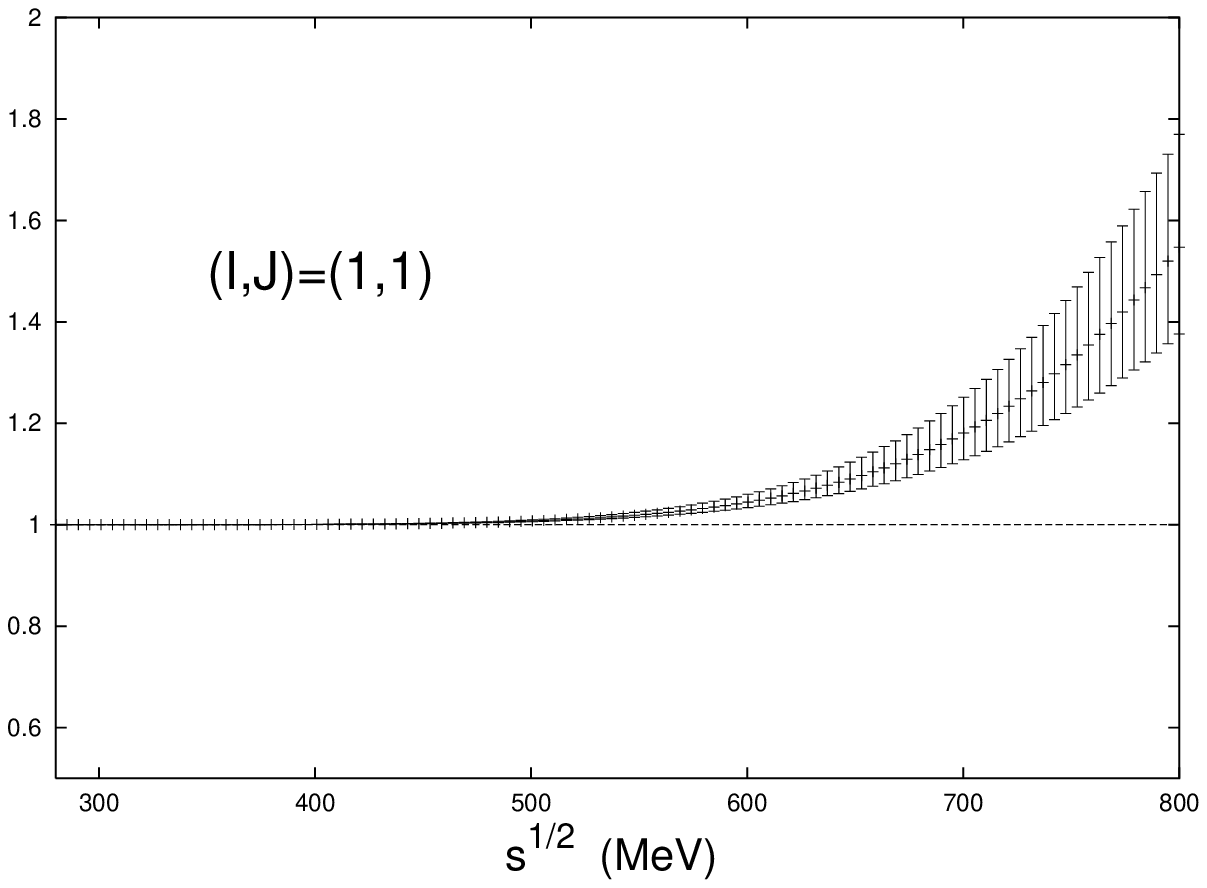,height=5.5cm,width=5.5cm}\epsfig{figure=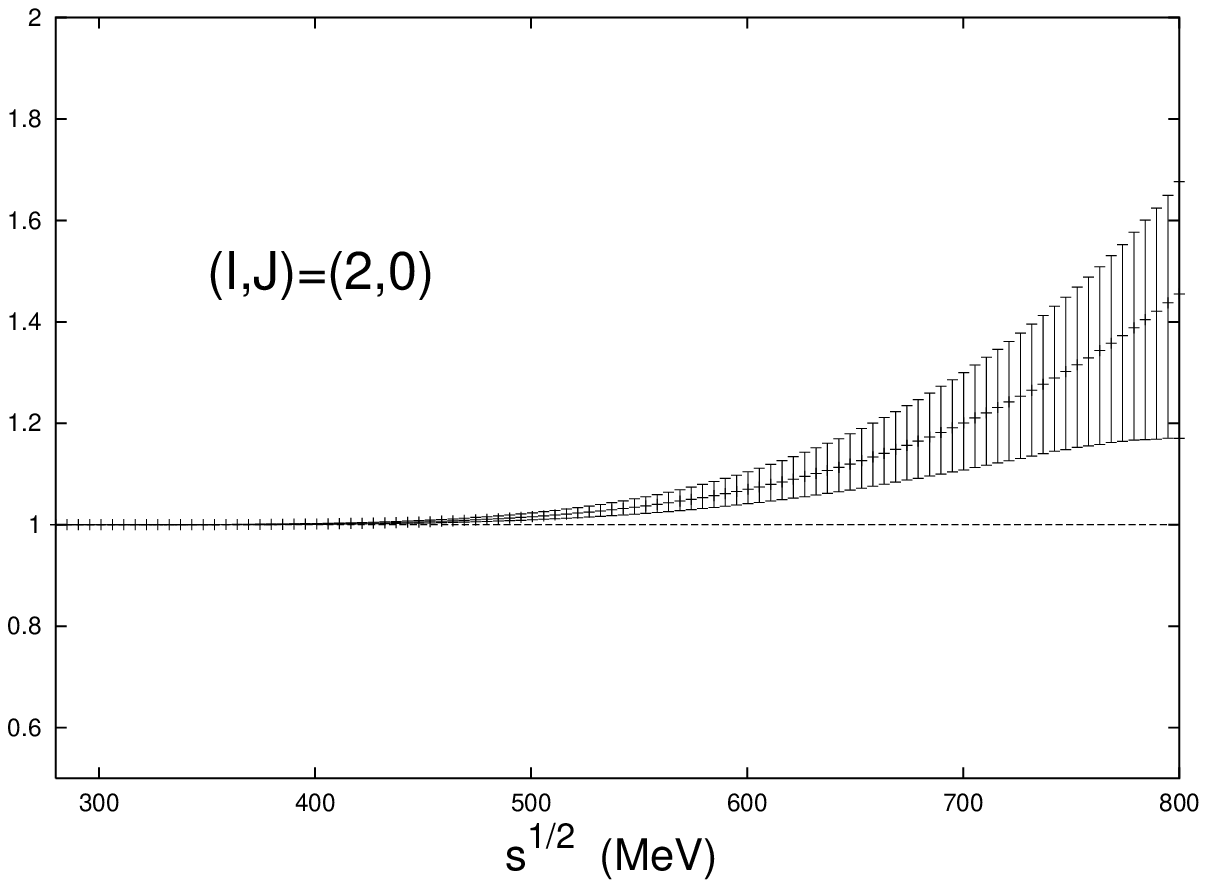,height=5.5cm,width=5.5cm}
\epsfig{figure=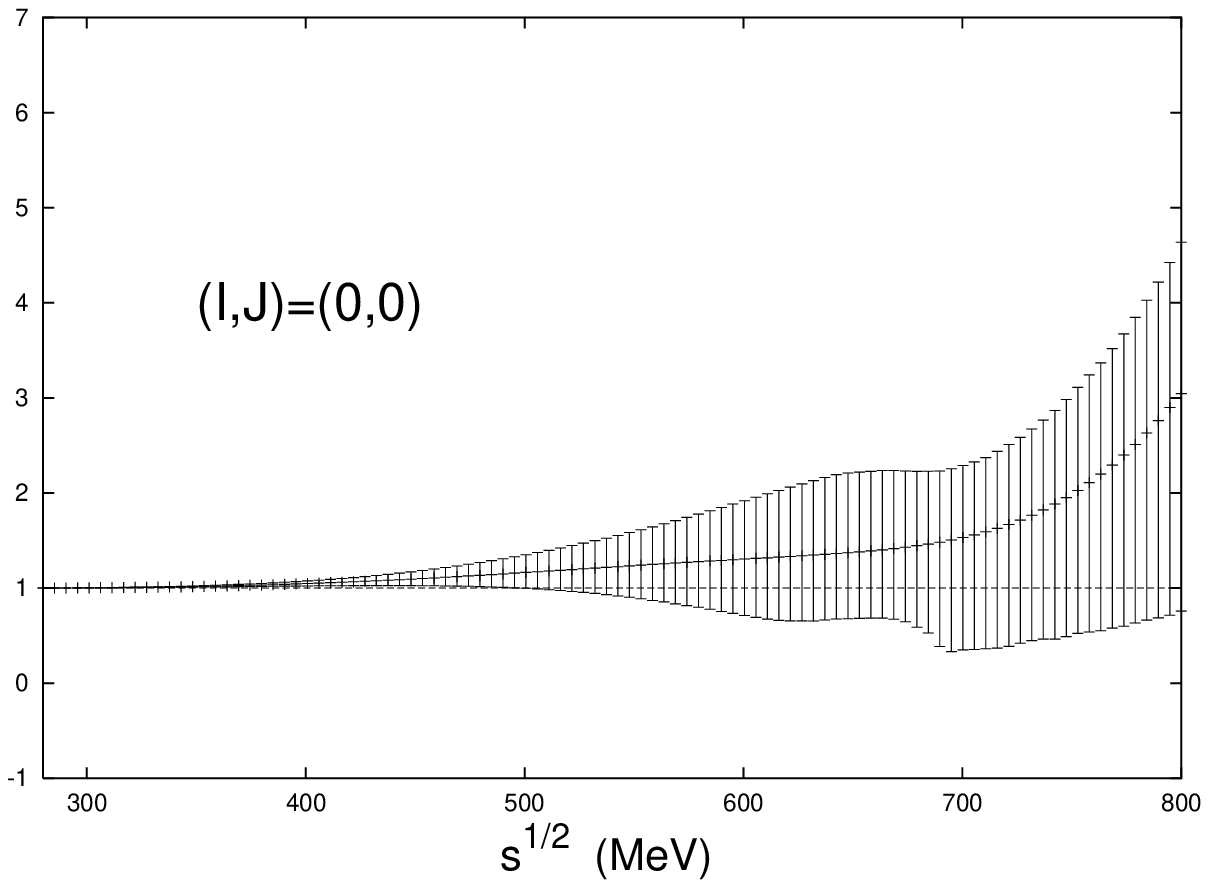,height=5.5cm,width=5.5cm}\epsfig{figure=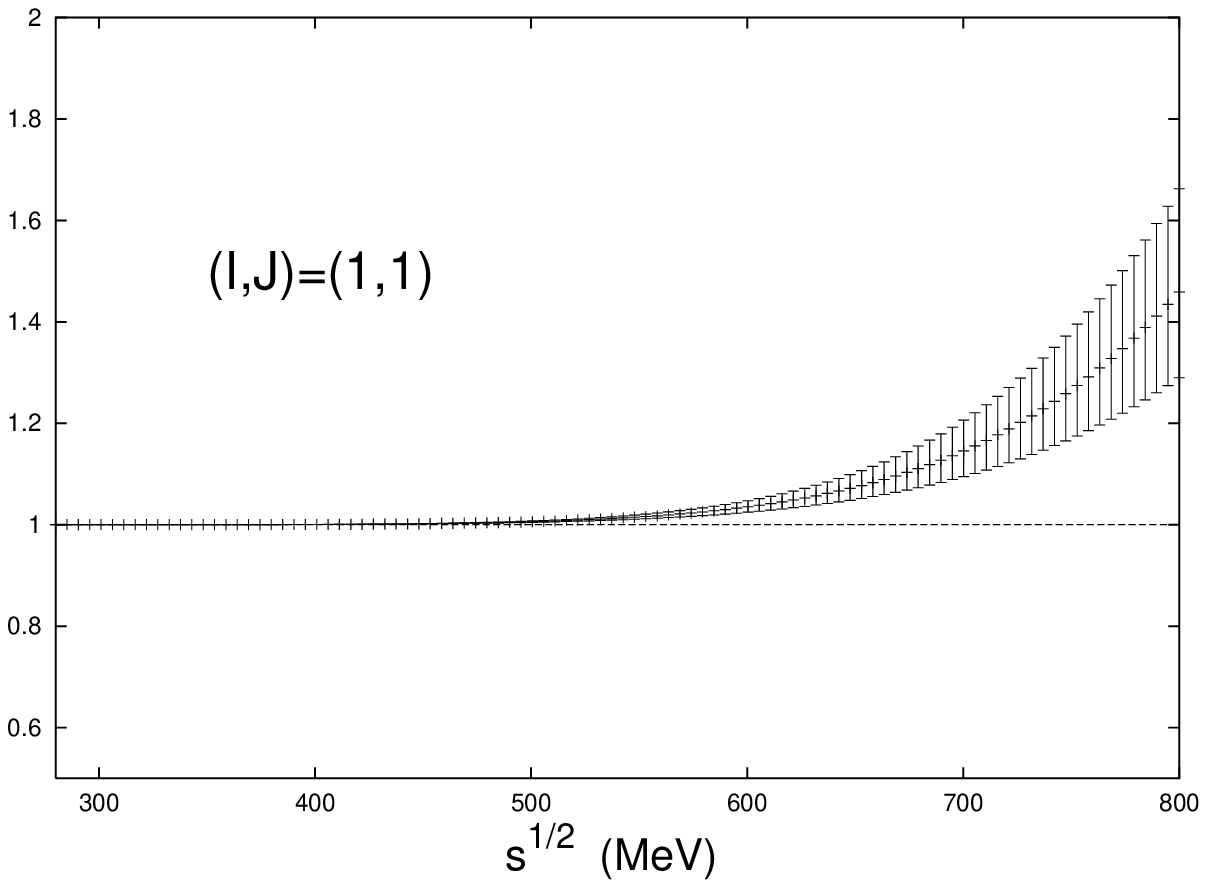,height=5.5cm,width=5.5cm}\epsfig{figure=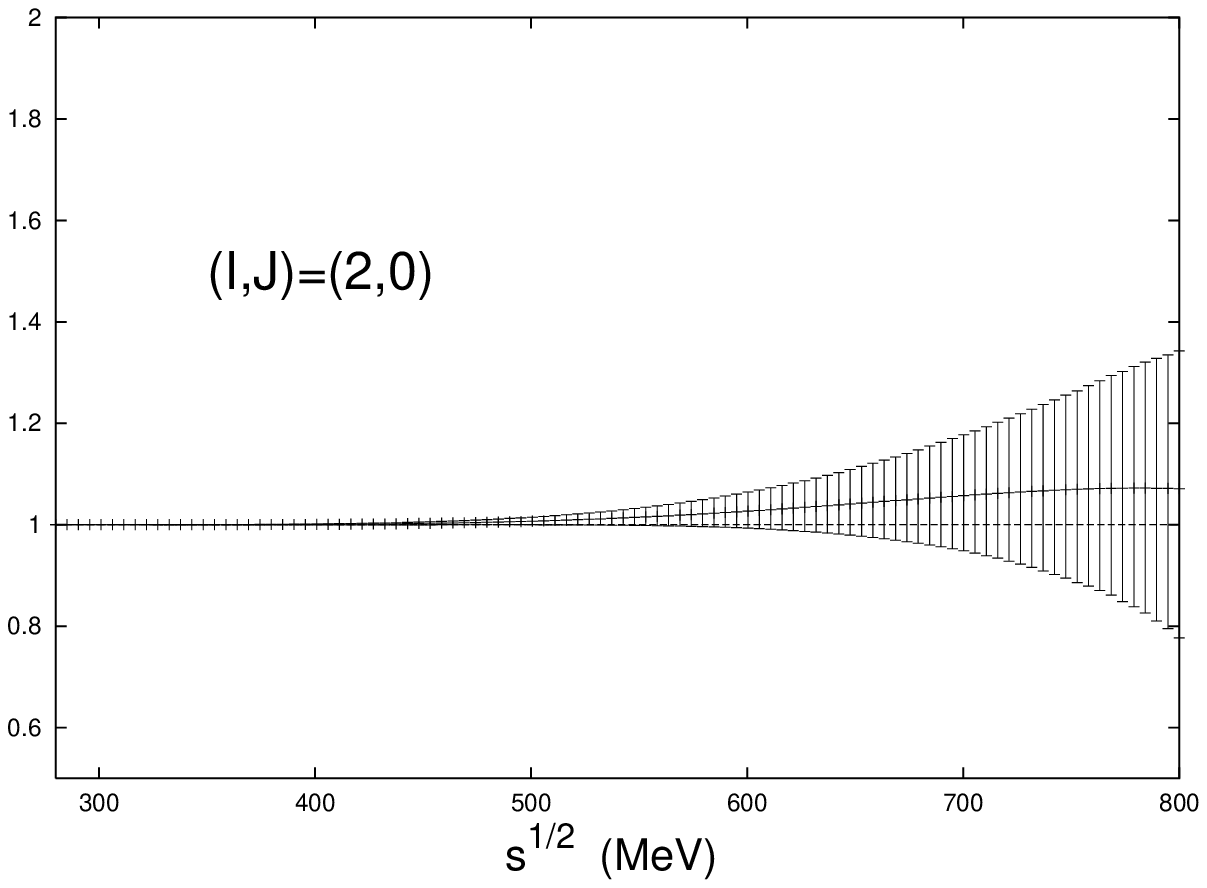,height=5.5cm,width=5.5cm}
\end{center} 
\caption{Unitarity condition for standard NNLO ChPT amplitudes in
$\pi\pi$ scattering for $S-$ and $P-$ waves defined by $U_{IJ} (s) = |
1 + 2 i \sigma(s) t_{IJ} (s) |$ . Upper panel: Set {\bf Ic} of
Ref.~\cite{EJ00a}. Lower panel: Set {\bf III} of Ref.~\cite{EJ00a}. In
the calculation the parameters $\bar b_i^0$ defined in
Eq.~(\ref{eq:b+db}) and given in Table~\ref{tab:bbar} have been used. The
unitarity condition requires $U_{IJ} (s)=1$.}
\label{fig:unitarity}
\end{figure}

\begin{figure}[t]
\begin{center}
\epsfig{figure=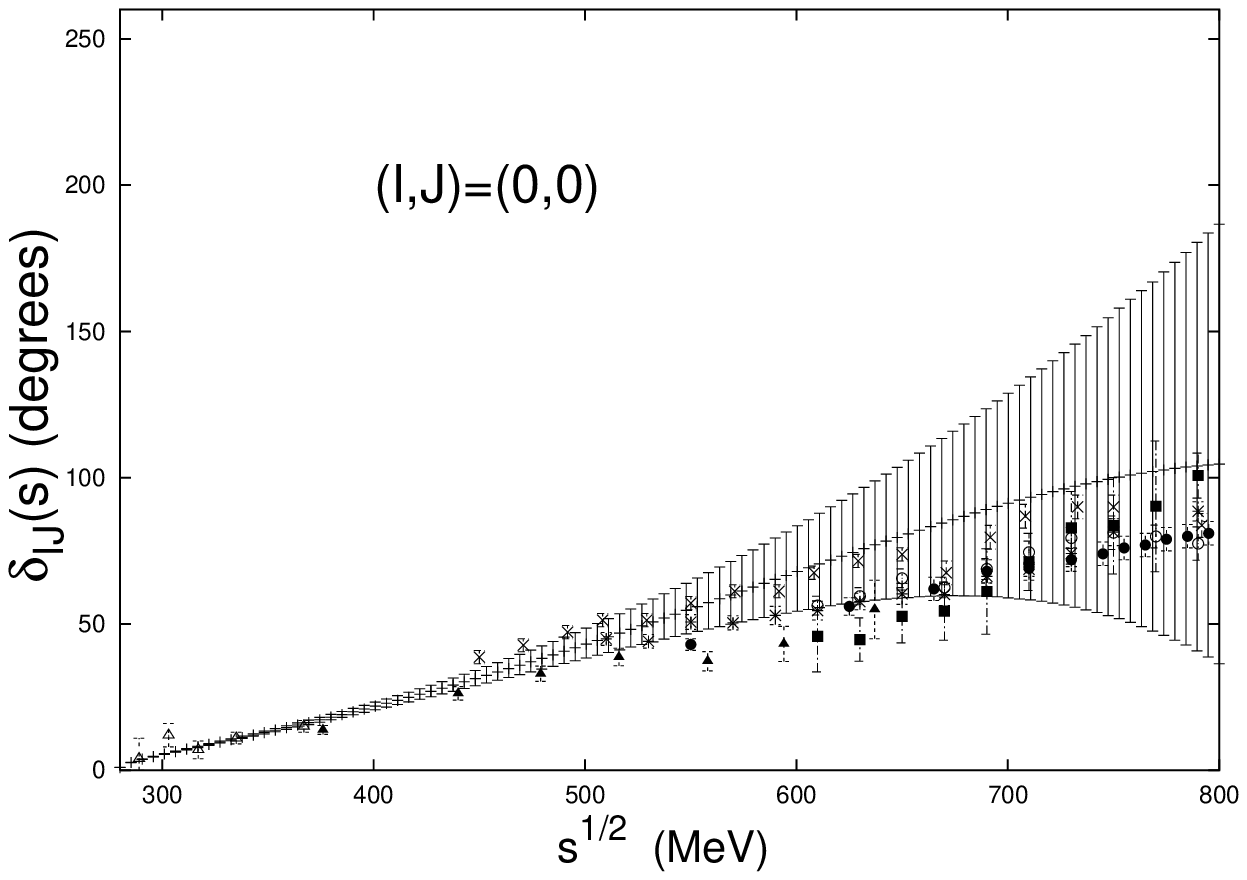,height=5.5cm,width=5.5cm}\epsfig{figure=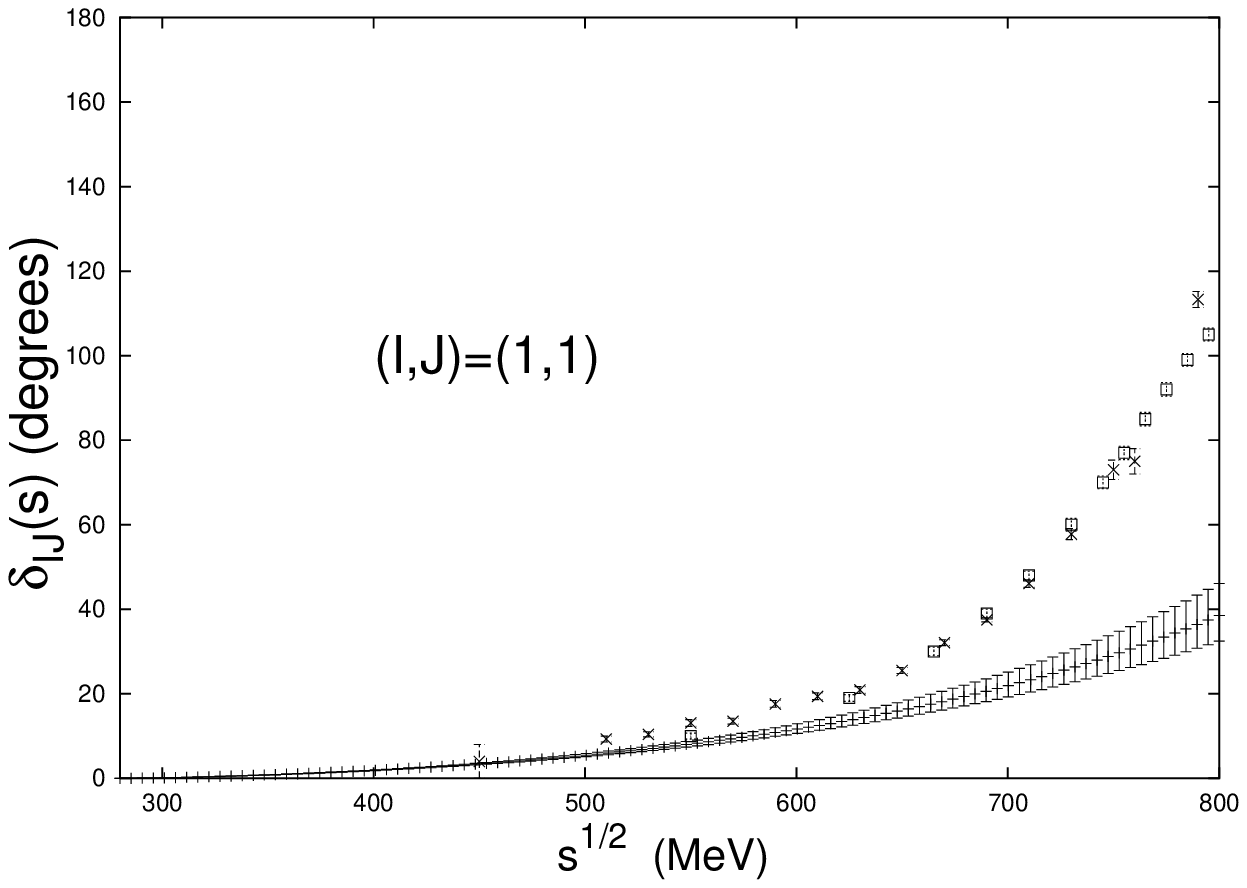,height=5.5cm,width=5.5cm}\epsfig{figure=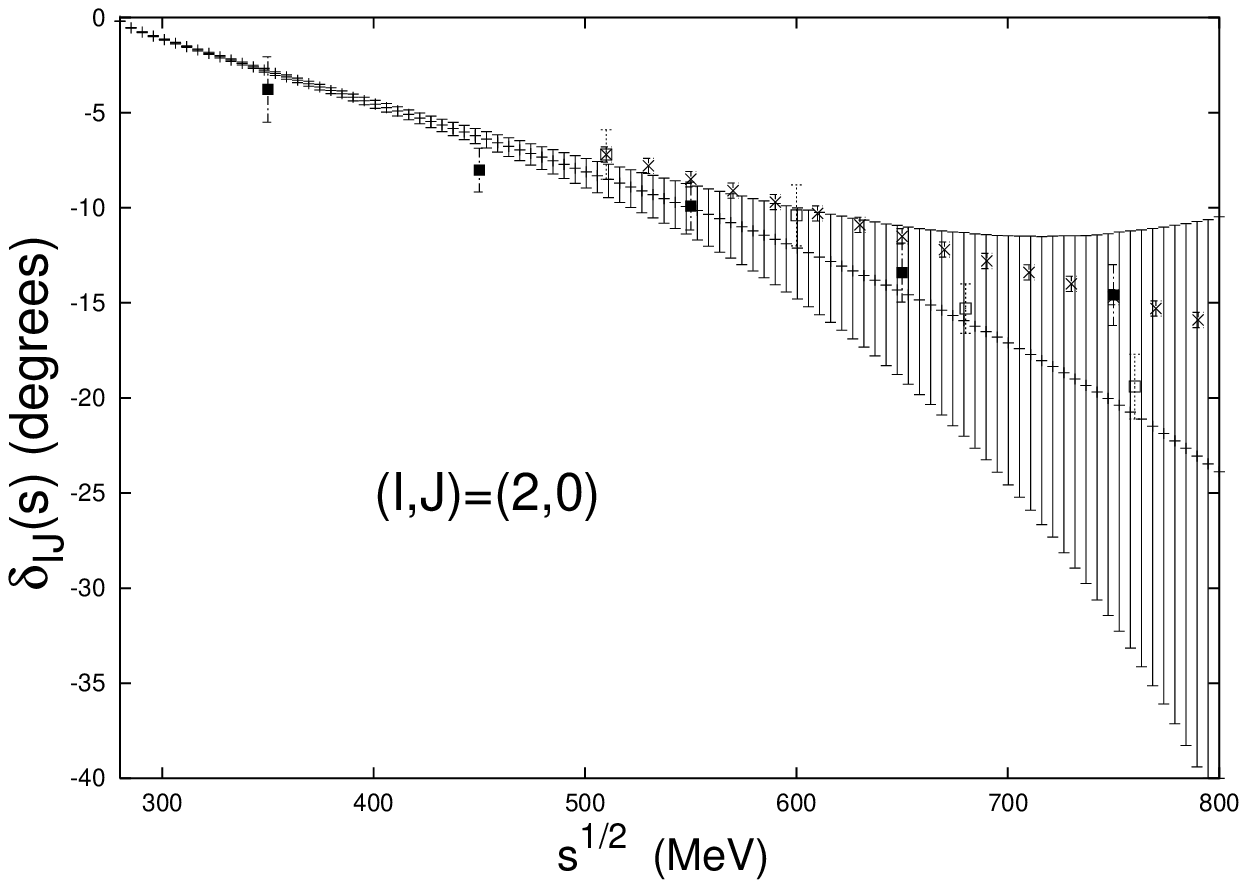,height=5.5cm,width=5.5cm}
\epsfig{figure=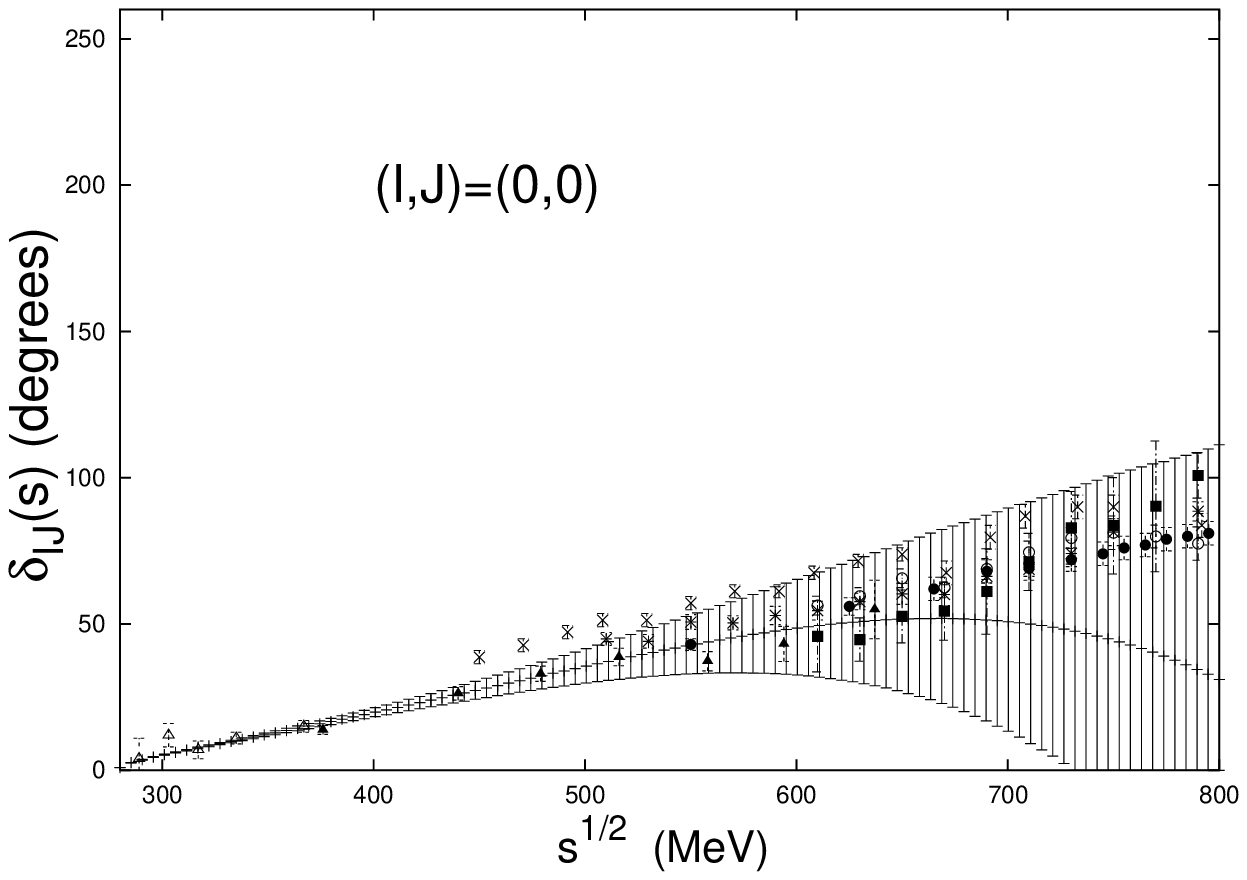,height=5.5cm,width=5.5cm}\epsfig{figure=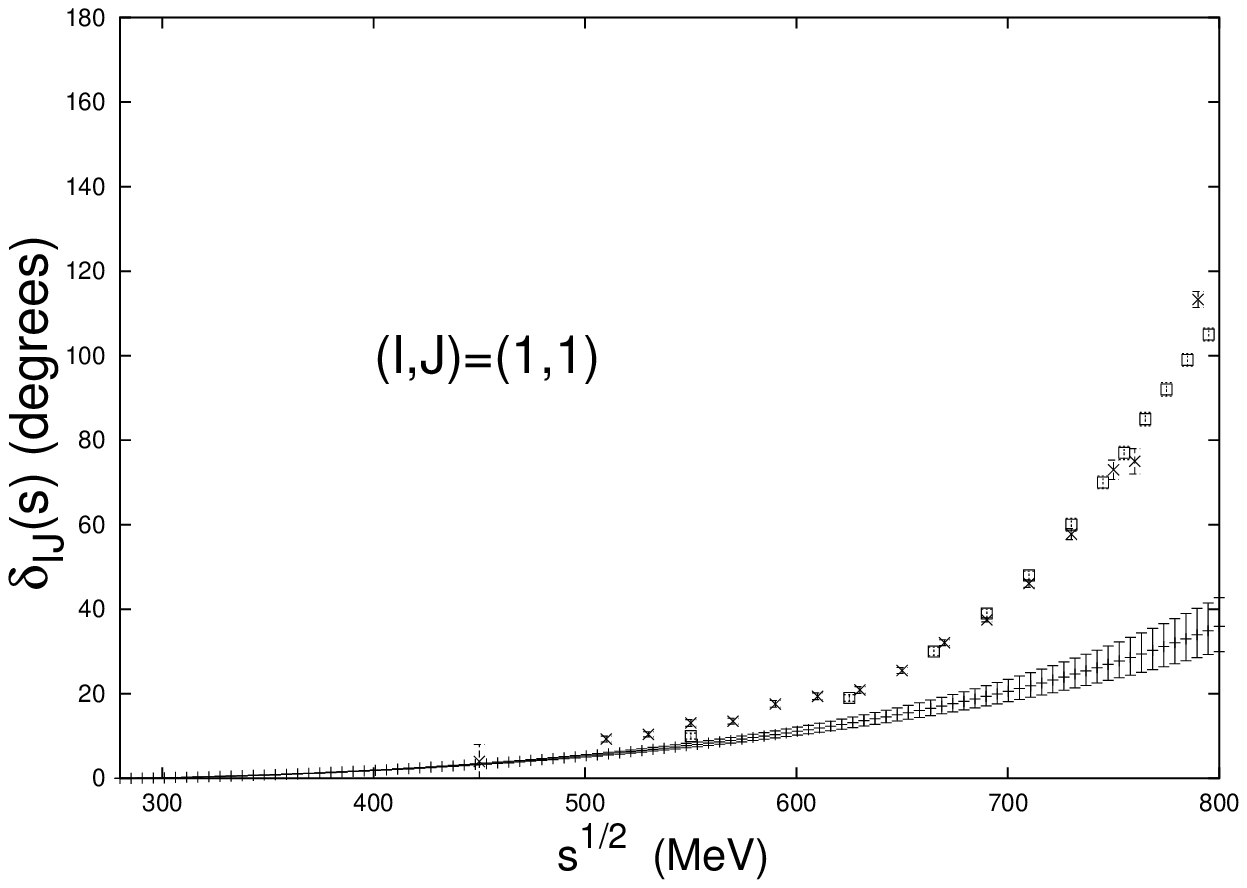,height=5.5cm,width=5.5cm}\epsfig{figure=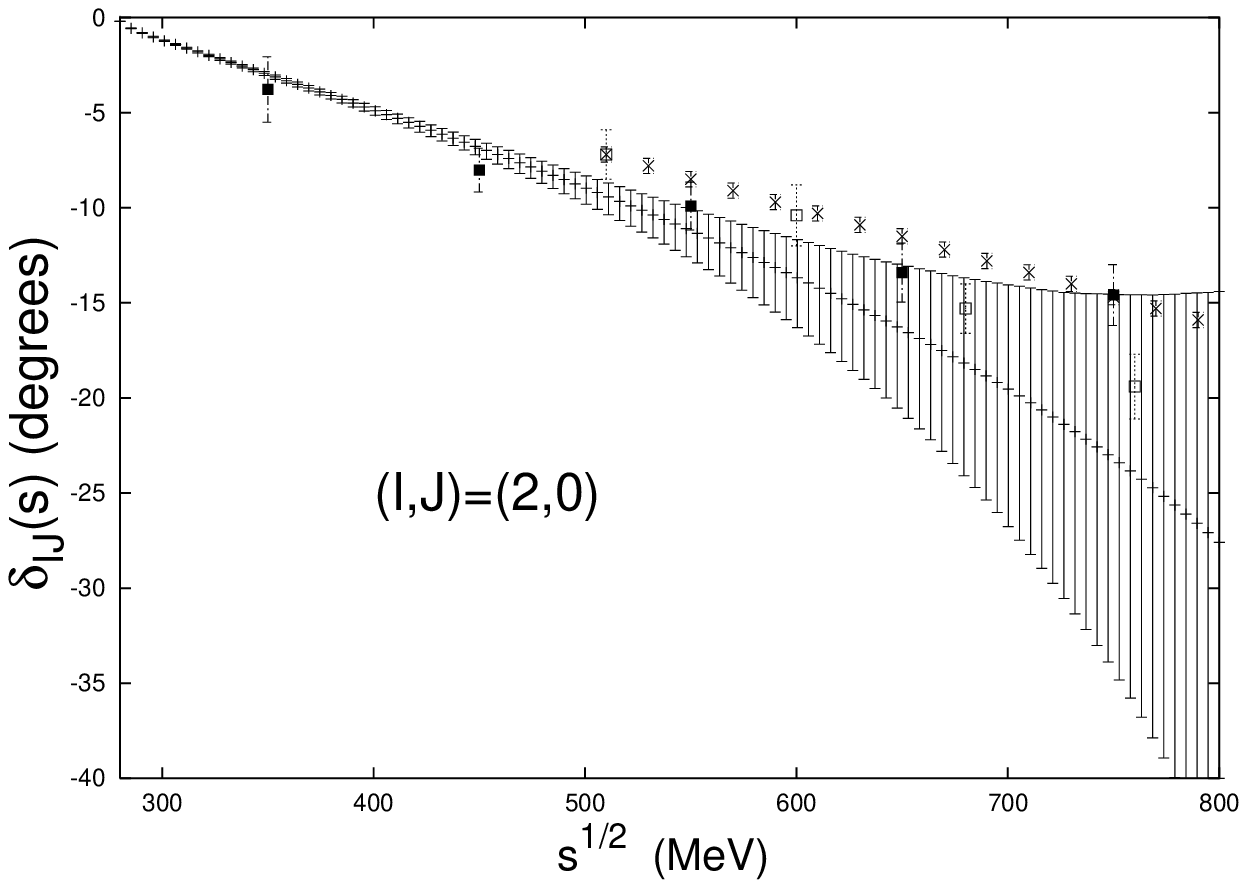,height=5.5cm,width=5.5cm}
\end{center} 
\caption{Standard NNLO ChPT phase shifts (in degrees) for $\pi\pi$
scattering for $S-$ and $P-$ waves after
Eq.~(\ref{eq:delta_chpt}). Upper panel: Set {\bf Ic} of
Ref.~\cite{EJ00a}. Lower panel: Set {\bf III} of Ref.~\cite{EJ00a}. In
the calculation the parameters $\bar b_i^0$ defined in
Eq.~(\ref{eq:b+db}) and given in Table~\ref{tab:bbar} have been
used. Combined data from Refs.~\cite{pa73}-\cite{fp77}.}
\label{fig:chpt}
\end{figure}

\begin{figure}[t]
\begin{center}
\epsfig{figure=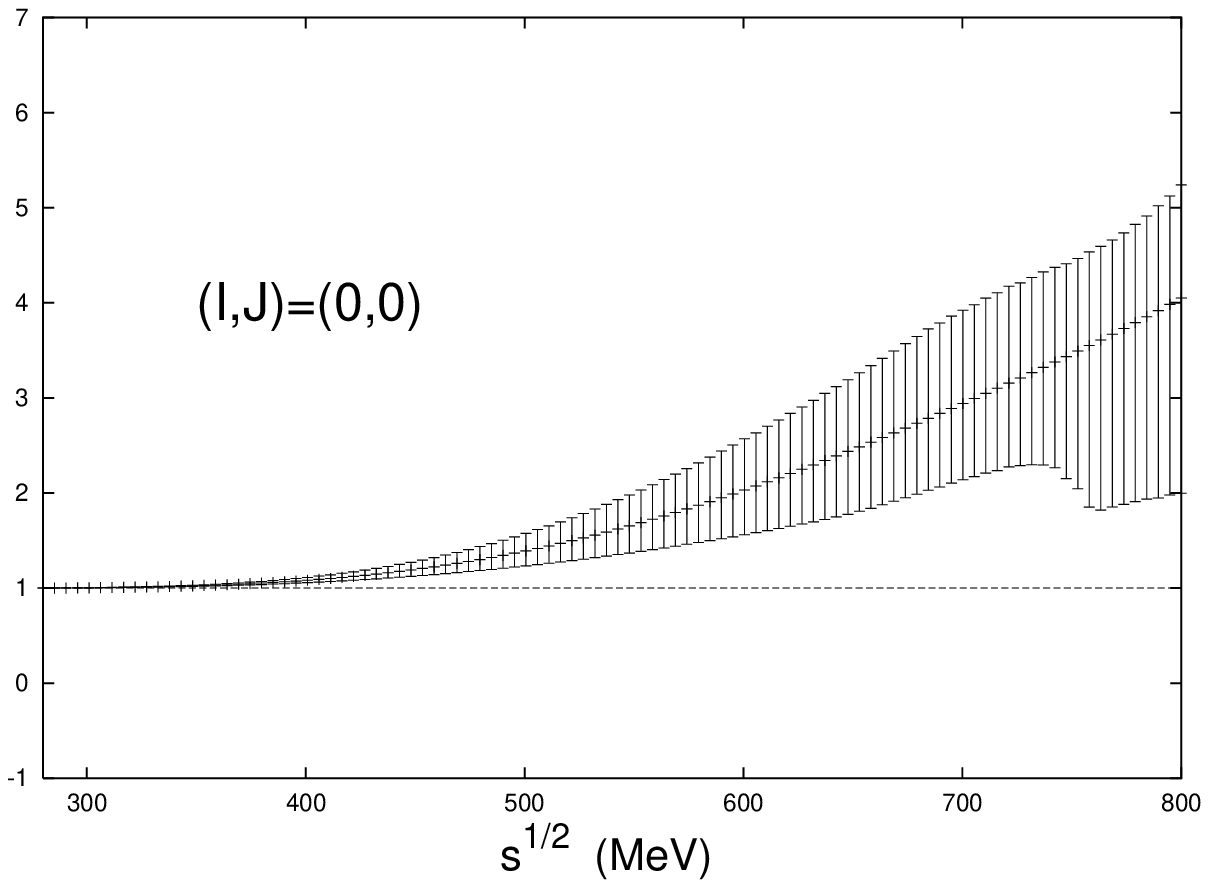,height=5.5cm,width=5.5cm}\epsfig{figure=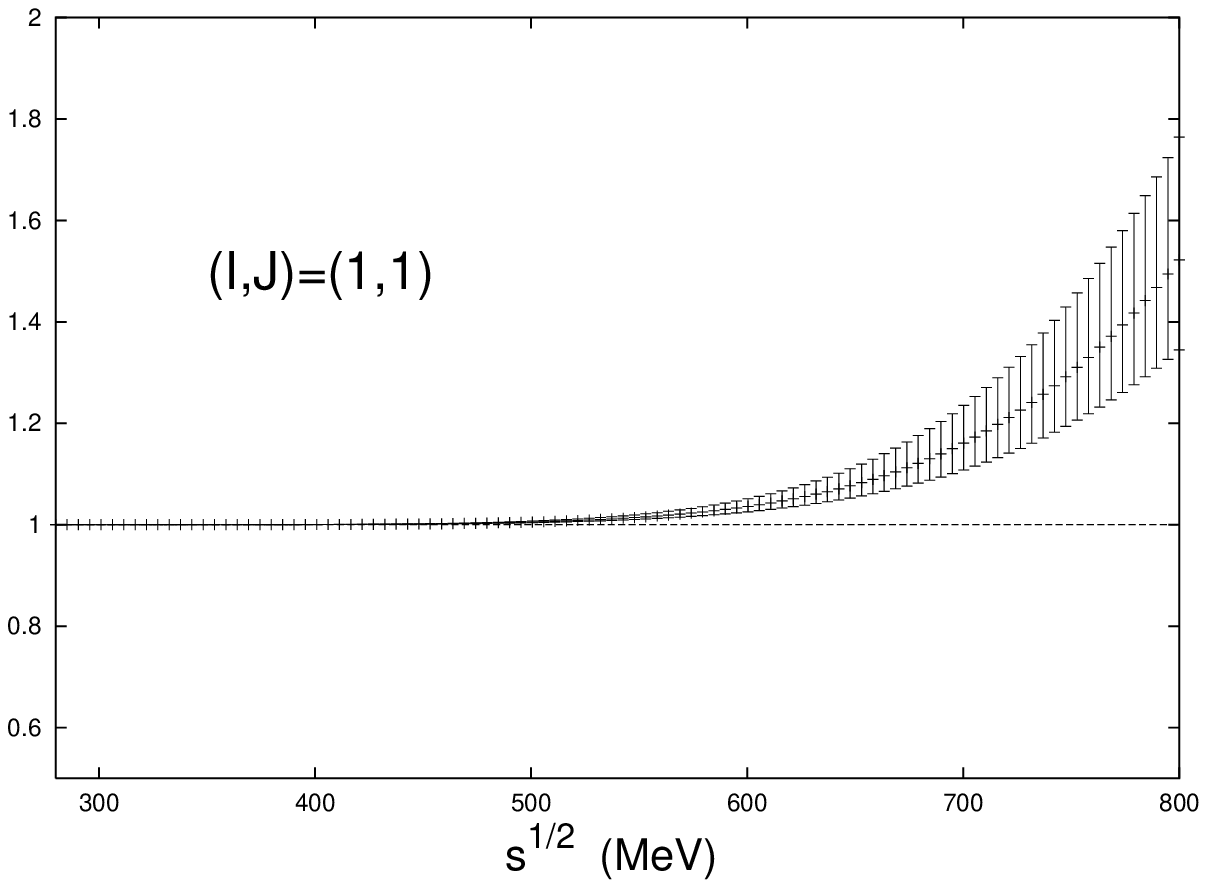,height=5.5cm,width=5.5cm}\epsfig{figure=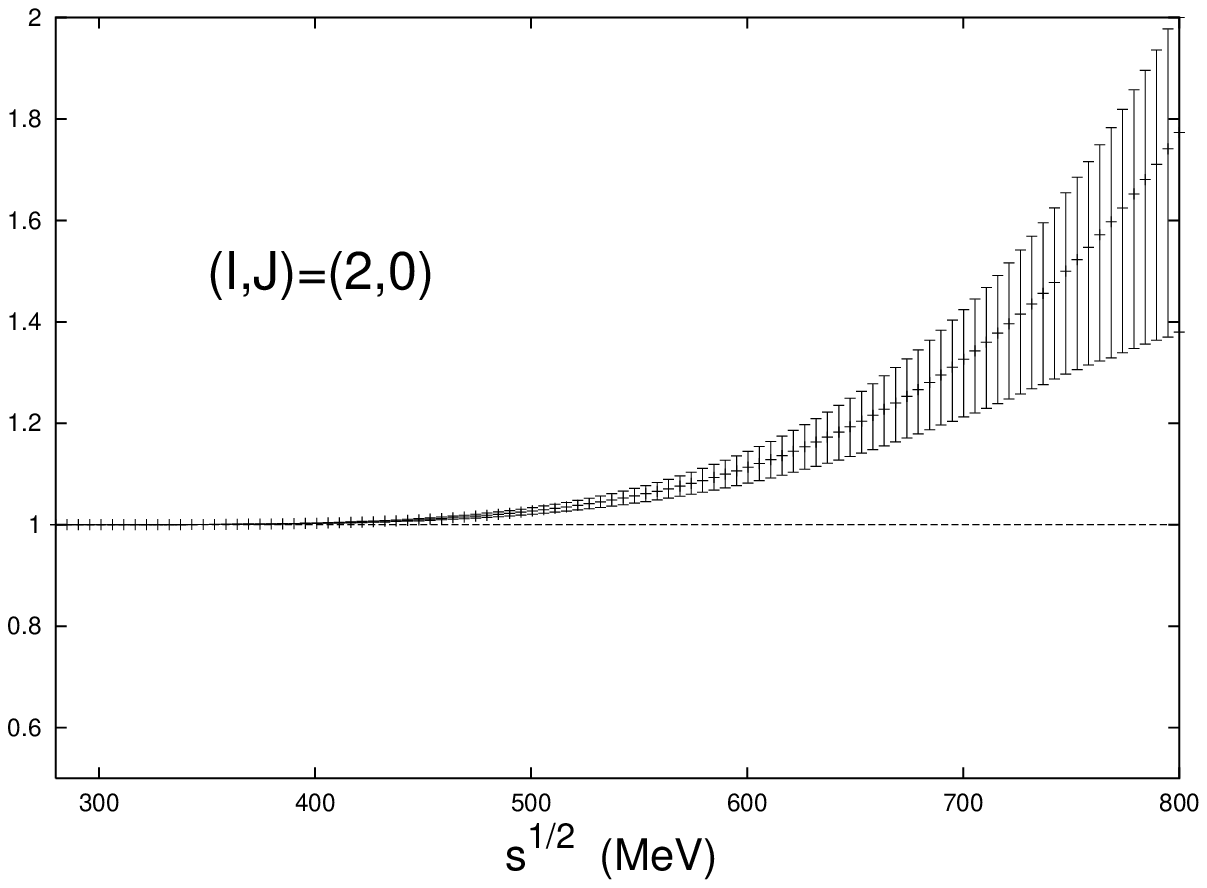,height=5.5cm,width=5.5cm}
\epsfig{figure=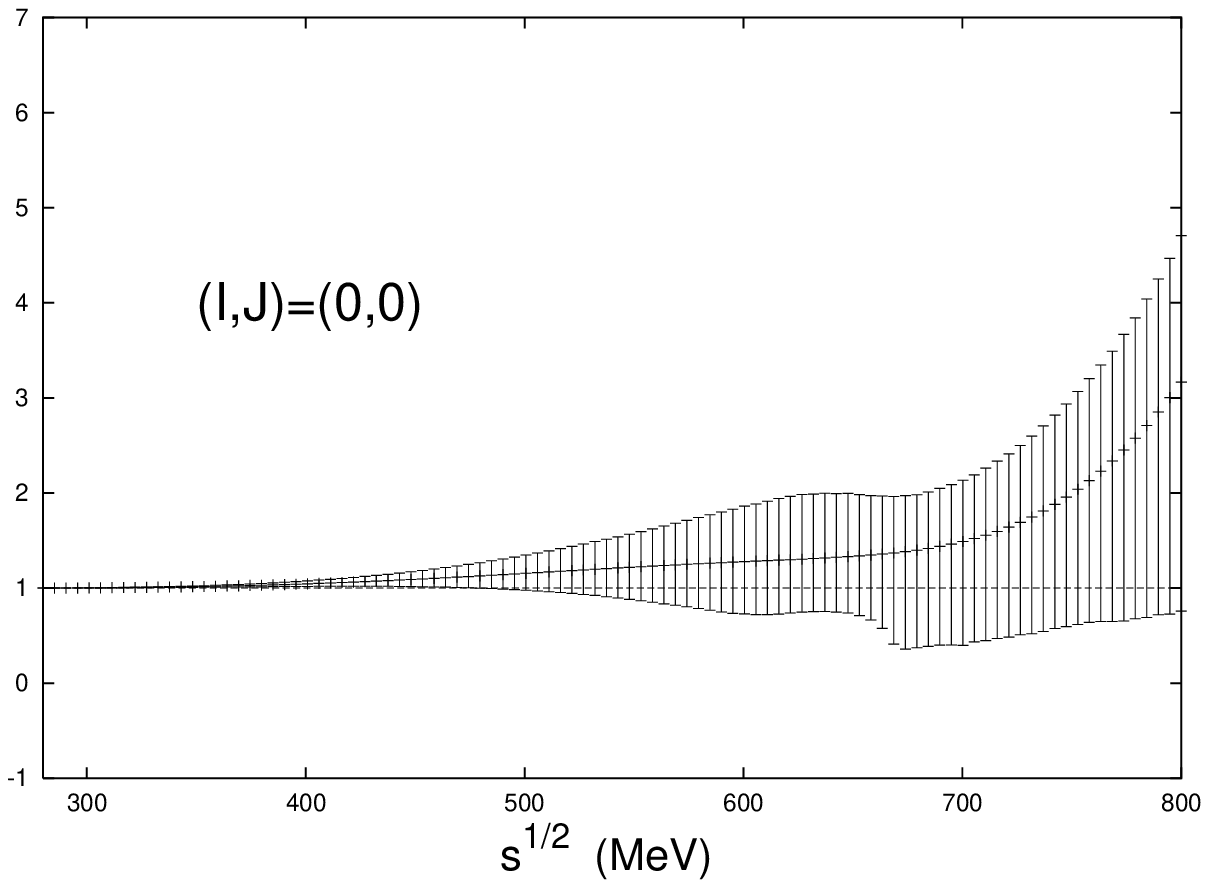,height=5.5cm,width=5.5cm}\epsfig{figure=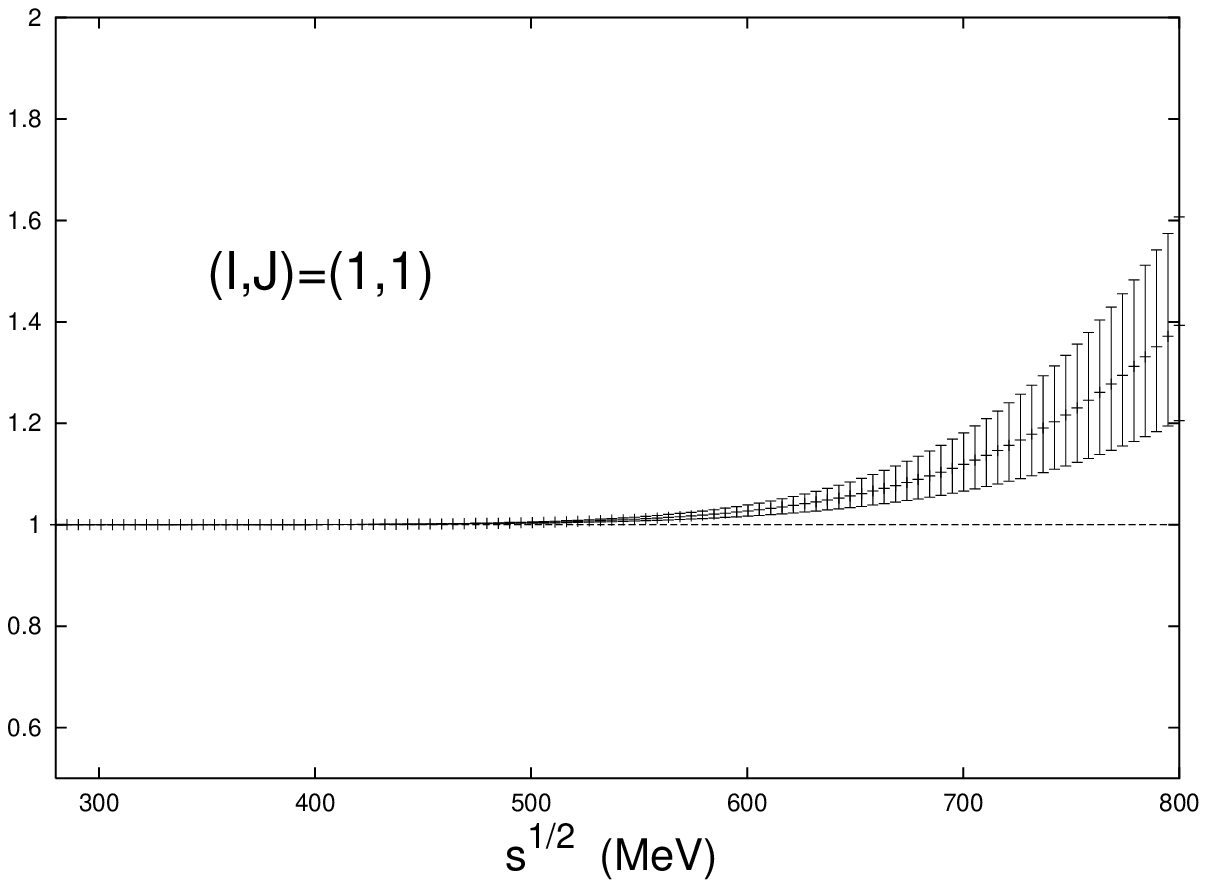,height=5.5cm,width=5.5cm}\epsfig{figure=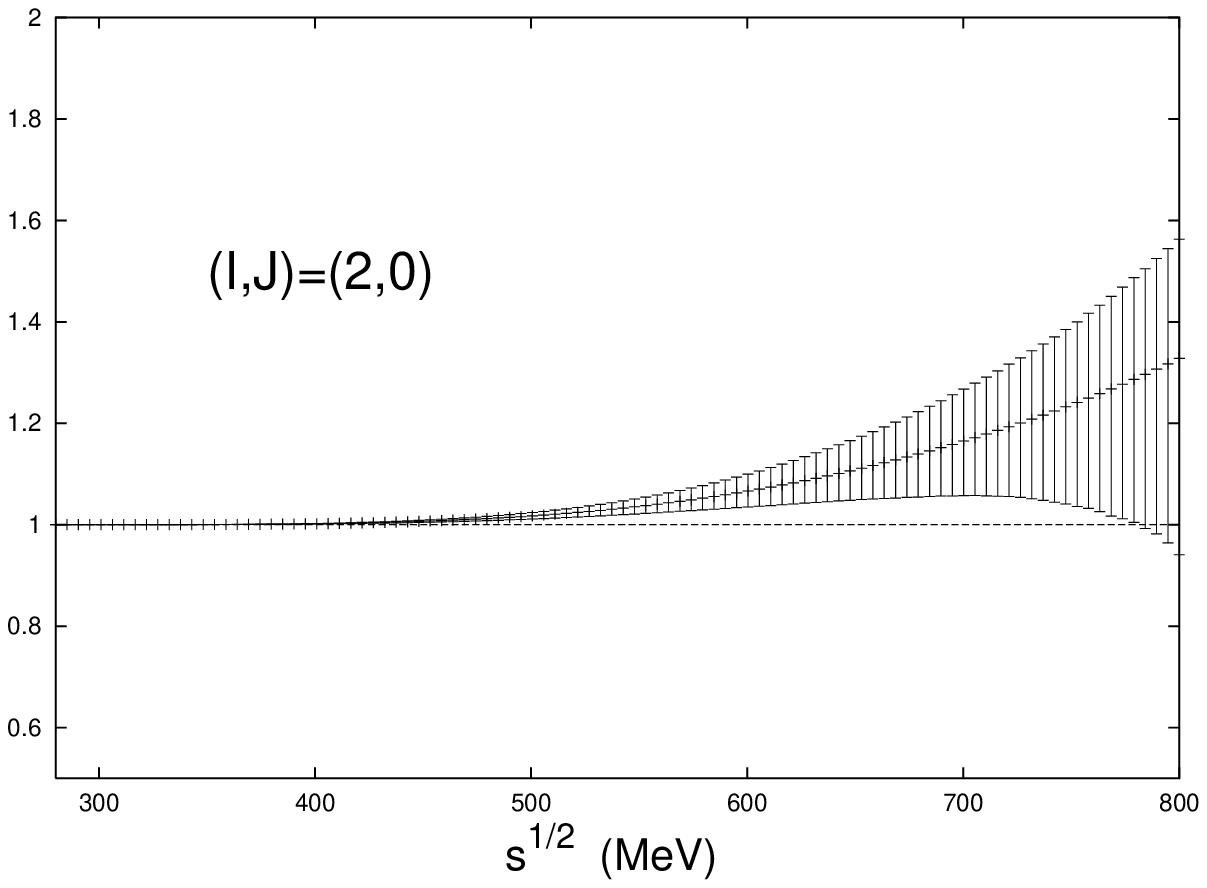,height=5.5cm,width=5.5cm}
\end{center} 
\caption{Unitarity condition for standard NNLO ChPT amplitudes in
$\pi\pi$ scattering for $S-$ and $P-$ waves defined by $U_{IJ} (s) = |
1 + 2 i \sigma(s) t_{IJ} (s) |$. Upper panel: Set {\bf Ic} of
Ref.~\cite{EJ00a}. Lower panel: Set {\bf III} of Ref.~\cite{EJ00a}. In
the calculation the parameters $\bar b_i$ defined in
Eq.~(\ref{eq:b+db}) and given in Table~\ref{tab:bbar} have been
used. The unitarity condition requires $U_{IJ} (s)=1$.}
\label{fig:naive-unitarity}
\end{figure}

\begin{figure}[t]
\begin{center}
\epsfig{figure=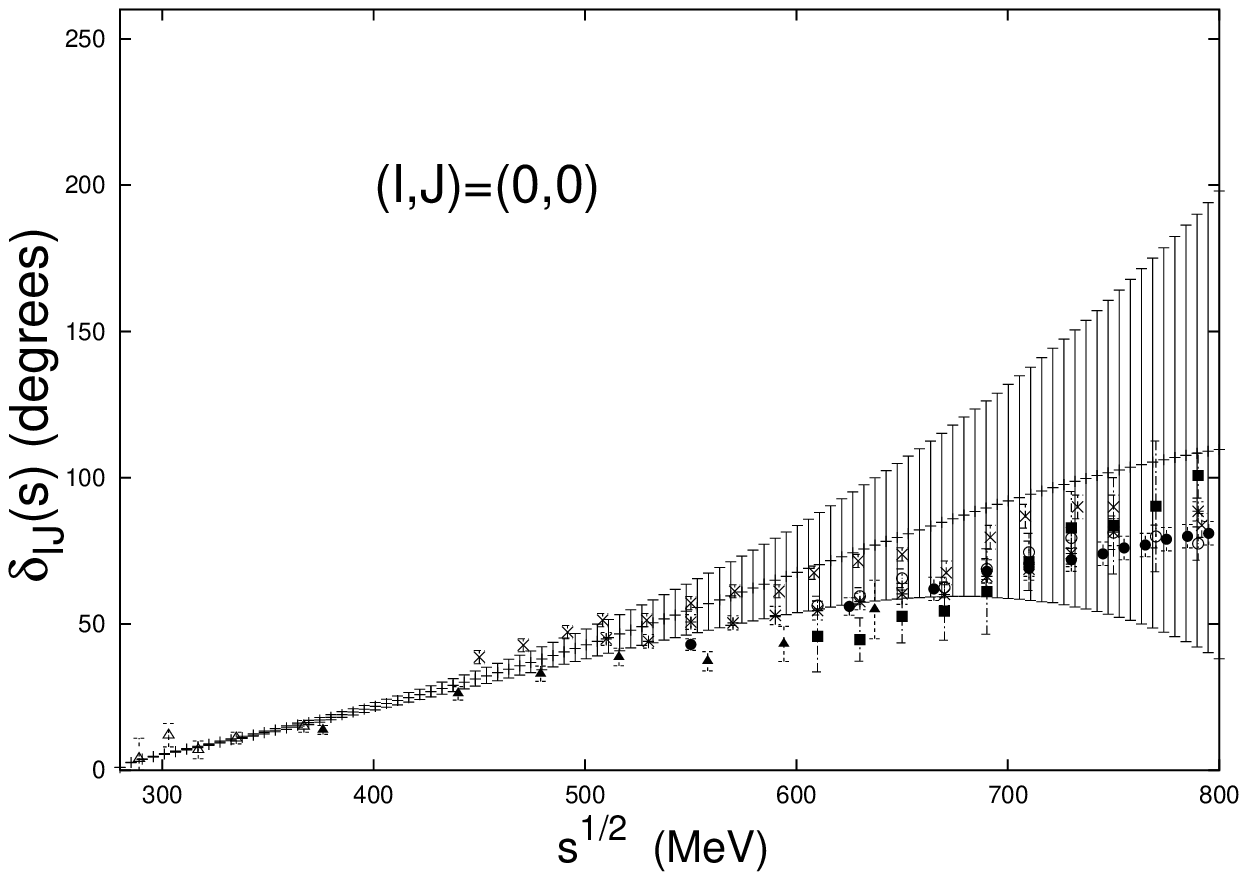,height=5.5cm,width=5.5cm}\epsfig{figure=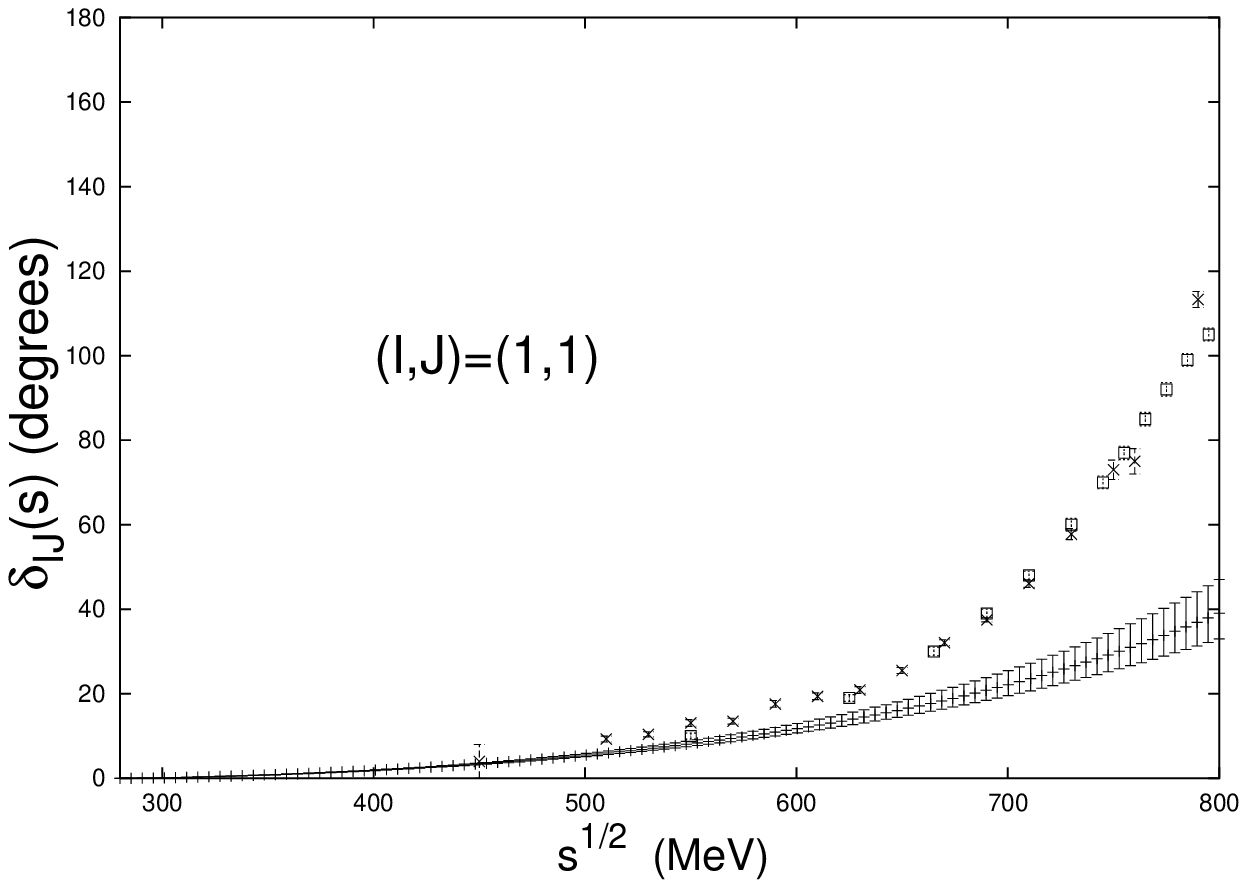,height=5.5cm,width=5.5cm}\epsfig{figure=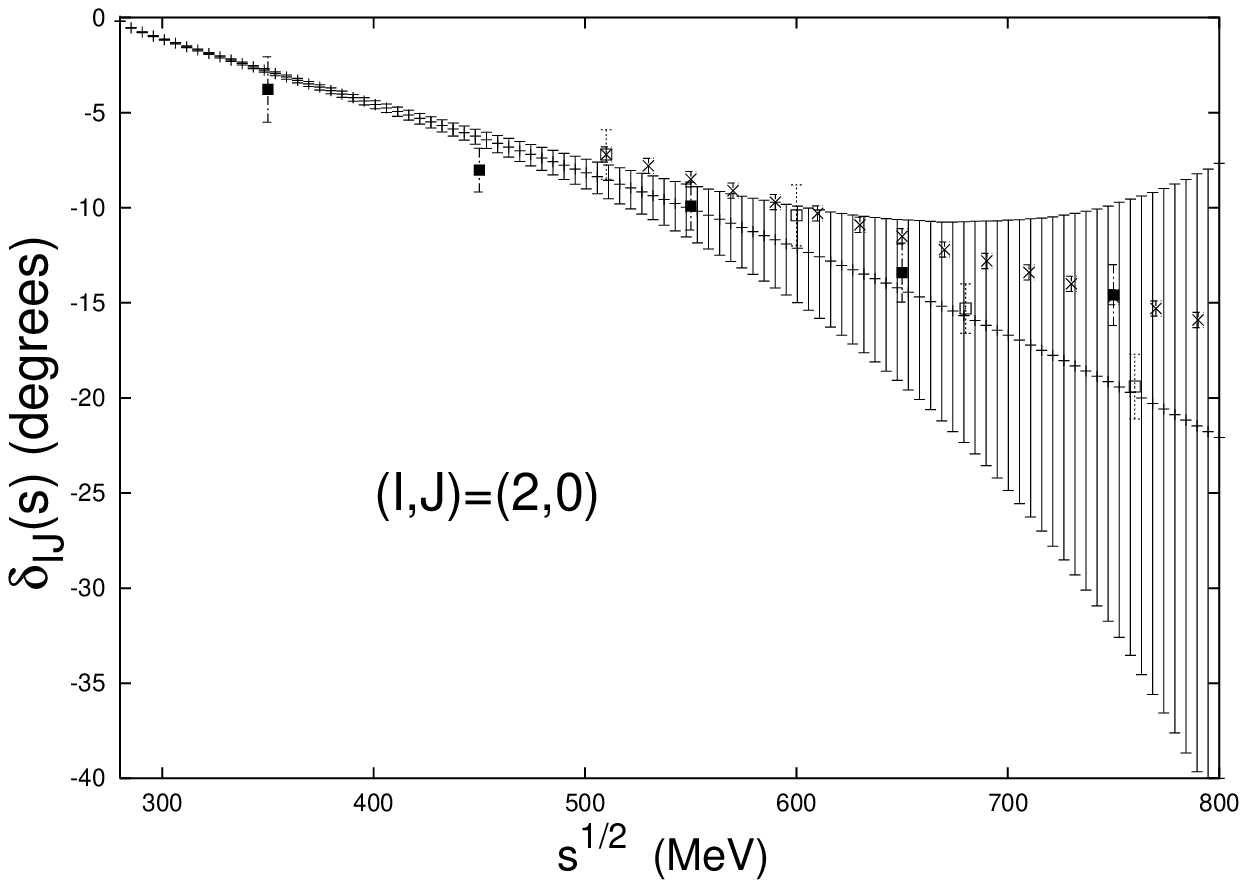,height=5.5cm,width=5.5cm}
\epsfig{figure=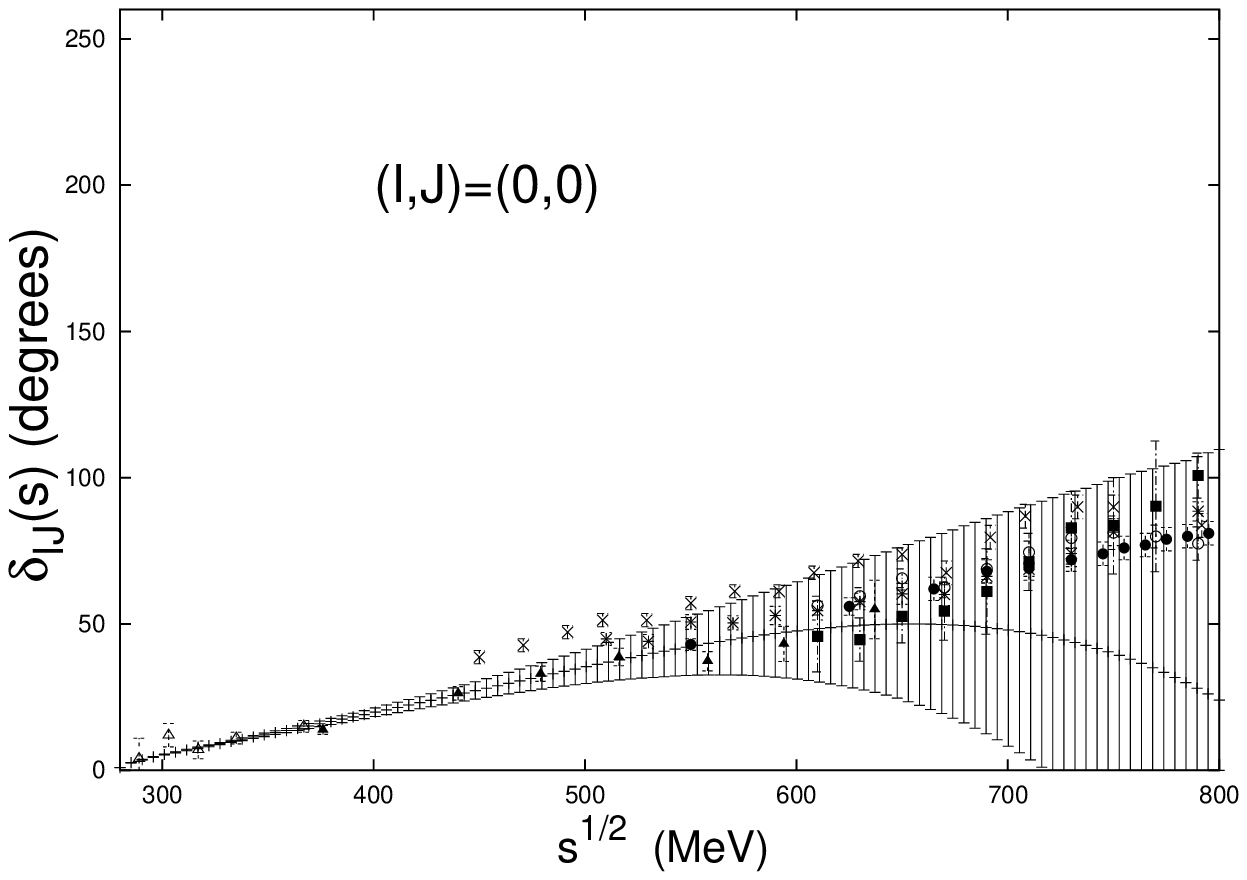,height=5.5cm,width=5.5cm}\epsfig{figure=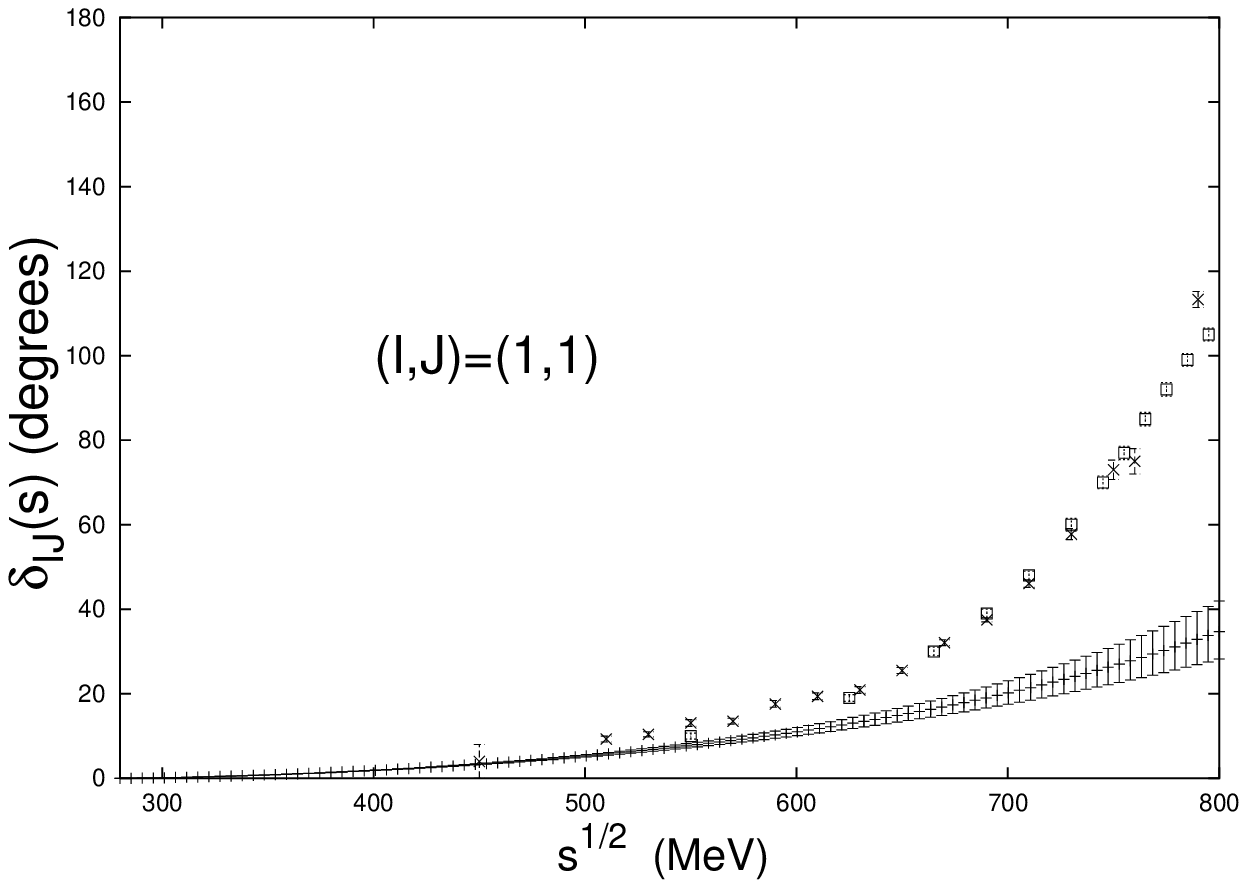,height=5.5cm,width=5.5cm}\epsfig{figure=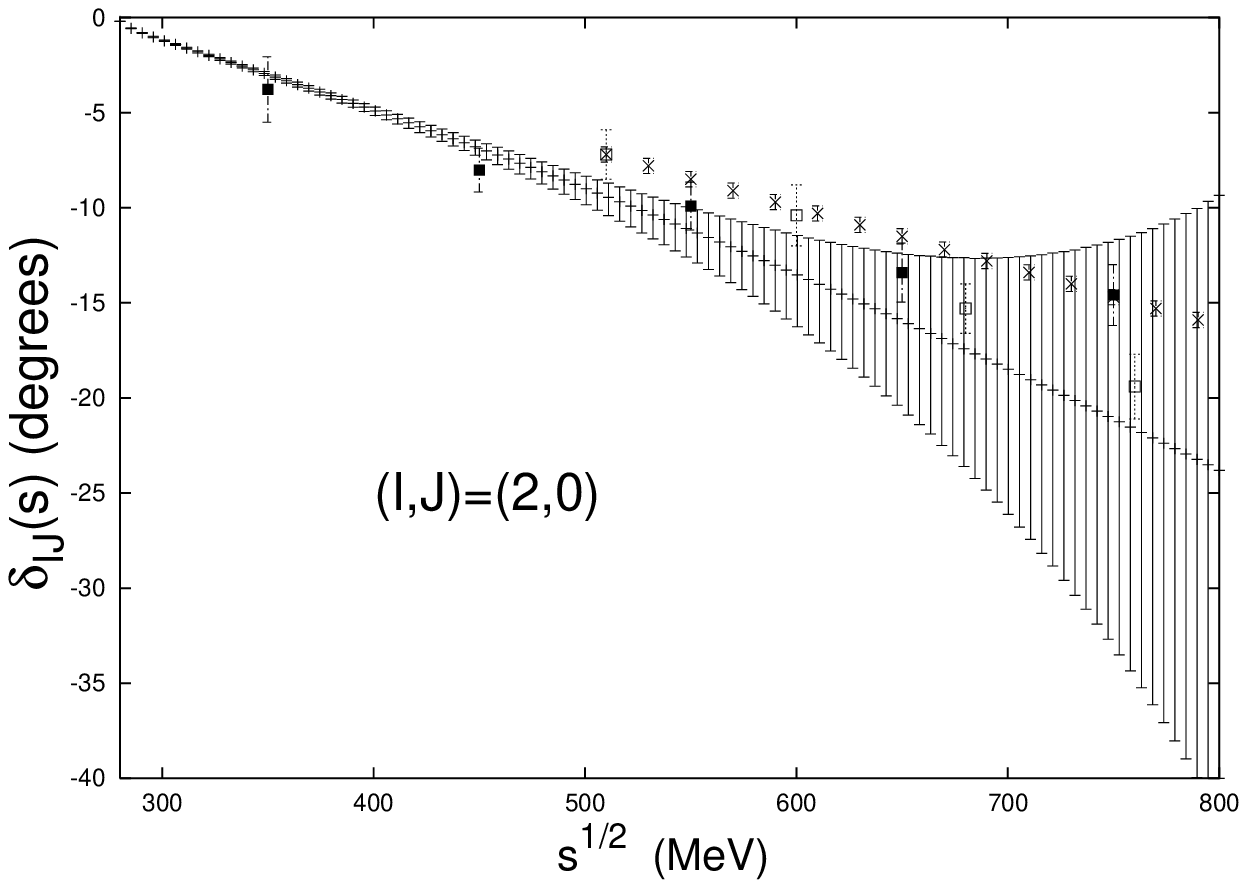,height=5.5cm,width=5.5cm}
\end{center} 
\caption{Standard NNLO ChPT phase shifts (in degrees) for $\pi\pi$
scattering for $S-$ and $P-$ waves after
Eq.~(\ref{eq:delta_chpt}). Upper panel: Set {\bf Ic} of
Ref.~\cite{EJ00a}. Lower panel: Set {\bf III} of Ref.~\cite{EJ00a}. In
the calculation the parameters $\bar b_i$ defined in
Eq.~(\ref{eq:b+db}) and given in Table~\ref{tab:bbar} have been
used. Combined data from Refs.~\cite{pa73}-\cite{fp77}.}
\label{fig:naive-chpt}
\end{figure}

\begin{table}[t]
%\vspace{-0.3cm}
\begin{center}
\begin{tabular}{c|c|c|c|c|c|c}  
 & $ \bar b_1 $ & $\bar b_2 $ & $\bar b_3 $ & $\bar b_4 $ & $\bar b_5$
& $\bar b_6$ \\
 
\hline\tstrut 

Set {\bf Ic \,} & $-11.6 \er{2.4}{2.5}$ &
$11.2 \pm {1.8} $ & $-0.2 \pm {0.3} $ & $0.8 \pm{0.1} $ & $5.7 
\er{3.2}{3.9}$ & $ 2.6 \er{0.8}{1.0} $ \\

Set {\bf III} & $-13.2 \er{2.5}{2.3}$ &
$12.4 \er{1.7}{1.8} $ & $-0.4 \er{0.4}{0.2} $ & $0.74 \pm{0.06} $ & $1.6 
\er{3.7}{3.8}$ & $ 2.0 \er{0.8}{0.9} $ \\

\hline\tstrut 

 & $ \bar b_1^0 $ & $\bar b_2^0 $ & $\bar b_3^0 $ & $\bar b_4^0 $ & $\bar b_5^0$ & $\bar b_6^0 $ \\
 
\hline\tstrut 

Set {\bf Ic \,} & $-9.1 \pm{2}$ &
$8.2 \pm {1.7} $ & $0.3 \pm {0.3} $ & $0.66 \pm{0.07} $ & $5.7 
\er{3.2}{3.9}$ & $ 2.6 \er{0.8}{1.0} $ \\

Set {\bf III} & $-10.7 \er{2.1}{2.0}$ &
$9.8 \pm{1.7} $ & $-0.16 \pm{0.40} $ & $0.58 \pm{0.04} $ & $1.6 
\er{3.7}{3.8}$ & $ 2.0 \er{0.8}{0.9} $ \\

\hline\tstrut 

 & $ \Delta \bar b_1 $ & $ \Delta \bar b_2 $ & $ \Delta \bar b_3 $ & $ \Delta \bar b_4 $ & $ \Delta \bar b_5$
& $\Delta \bar b_6$ \\
 
\hline\tstrut 

Set {\bf Ic \,} & $-2.4 \er{0.5}{0.4}$ &
$3.0 \pm {0.3} $ & $-0.5 \er{0.2}{0.1} $ & $0.19 \pm{0.04} $ &  $0$ & $0$ \\

Set {\bf III} & $-2.4 \er{0.5}{0.4}$ & $2.6 \er{0.4}{0.3} $ & $-0.4
\pm{0.2} $ & $0.15 \er{0.15}{0.04} $ & $0$ & $0$ \\

\end{tabular}
\end{center}
\caption{\footnotesize Two loop $\bar b_{1,2,3,4,5,6} $ low energy
parameters in ChPT for the parameter Sets {\bf Ic} and {\bf III} of
Ref.~\cite{EJ00a} used in the present paper. The relation to the
$b_{1,2,3,4,5,6}$ parameters used in the work is $\bar b_i = 16\pi^2
b_i$. We also show explicitly the decomposition $\bar b_i = \bar b_i^0
+ \Delta b_i $ referred to in Eq.~(\ref{eq:b+db}).}
\label{tab:bbar} 
\end{table}

From the formulas above, direct application of ChPT requires a
scrupulous separation between different chiral orders.  As we have
already mentioned above, the NLO amplitudes $t_{IJ}^{(4)}(s)$ depend
linearly on four dimensionless parameters $\bar l_{1,2,3,4} $. These
parameters are supposed to be independent of $f_\pi$ and $m_\pi$ and
therefore they are zeroth order in the chiral
expansion\footnote{Practical calculations require, however, a
truncation of the chiral expansion and confrontation to experimental
data, and hence some higher order systematic uncertainties remain in
the one loop parameters besides the experimental
uncertainties}. Finally, at NNLO the amplitude can be expressed in
terms of six independent parameters $\bar b_{1,2,3,4,5,6}$. As it has
been shown in the original two loop calculation works
\cite{mksf95,bc97} these $\bar b-$coefficients contain a zeroth order
piece and a second order piece
\begin{eqnarray}
\bar b_i = \bar b_i^0  +  \Delta \bar b_i 
\label{eq:b+db} 
\end{eqnarray} 
This makes the separation a bit subtle because, when plugged into
Eq.(\ref{eq:chpt}), a unwanted eighth order correction is
induced. From the point of view of ChPT this contribution has to be
dropped out as far as the complete eighth order calculation is not
available. On the other hand, these higher order corrections are
numerically small as can be deduced from Table \ref{tab:bbar}.

\subsection{Numerical results} 

The CM energy dependent figures with error-bars presented in this work
are generated as follows. If we have an energy dependent function,
$F$, in terms of a set of random parameters, $(a_1 \dots a_n)$,
distributed according to some statistical law, a random variable for
any fixed value of $s$, $F(s;a_1 , \dots , a_n )$ is generated.
Obviously, for a non-linear parameter dependent function the mean
value of the curve is not equal to the function of the mean values, $
\langle F(s;a_1 , \dots , a_n ) \rangle \neq F(s; \langle a_1 \rangle,
\dots , \langle a_n \rangle )$. There is nothing wrong with this and
one could simply bin the distributions for any fixed $s$
value. Nevertheless, to make the results a bit more portable we wish
to quote such a function of the mean values, $F(s; \langle a_1
\rangle, \dots , \langle a_n \rangle )$ as our central curves. To
assign an upper and lower error bar (the distribution may in general
be asymmetric) relative to this central value, we bin the distribution
and firstly exclude the $16 \%$ top values and the $16 \%$ bottom
values of the distribution. The remaining bins comprise the $68 \%$ of
the distribution values, the distance from the upper and lower values
to our central value provide the upper and lower error-bars
respectively. Evidently, our bands correspond to a $68 \%$ confidence
level. This procedure of assigning errors fails for extremely
asymmetric distributions, such that the central value turns out to be
within the discarded upper or lower $16 \%$ intervals. Although we
find that this situation seldom takes place in our calculation, in
such a case we proceed in a different way. We first discard the $16
\%$ upper and lower intervals and then compute the arithmetic mean,
which we assign as the central value. To control on the quality of
this second definition of central value we always compute, whenever
possible, both definitions and find that the differences are
numerically less significant than the error-bars.

The results for the unitarity condition, $U_{IJ} (s) $ as defined in
Eq.~(\ref{eq:uni_viol}), can be seen in Figs.~\ref{fig:unitarity} and
\ref{fig:naive-unitarity} in terms of the $\bar b_i^0 $ and the $\bar
b_i = \bar b_i^0 + \Delta \bar b_i $ coefficients respectively and for
the parameter Sets {\bf Ic} and {\bf III}. As we see, standard ChPT
theory violates unitarity in a systematic manner well below the
resonance region including uncertainties for Set {\bf Ic}. For Set
{\bf III} only the $P-$wave exhibits this behaviour, due to large
uncertainties in the $S-$wave unitarity violation. The perturbatively
defined phase-shifts, $\delta_{IJ}^{\rm ChPT} (s)$ as given by
Eq.~(\ref{eq:delta_chpt}) can be seen in Figs.~\ref{fig:chpt} and
\ref{fig:naive-chpt} in terms of the $\bar b_i^0 $ and the $\bar b_i =
\bar b_i^0 + \Delta \bar b_i $ coefficients respectively and for the
parameter Sets {\bf Ic} and {\bf III}. As we see from the figures, the
perturbatively defined phase shifts seem compatible with experimental
data whenever the corresponding unitarity condition is compatible
within uncertainties with elastic unitarity. Let us remind that
threshold parameters for Set {\bf III} have uncertainties similar or a
bit smaller than Set {\bf Ic} \cite{EJ00a}. This also holds for higher
energies but there appear some systematic discrepancies with the data,
slightly favouring Set {\bf III}.

\begin{table}[t]
%\vspace{-0.3cm}
\begin{center}
\begin{tabular}{c|c|c}  
$s_A / m_\pi^2$  & $(I,J)=(0,0)$ & $ (I,J)=(2,0) $ \\ 

\hline\tstrut 

ChPT-{\bf Ic \,} Fig.~\ref{fig:chpt} & $0.38(6)$ & $2.03(5)$  \\

ChPT-{\bf III} Fig.~\ref{fig:chpt} & $0.43(6) $ &  $2.00(5)$  \\

\hline\tstrut 

ChPT'-{\bf Ic \,} Fig.~\ref{fig:naive-chpt} & $0.38(6)$ & $2.03(5)$  \\

ChPT'-{\bf III} Fig.~\ref{fig:naive-chpt} & $0.43(6) $ & $2.00(5) $  \\

\end{tabular}
\end{center}
\caption{\footnotesize Non-kinematical Adler zeros for $S-$wave $I=0$
and $I=2$ amplitudes for ChPT and the parameter sets {\bf Ic} and {\bf
III} of Ref.~\cite{EJ00a}. We also indicate the phase-shift figures
which correspond to these zeros. Errors are given in brackets.}
\label{tab:adler} 
\end{table}

Threshold parameters for Set {\bf Ic} and Set {\bf III} at the two
loop level have been computed in our previous work \cite{EJ00a}.
There, a separation of the tree-level, one loop and two loop
contributions with their corresponding error estimates have been
presented. 

\begin{figure}[t]
\begin{center}
\epsfig{figure=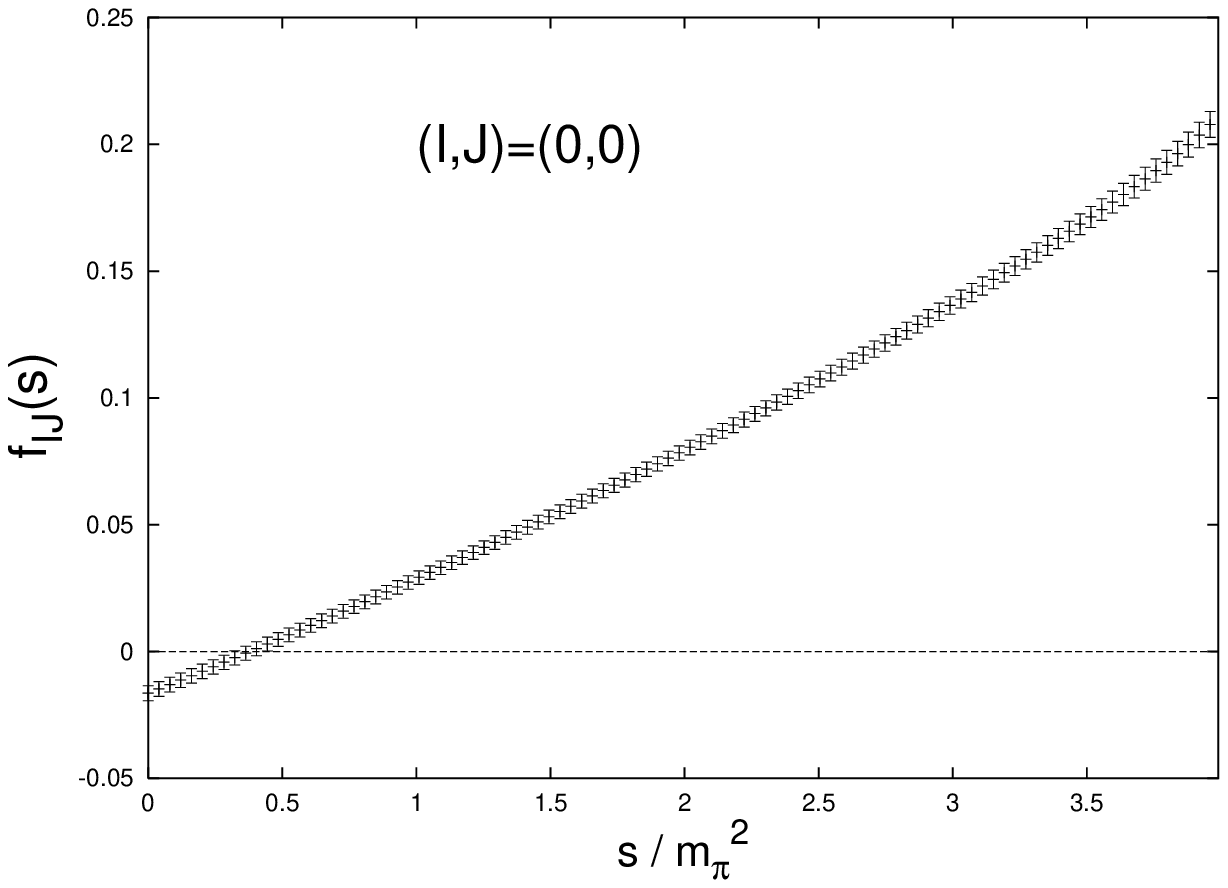,height=5.5cm,width=5.5cm}\epsfig{figure=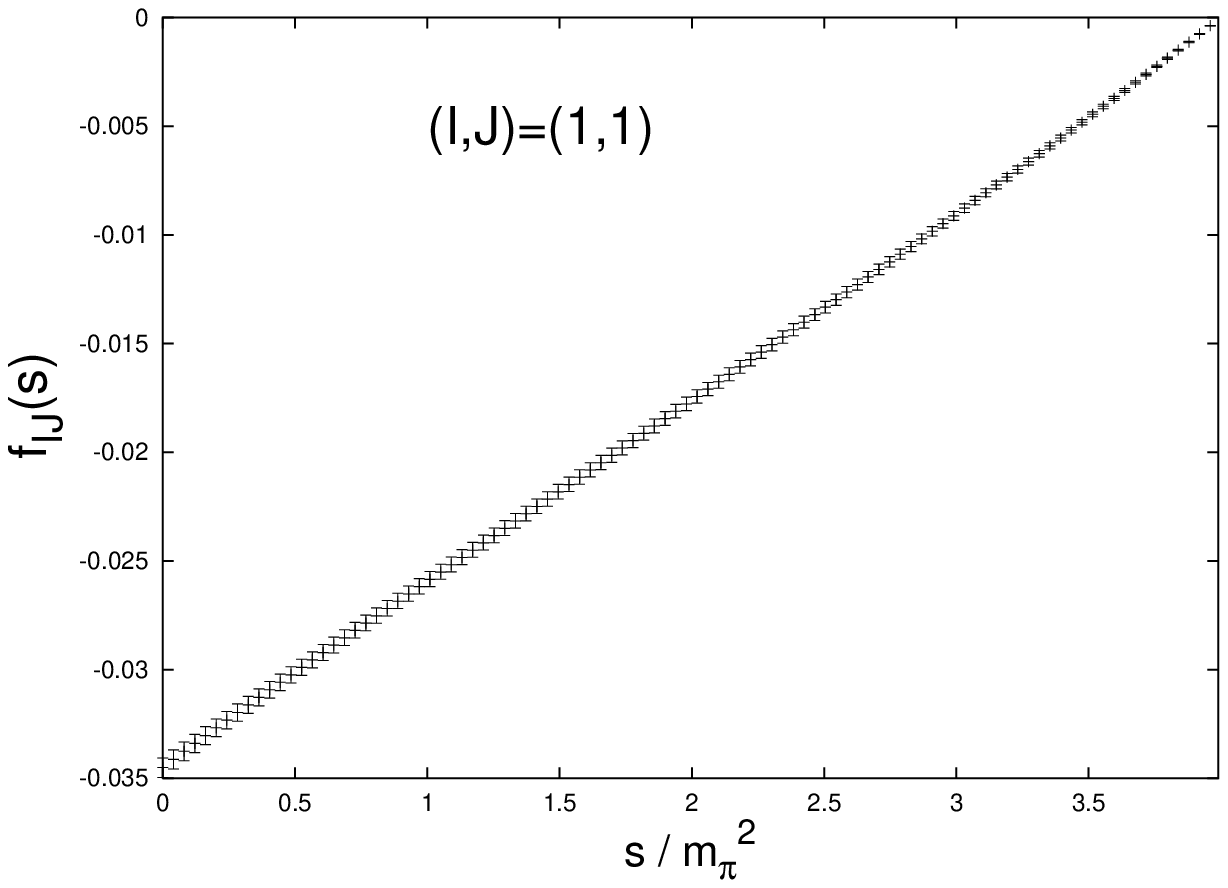,height=5.5cm,width=5.5cm}\epsfig{figure=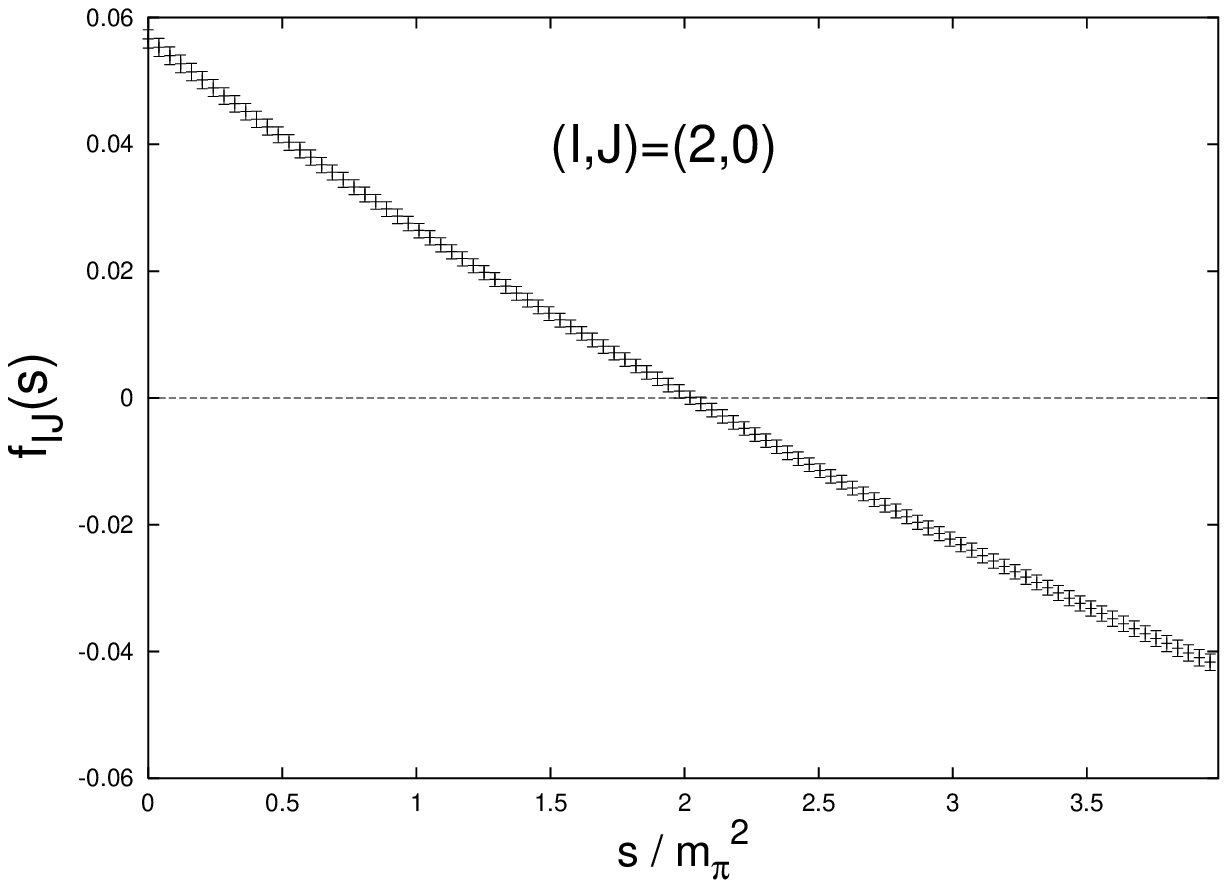,height=5.5cm,width=5.5cm}
\epsfig{figure=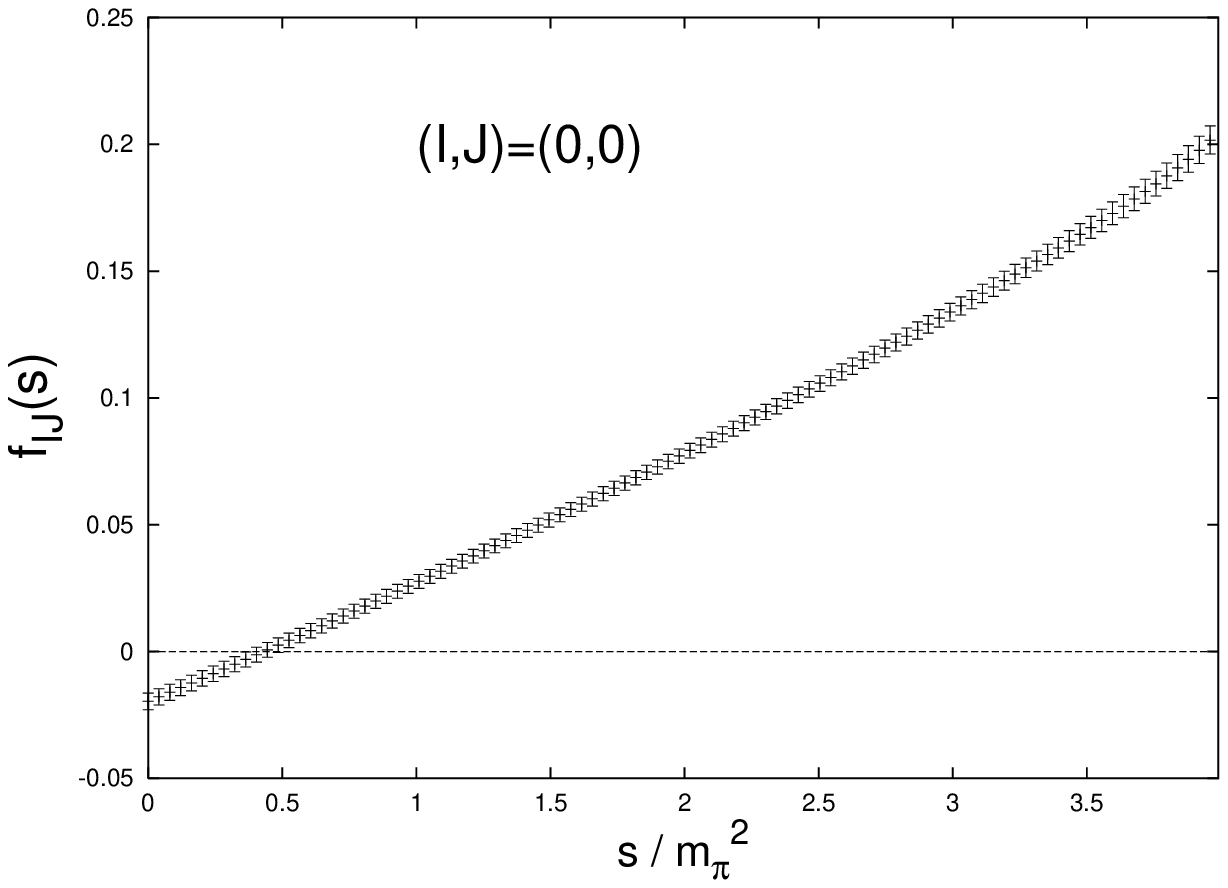,height=5.5cm,width=5.5cm}\epsfig{figure=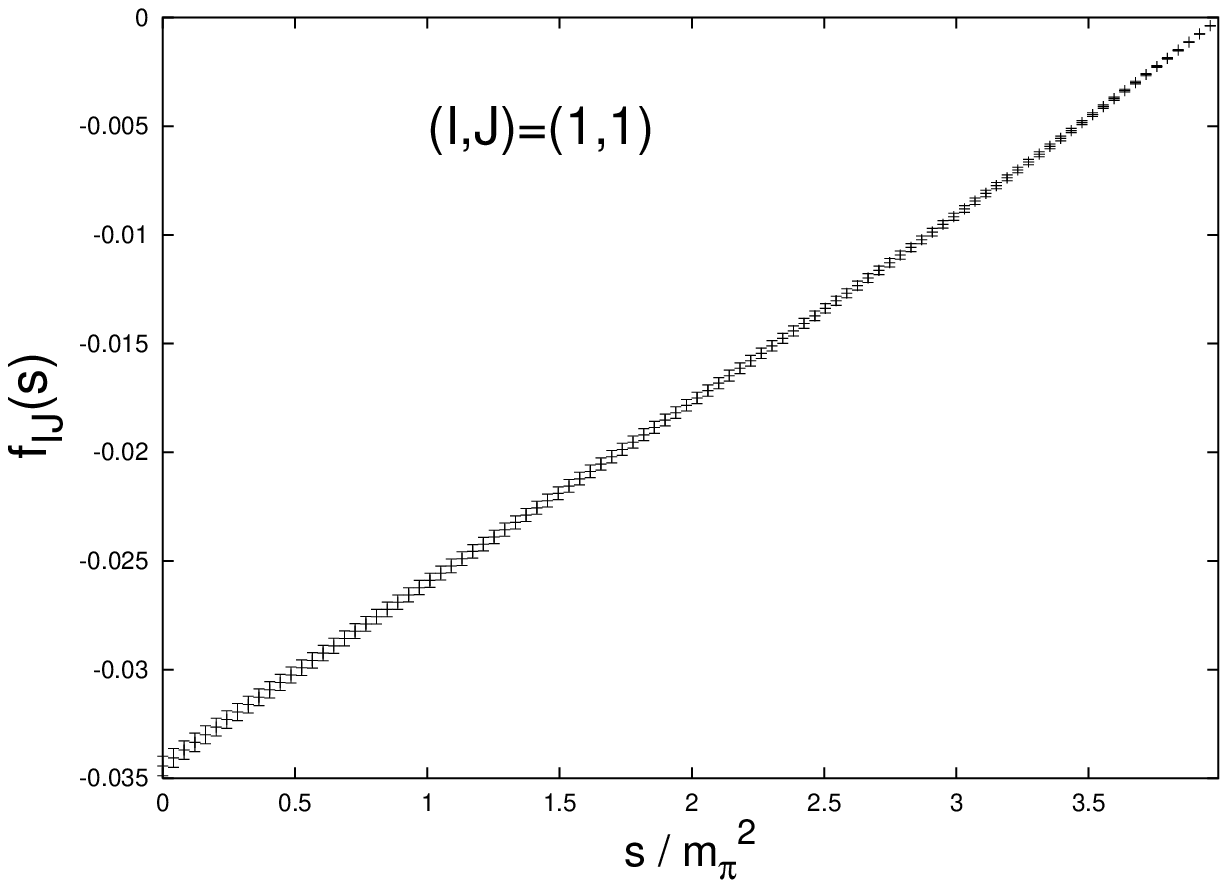,height=5.5cm,width=5.5cm}\epsfig{figure=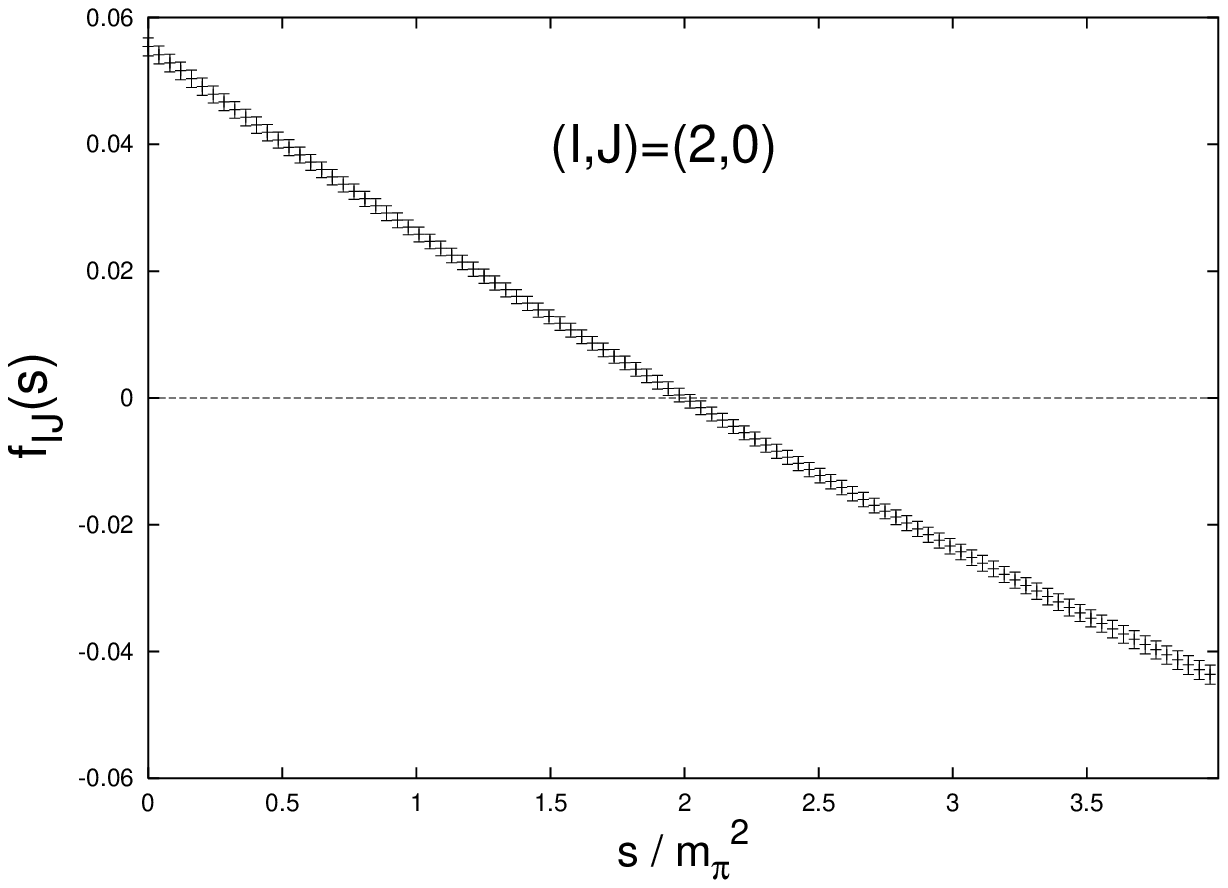,height=5.5cm,width=5.5cm}
\end{center} 
\caption{$S-$ and $P-$ partial wave amplitudes $ f_{IJ} (s) = \sqrt{s}
\, t_{IJ}(s) $ (in fm) for $\pi\pi$ scattering in standard ChPT to two
loops in the unphysical region, $ 0 \le s \le 4 m_\pi^2$. Upper panel:
Set {\bf Ic} of Ref.~\cite{EJ00a}. Lower panel: Set {\bf III} of
Ref.~\cite{EJ00a}. Normalization is such that at $s=4m_\pi^2$ one has
for the $S-$ waves the scattering length. In the scattering region ( $
s > 4 m_\pi^2 $ ) this figure corresponds to Fig.~\ref{fig:chpt}.}
\label{fig:unphys.chpt}
\end{figure}

The partial $S-$ and $P-$ wave amplitudes in the unphisical region
below threshold and above the left cut, $ 0 \le 0 \le 4 m_\pi^2 $, are
depicted in Fig.~\ref{fig:unphys.chpt}. In this region partial wave
amplitudes are real and present real single zeros. The single zero at
$4 m_\pi^4$ in the $P-$ wave is of kinematical origin as can be seen
from Eq.~(\ref{eq:threshold}). However, zeros in the $S-$waves are
dynamical consequences of chiral symmetry. As we see, the agreement
between the parameter Sets {\bf Ic} and {\bf III} is very good within
uncertainties. The two-loop location of Adler zeros with error
estimates is given in Table~\ref{tab:adler}. The additive structure of
the ChPT amplitude makes a numerically small distinction between
making the separation $\bar b_i= \bar b^0_i + \Delta \bar b_i$. As we
see, the isotensor $S-$wave chiral zero does not move within
uncertainties from its tree level value of Eq.~(\ref{eq:adler0}) both
for Set {\bf Ic} and Set {\bf III}. For the parameter Set {\bf III},
the two-loop shift of the $S-$wave isoscalar Adler zero this is almost
the case. Nevertheless, for parameter Set {\bf Ic} there is a
systematic shift of about $20 \%$.

\section{IAM to two loops and perturbative matching} \label{sec:iam}

\subsection{Algebraic derivation} 

The idea of the method is quite simple and we review it here for the
sake of completeness.  If instead of expanding the amplitude, one
considers the inverse amplitude and expands according to the chiral
expansion (assuming $t^{(2)}_{IJ} (s) \neq 0 $ ), one gets
\begin{eqnarray} 
{1\over t_{IJ} (s) } = \sigma(s) \cot \delta_{IJ}^{IAM} (s) -i
\sigma(s) = {1\over  t^{(2)}_{IJ} (s)} - { t^{(4)}_{IJ} (s)
\over t^{(2)}_{IJ} (s)^2 } +  \left[ { t^{(4)}_{IJ} (s)^2
\over t^{(2)}_{IJ} (s)^3} - {t^{(6)}_{IJ} (s) \over t^{(2)}_{IJ}
(s)^2} \right] + \dots
\label{eq:iam} 
\end{eqnarray} 
One may check that the unitarity relation, Eq.~(\ref{eq:unitarity}),
is exactly preserved due to the perturbative relations of
Eq.~(\ref{eq:pert-uni}).  Note that a direct application of the IAM
including up to two loops can only unitarize $S-$ and $P-$ waves. To
unitarize $D-$ waves a three loops calculation would be needed. Since
such a calculation has not yet been done, we will restrict to $S-$ and
$P-$ waves in the present work. The structure of Eq.~(\ref{eq:iam})
makes possible to have poles in the second Riemann sheet, i.e. zeros
of $t^{-1}(s)$, but this is done at the expense of some fine-tuning
between several orders. Actually, the IAM method assumes that the
inverse amplitude, $1/t_{IJ}(s)$, is small, which is particularly true
in the neighbourhood of a resonance.

\subsection{IAM Phase-shifts} 

The best way to quantify the goodness of a unitarization scheme such
as the IAM is to check whether or not the information contained in the
low energy parameters, in conjunction with the unitarized amplitude
given by Eq.~(\ref{eq:iam}), predicts within acceptable errors the
phase shifts in the region above the threshold. To proceed further we
have to fix in some way our sets of parameters. An alternative, and
actually complementary point of view is to make a direct fit to the
data. Unfortunately, this involves a 10-parameter fit, and moreover
there are some parameters, like for instance $\bar l_3$ and $\bar
l_4$, for which $\pi\pi$ scattering is not very sensitive. Actually,
as it has been recognized in Ref.~\cite{CGL01} there are two kinds of
low energy parameters according to the properties of the corresponding
terms in the partial wave amplitudes,
\begin{itemize}
\item Class A {\it Terms that survive in the chiral
limit} comprising $\bar l_1$, $\bar l_2$, $r_5$ and $r_6$
\item Class B {\it Symmetry breaking terms} corresponding to the remaining low
energy parameters $\bar l_3$, $\bar l_4$, $r_1$, $r_2$, $r_3$ and
$r_4$.
\end{itemize} 
where the $r_i$ parameters determine the pure two loop contribution to
the amplitude.  On the basis that chiral symmetry breaking is a small
effect, we expect a higher sensitivity of the scattering data to
variations of the class A parameters.

\subsubsection{Naive scheme} 

\begin{figure}[t]
\begin{center}
\epsfig{figure=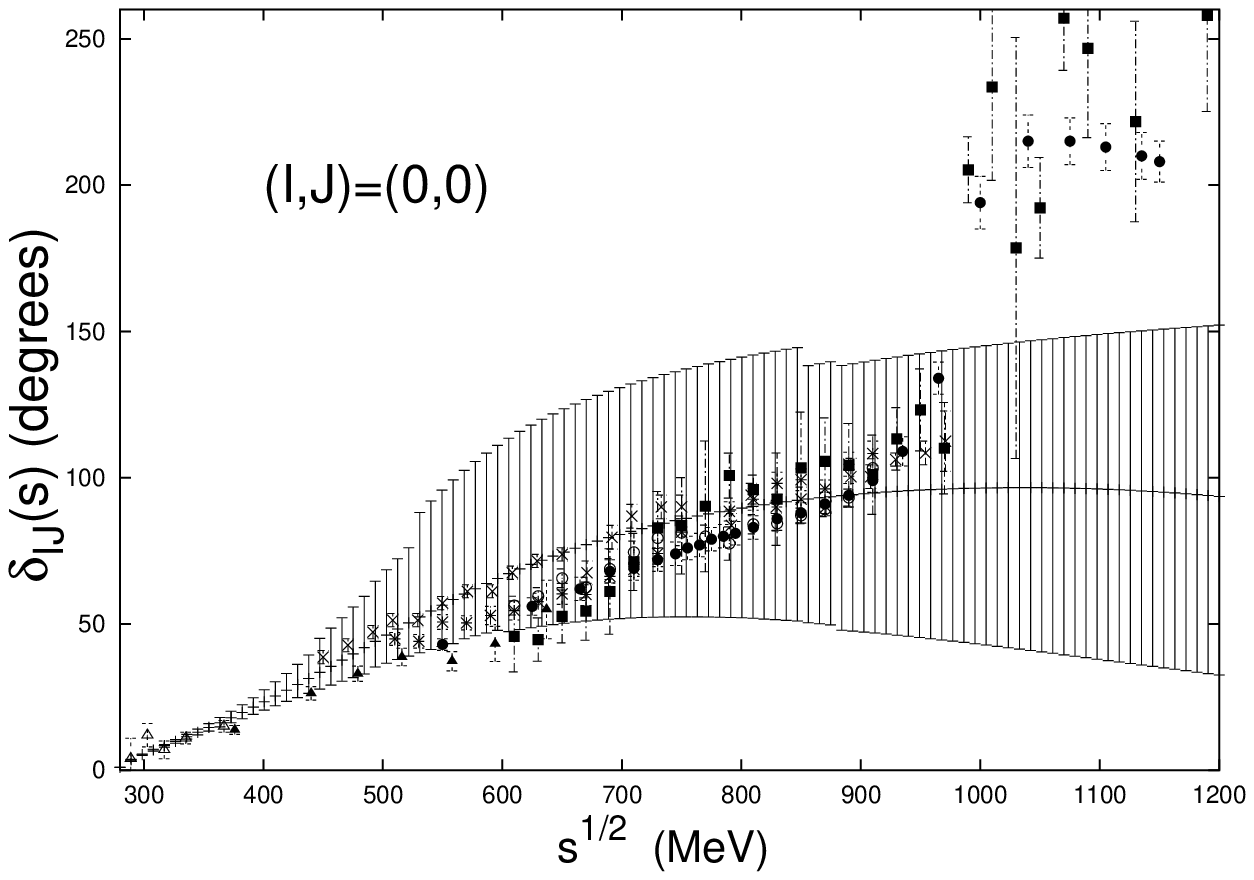,height=5.5cm,width=5.5cm}\epsfig{figure=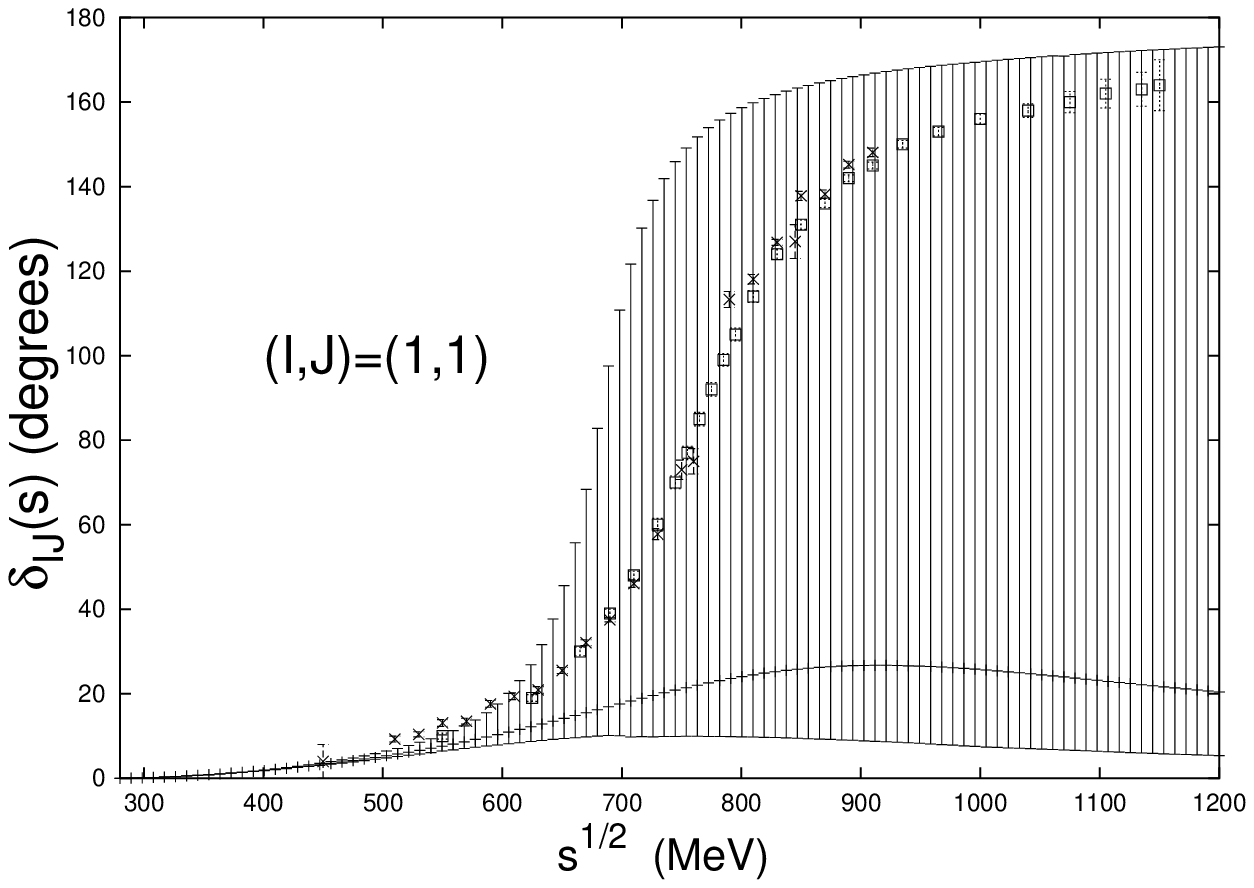,height=5.5cm,width=5.5cm}\epsfig{figure=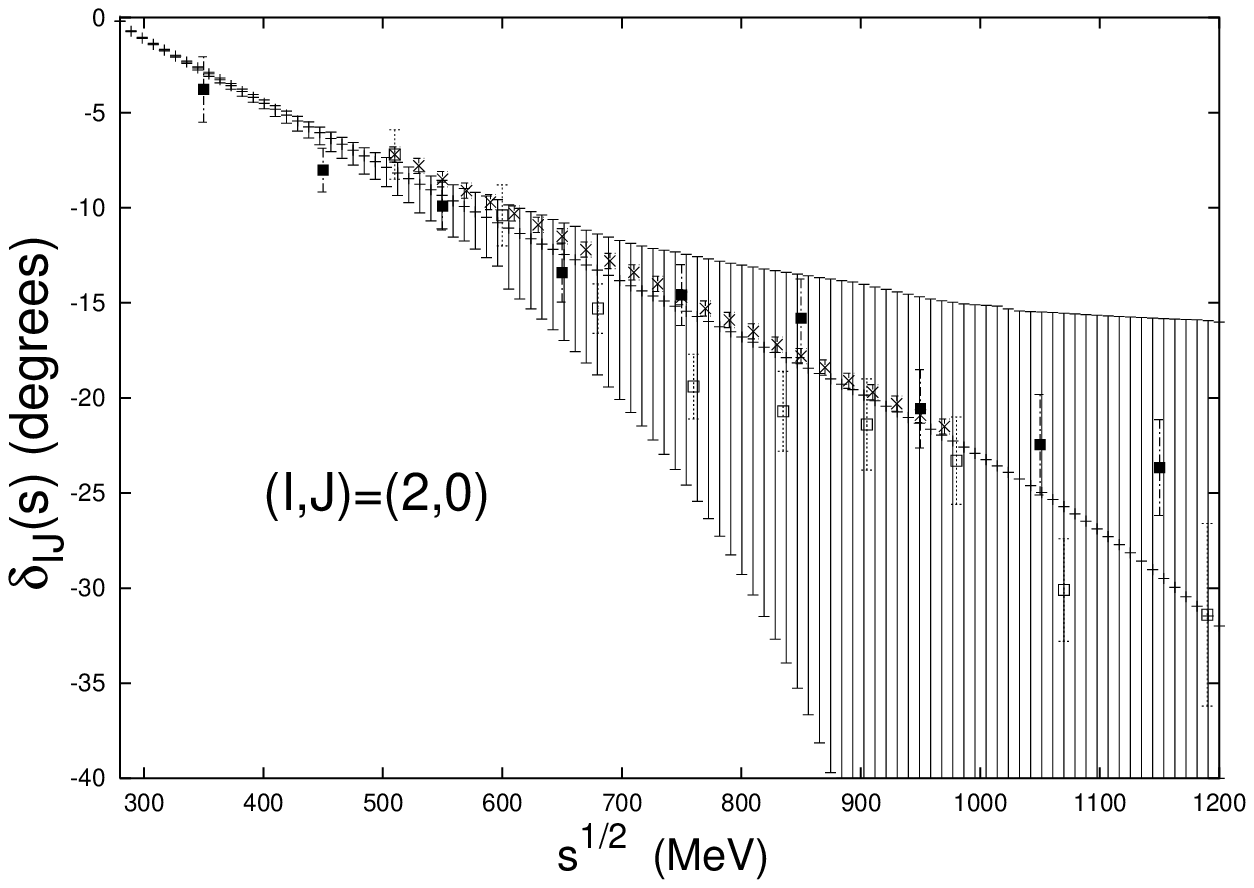,height=5.5cm,width=5.5cm}
\epsfig{figure=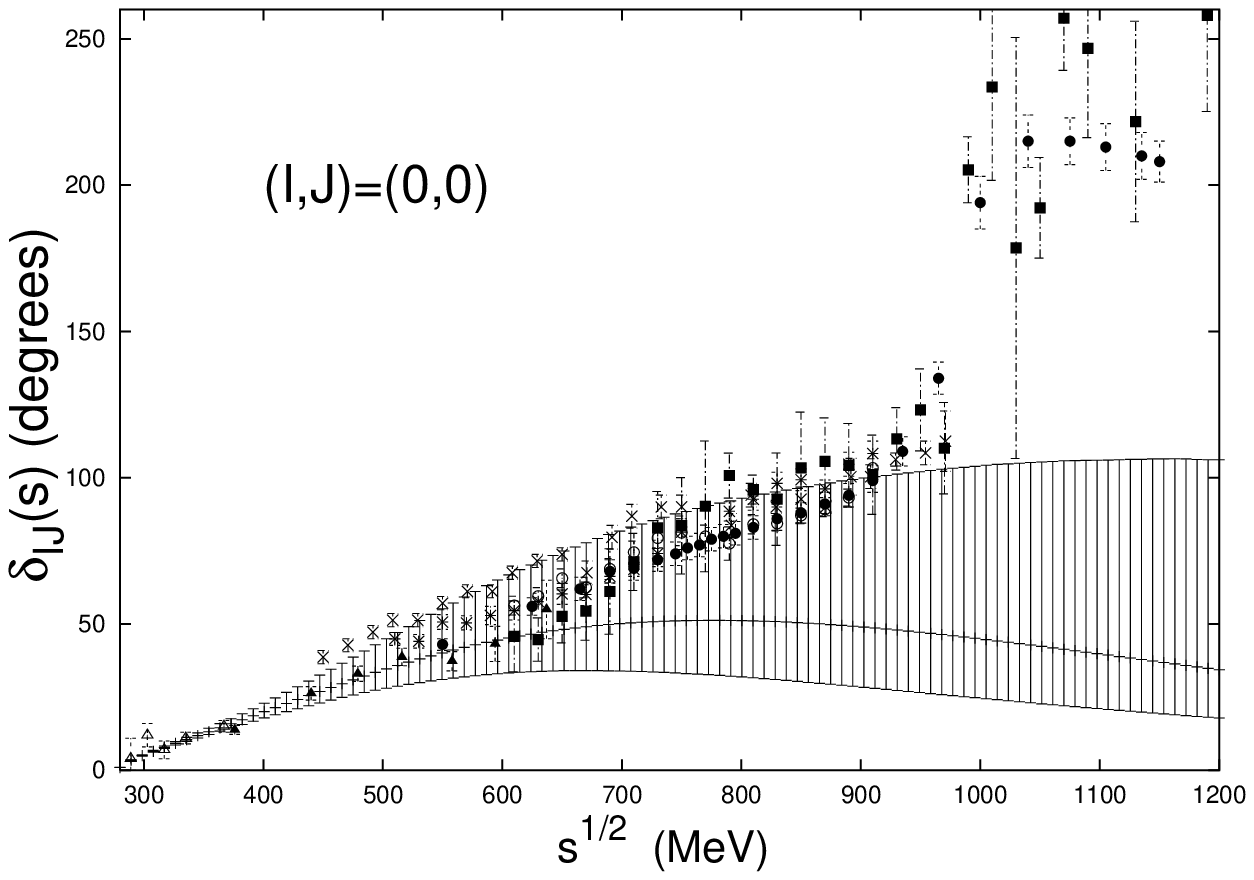,height=5.5cm,width=5.5cm}\epsfig{figure=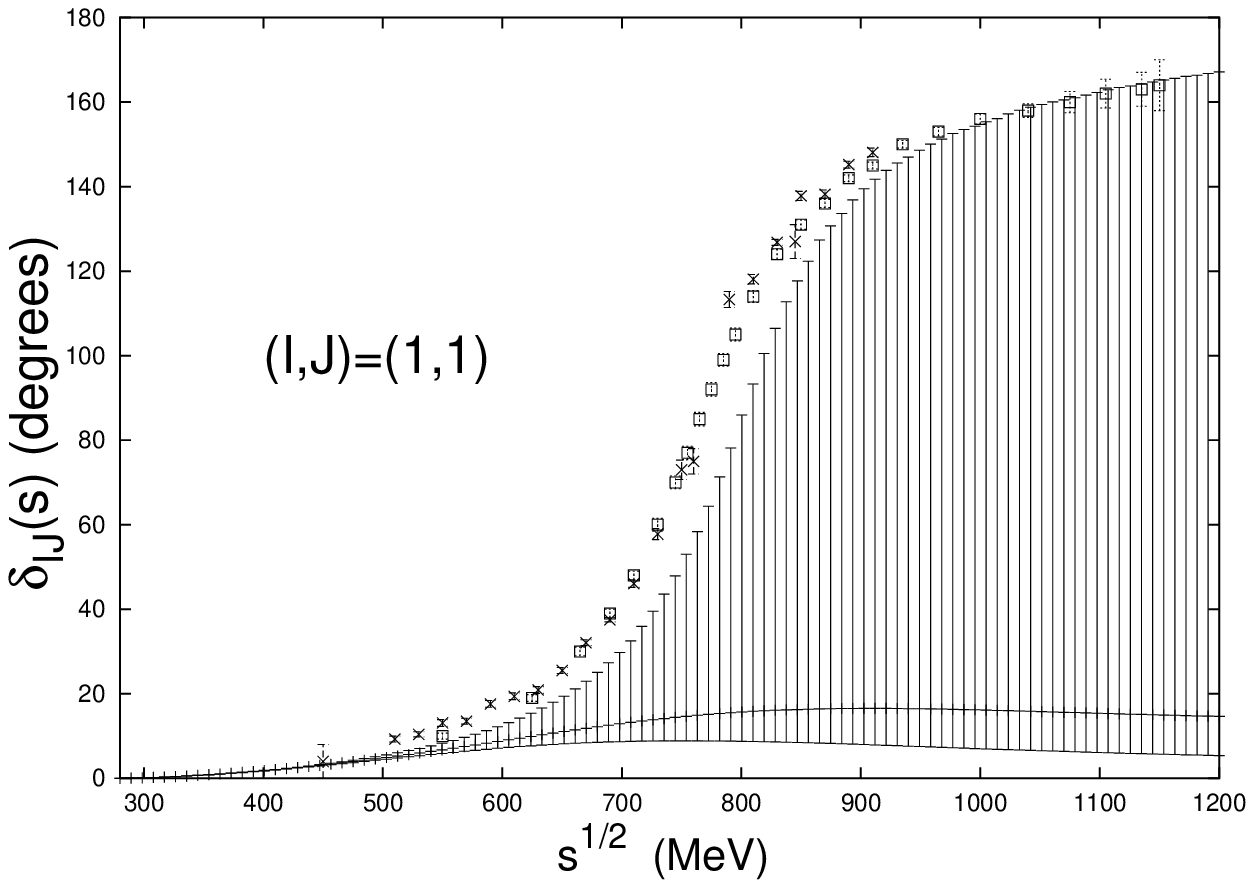,height=5.5cm,width=5.5cm}\epsfig{figure=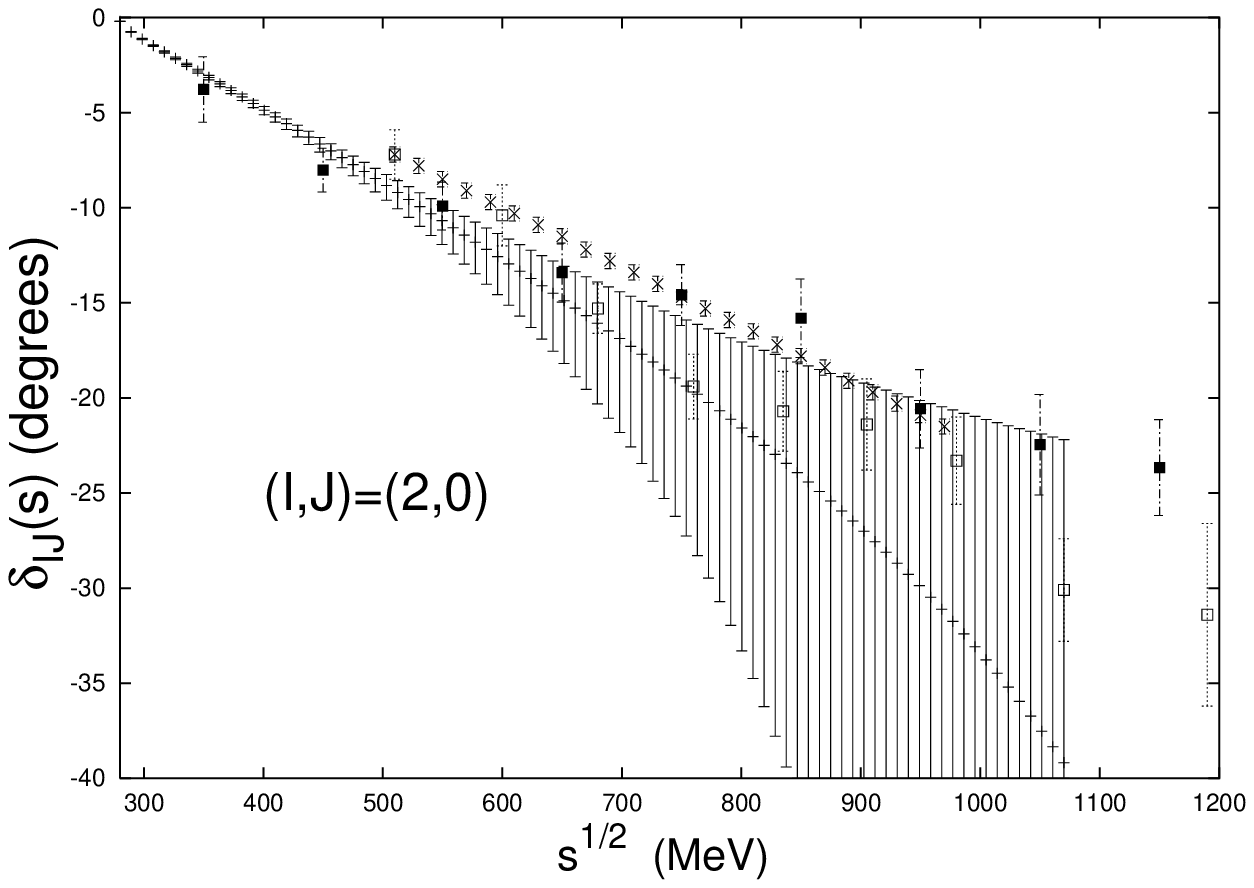,height=5.5cm,width=5.5cm}
\end{center} 
\caption{IAM Unitarized phase shifts (in degrees) for $\pi\pi$
scattering for $S-$ and $P-$ waves. Naive scheme (see main text).
Upper panel: Set {\bf Ic} of Ref.~\cite{EJ00a}. Lower panel: Set {\bf
III} of Ref.~\cite{EJ00a}. Combined data from
Refs.~\cite{pa73}-\cite{fp77}. }
\label{fig:naive}
\end{figure}

The simplest and most direct way to look at the quantitative
predictions of the IAM is to take the unitarized amplitude
Eq.~(\ref{eq:iam}) for all partial waves and propagate the errors in
the one- and two-loop parameters $\bar l_{1,2,3,4} $ and $\bar
b_{1,2,3,4,5,6}$ respectively. As we have already pointed out before,
this may be a dangerous procedure since the two loop parameters
contain a higher order piece, but one might argue that since the
numerical effect on the $\bar b$'s should be small (see
Table~\ref{tab:bbar}), one might expect an overall small effect
anyhow. Along these lines, we present in Fig.~\ref{fig:naive} the
results obtained by using the parameter Sets {\bf Ic} and {\bf III} of
Ref.~\cite{EJ00a}. As can be seen from the figures, the errors are
huge and there is even a trend to discrepancy in the $\rho$ channel
for Set {\bf III}.

\begin{figure}[t]
\begin{center}
\epsfig{figure=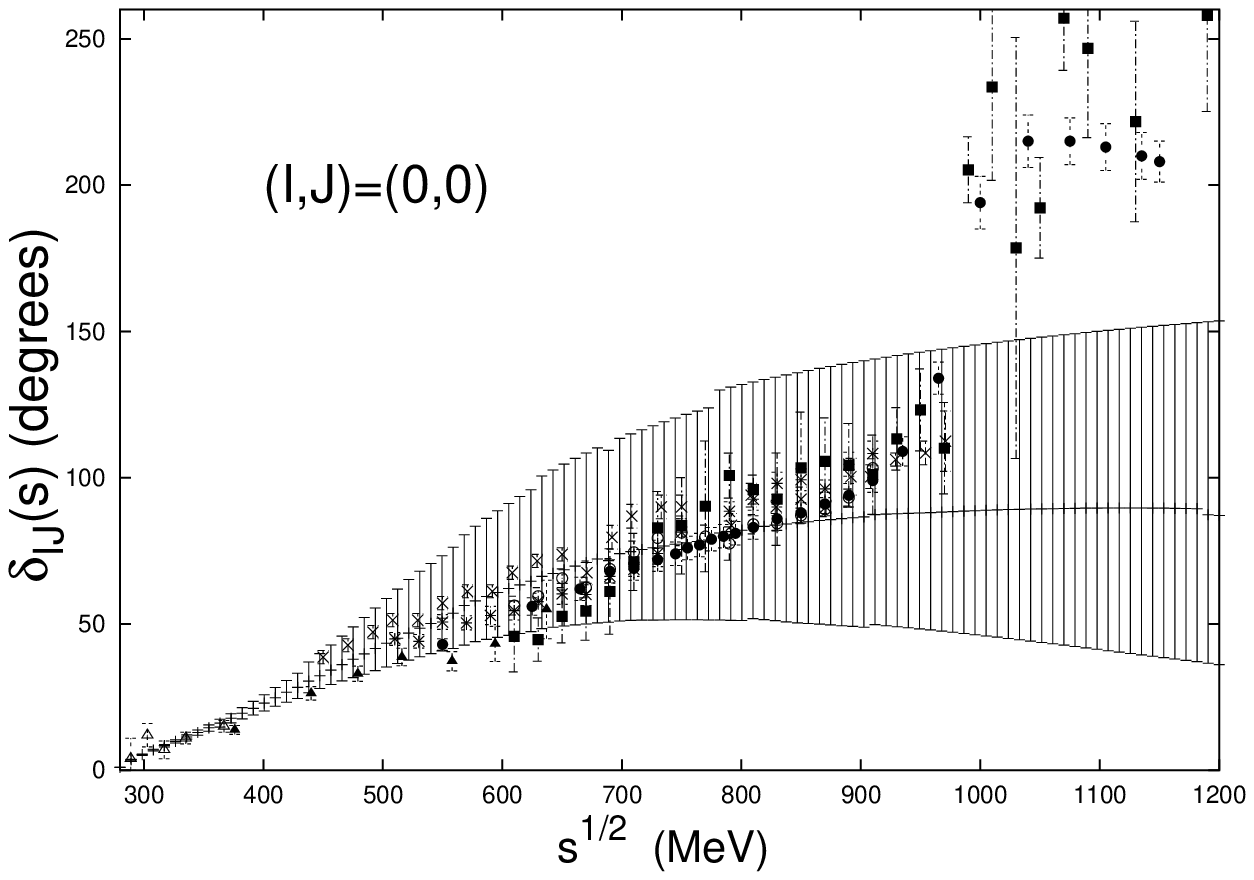,height=5.5cm,width=5.5cm}\epsfig{figure=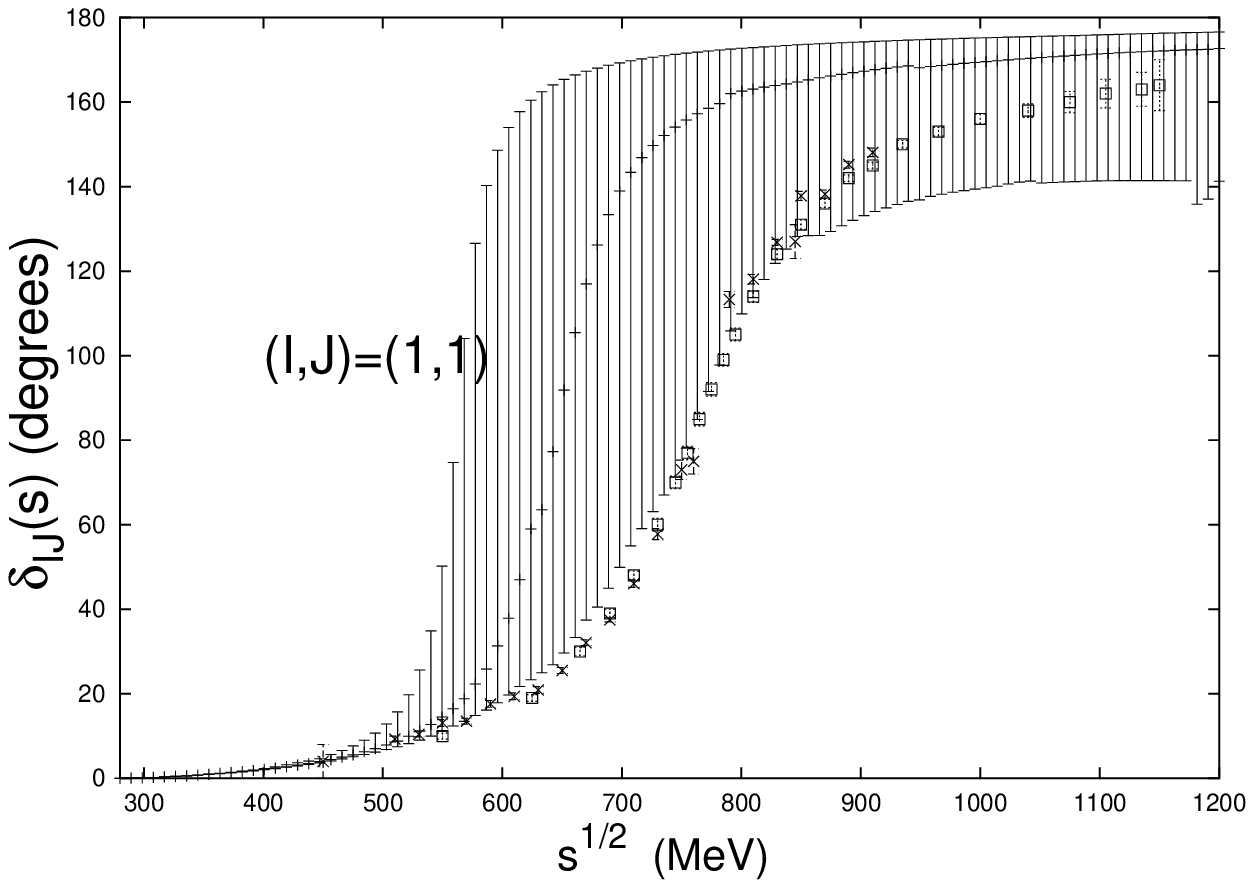,height=5.5cm,width=5.5cm}\epsfig{figure=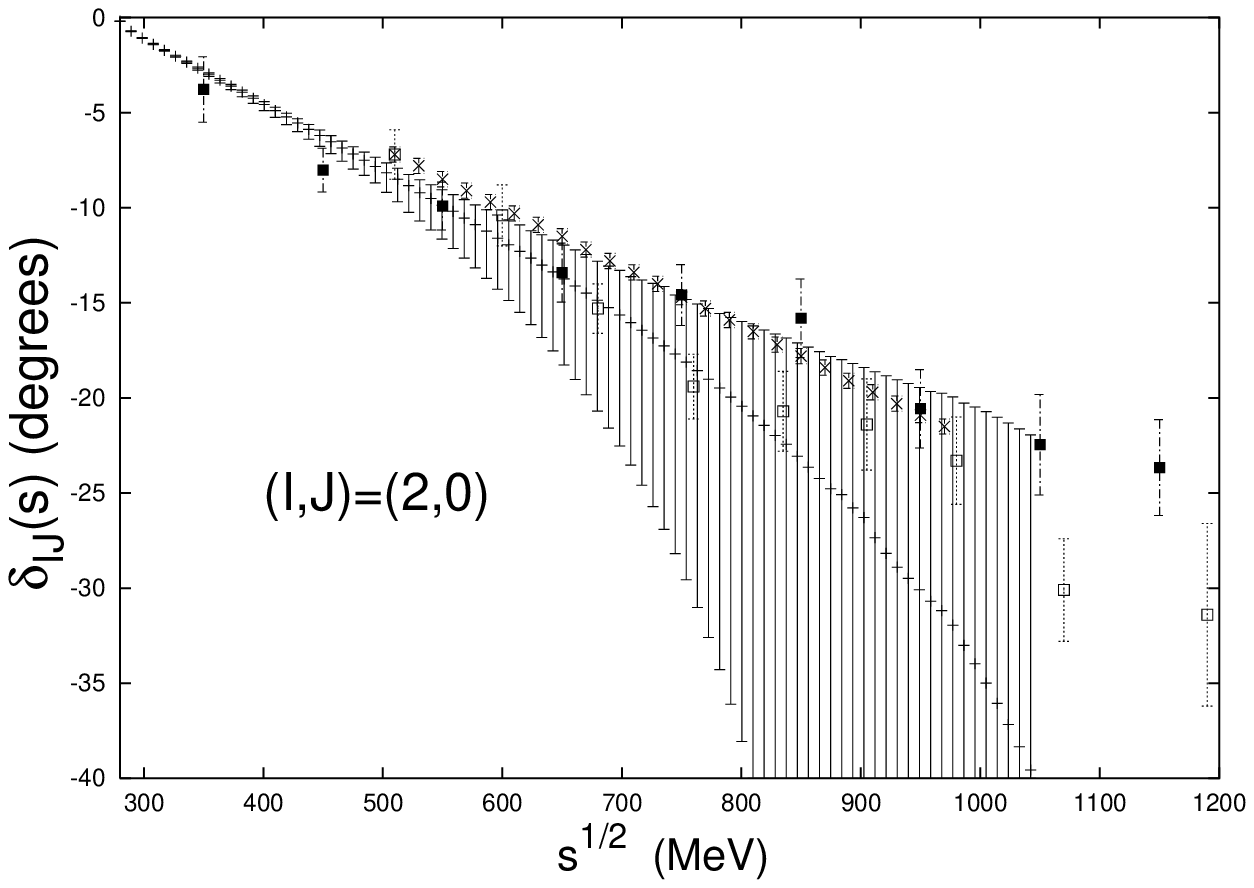,height=5.5cm,width=5.5cm}
\epsfig{figure=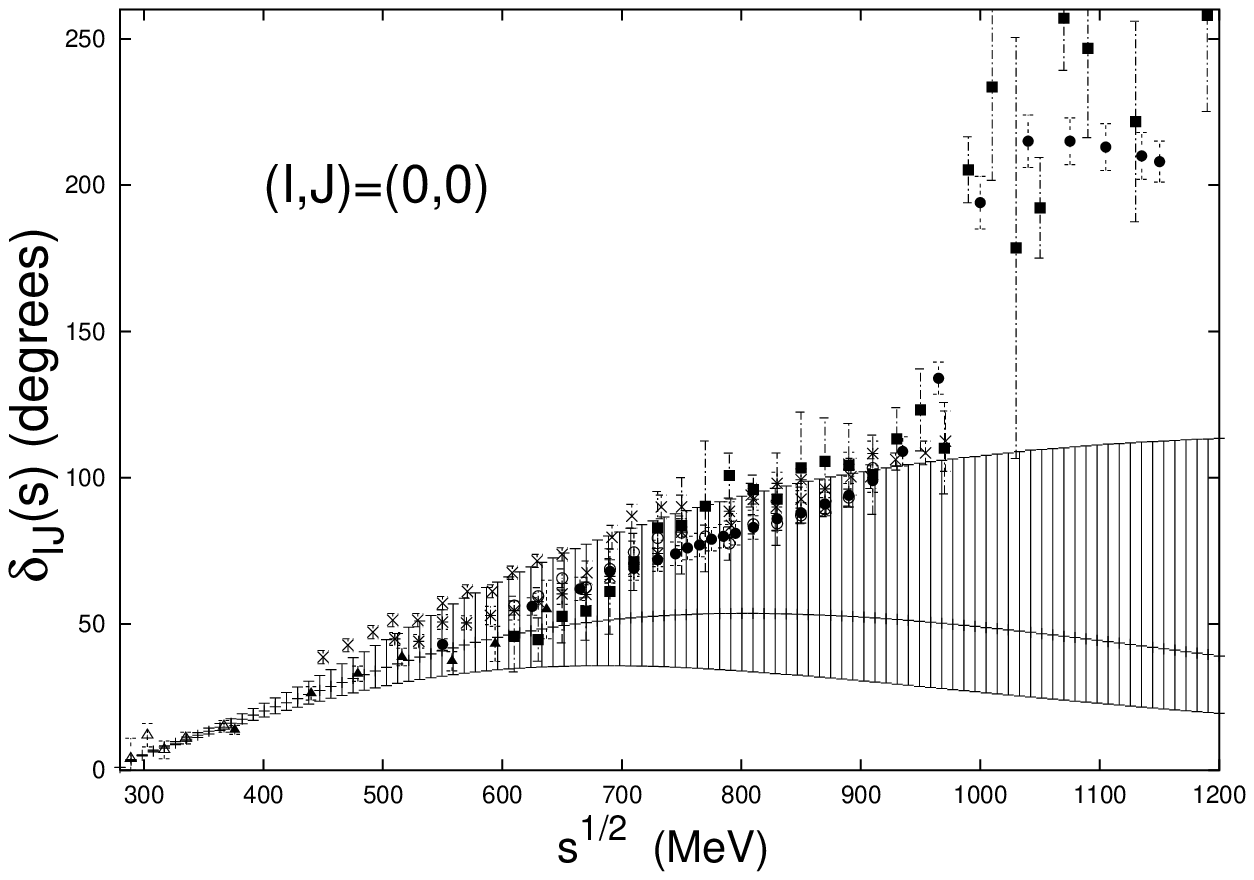,height=5.5cm,width=5.5cm}\epsfig{figure=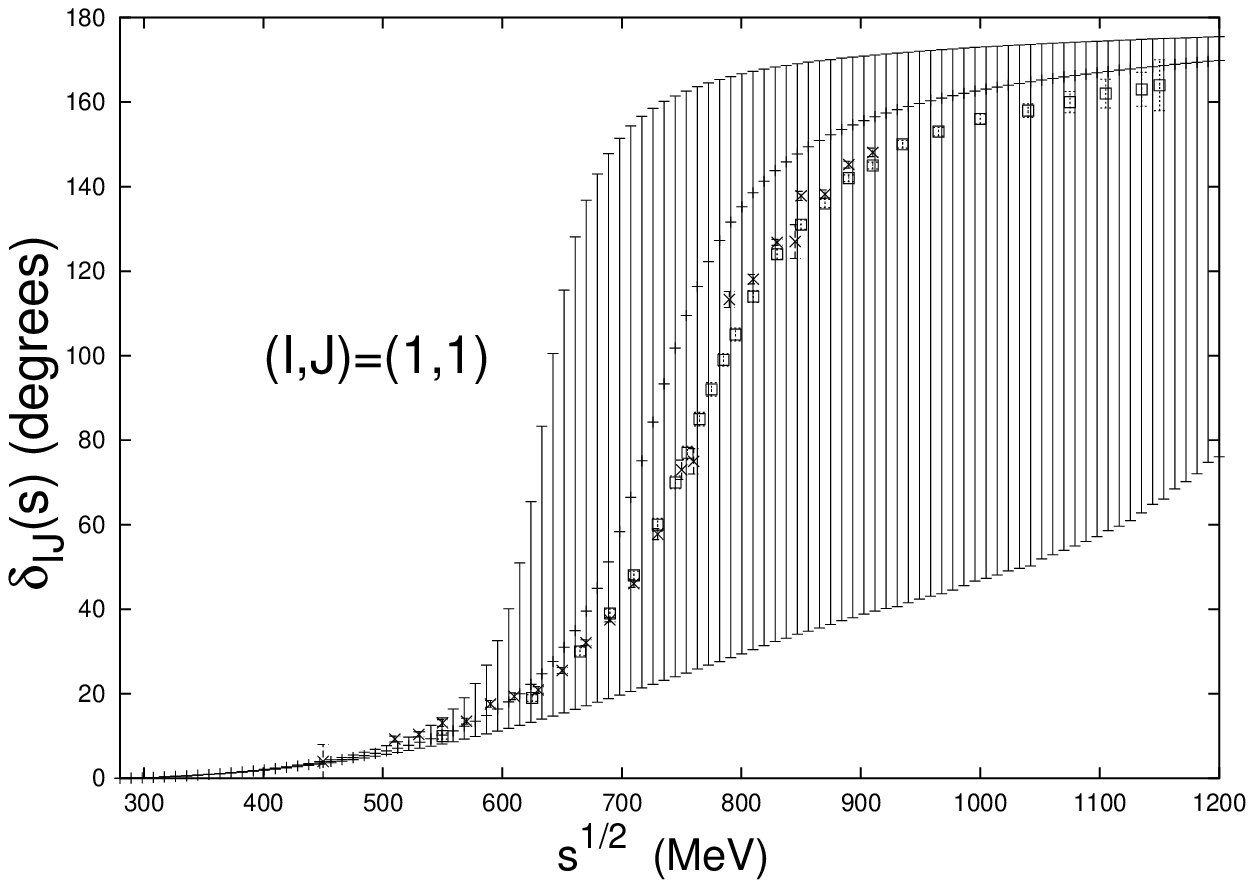,height=5.5cm,width=5.5cm}\epsfig{figure=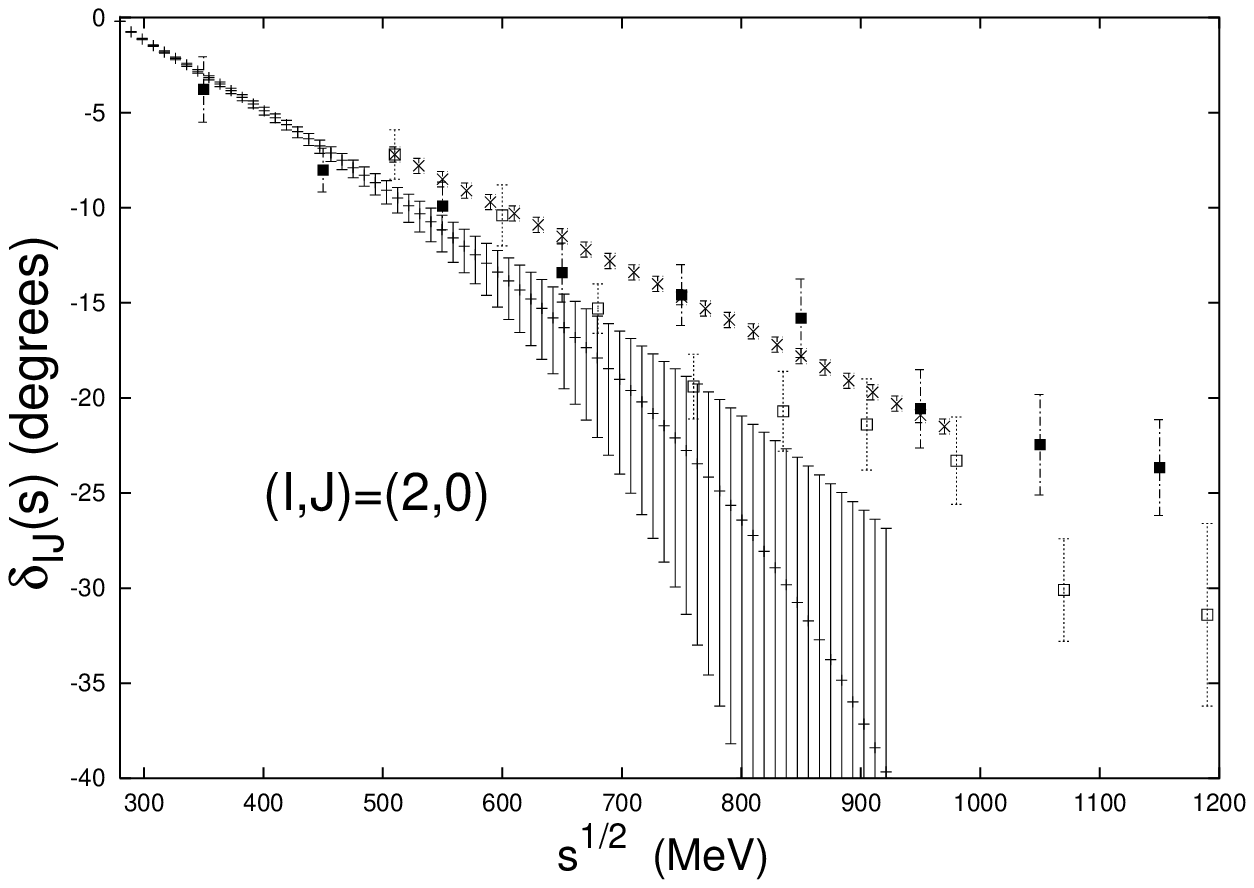,height=5.5cm,width=5.5cm}
\end{center} 
\caption{IAM Unitarized phase shifts (in degrees) for $\pi\pi$
scattering for $S-$ and $P-$ waves. Monte-Carlo scheme (see main
text). Upper panel: Set {\bf Ic} of Ref.~\cite{EJ00a}. Lower panel:
Set {\bf III} of Ref.~\cite{EJ00a}.Combined data from
Refs.~\cite{pa73}-\cite{fp77}.}
\label{fig:separated}
\end{figure}

\begin{figure}[t]
\begin{center}
\epsfig{figure=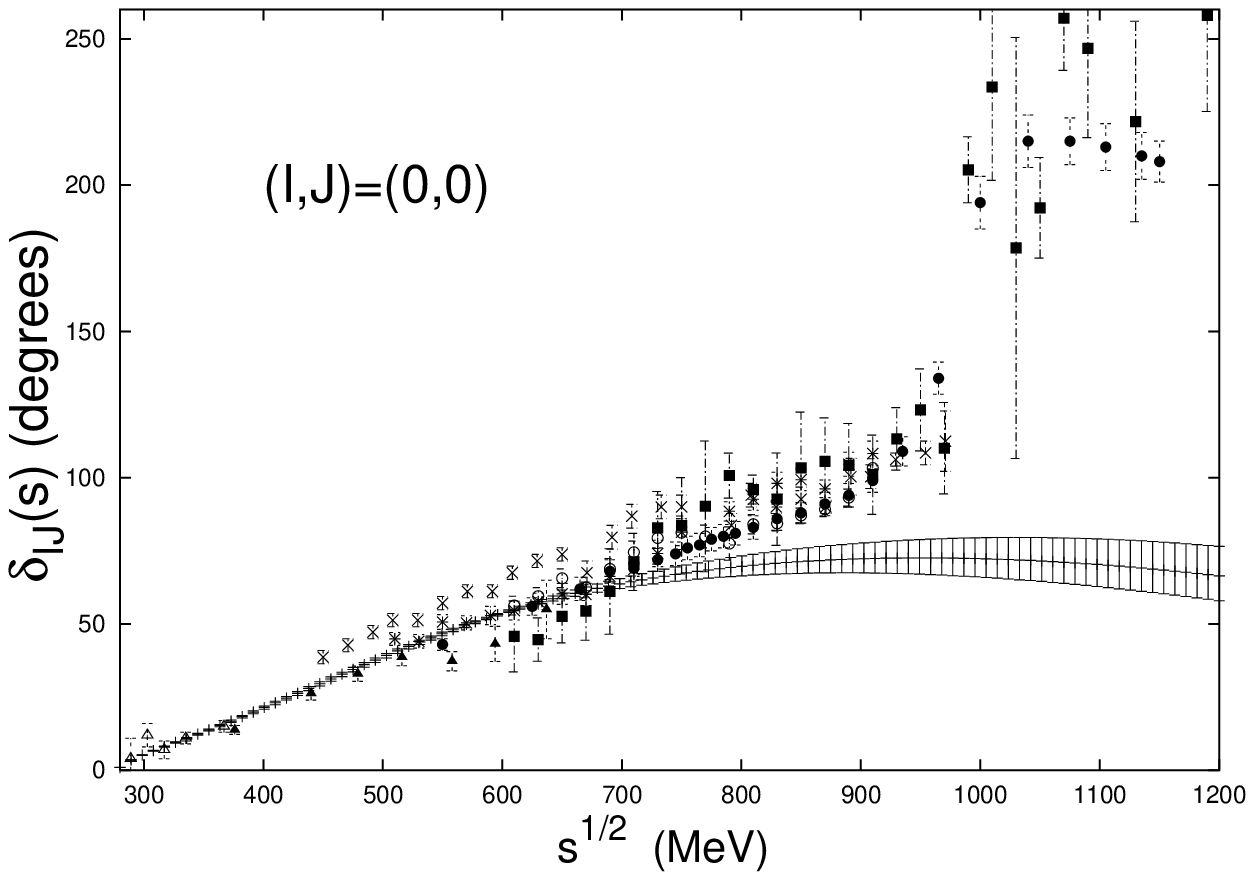,height=5.5cm,width=5.5cm}\epsfig{figure=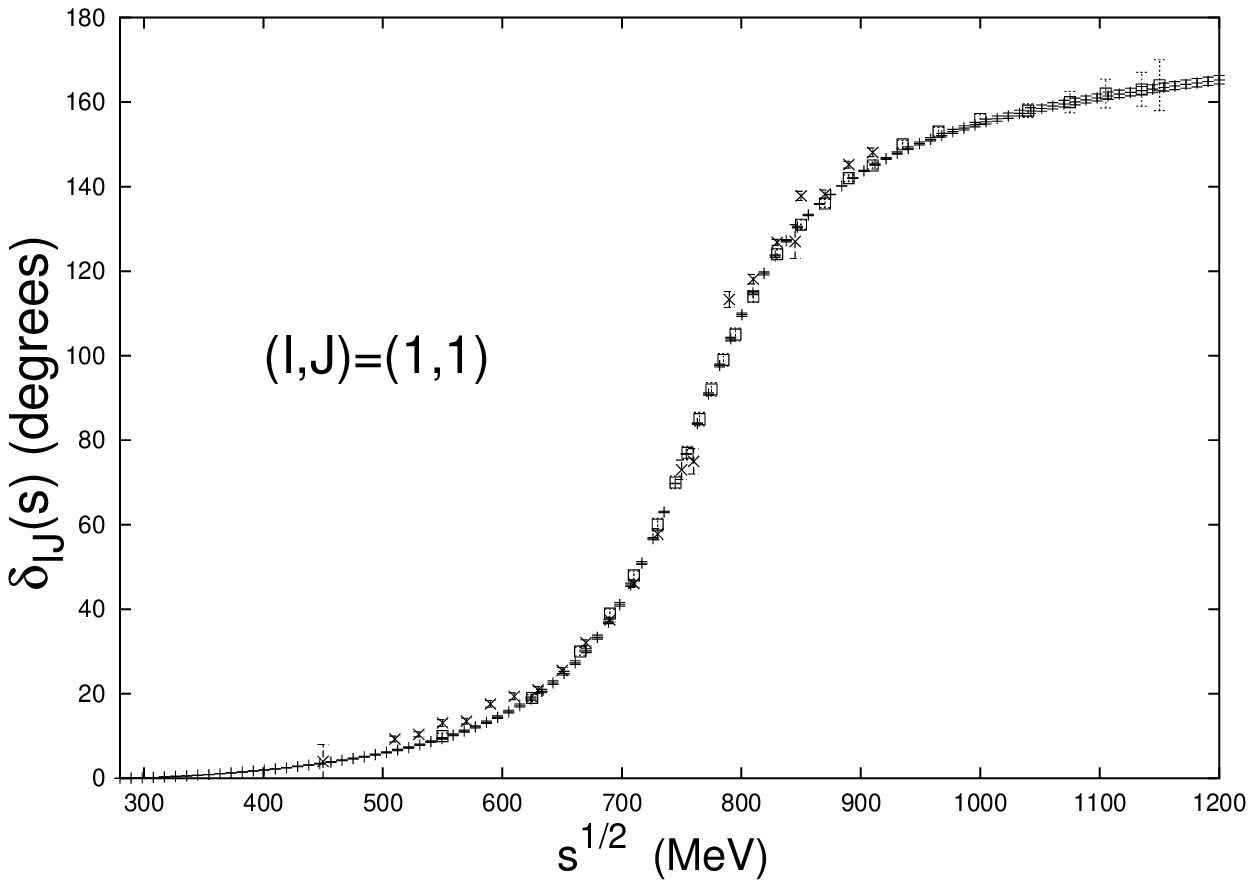,height=5.5cm,width=5.5cm}\epsfig{figure=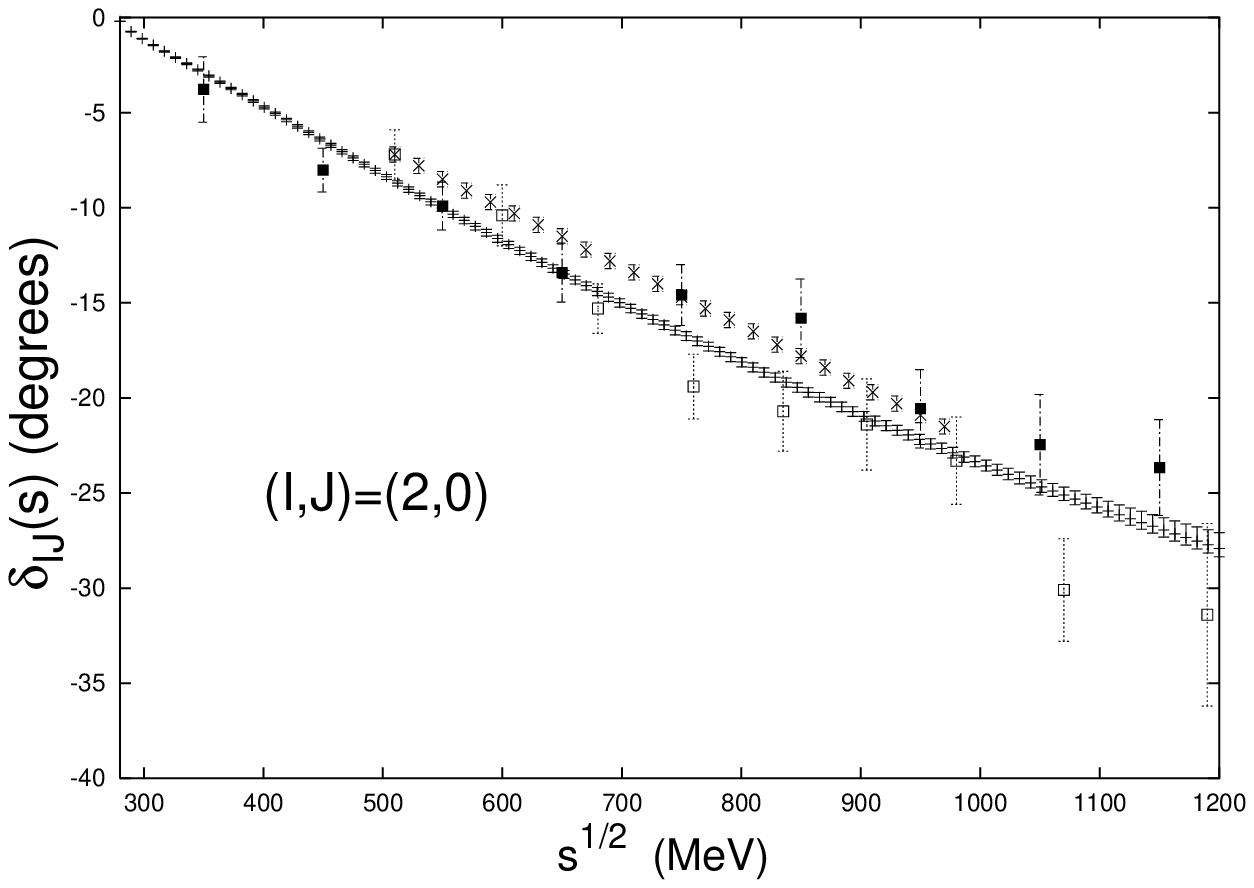,height=5.5cm,width=5.5cm}
\end{center} 
\caption{IAM Unitarized phase shifts (in degrees) for $\pi\pi$
scattering for $S-$ and $P-$ waves. Partial fit scheme (see main
text). There are no Set {\bf Ic} and Set {\bf III} labels because
$\bar l_1$ and $\bar l_2$, which provide this label, are determined
from the fit. Combined data from
Refs.~\cite{pa73}-\cite{ho77}. Uncertainties in the curves stem from
those of class B parameters and the scale for which resonance
saturation is assumed, $\mu = 750 \pm 250 {\rm MeV} $.}
\label{fig:mc-fit.rl}
\end{figure}

\subsubsection{Monte-Carlo scheme} 

Having one realized of the dangers of making a naive Monte Carlo error
propagation from the analysis of Ref.~\cite{EJ00a}, we proceed now in
a different manner.  We consider Sets {\bf Ic} and {\bf III} of
Ref.~\cite{EJ00a} for both the one loop $\bar l_i$-parameters and the
zeroth order two loop parameters $\bar b^0_i$ as explicitly given in
the Appendix of Ref.~\cite{bc97}. The results are shown in
Fig.~\ref{fig:separated}. Although in the $S-$ waves there are no
big differences as compared to Fig.~\ref{fig:naive}, in the $\rho$
channel, the effect at intermediate energies makes the predicted
phase-shifts not only compatible with data but also the error-bars get
substantially reduced.

\begin{figure}[t]
\begin{center}
\epsfig{figure=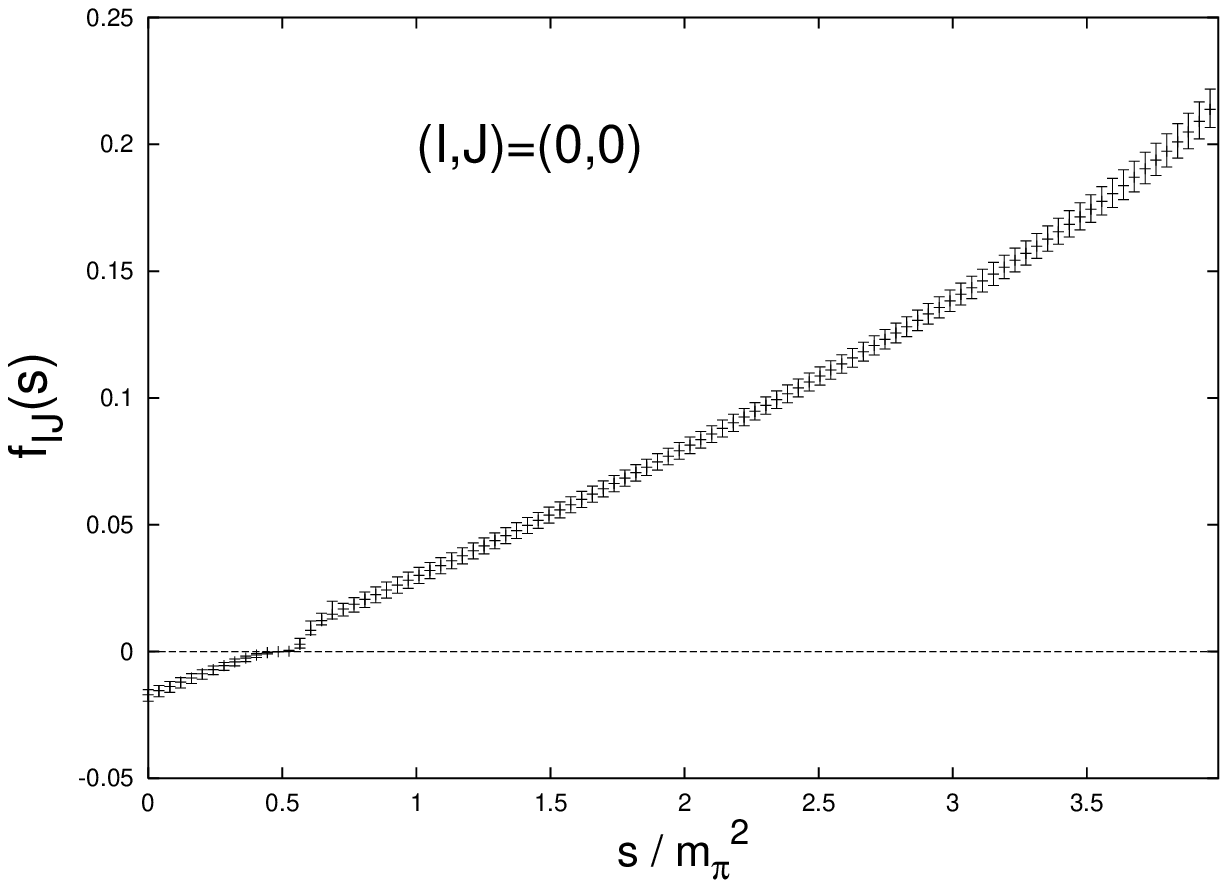,height=5.5cm,width=5.5cm}\epsfig{figure=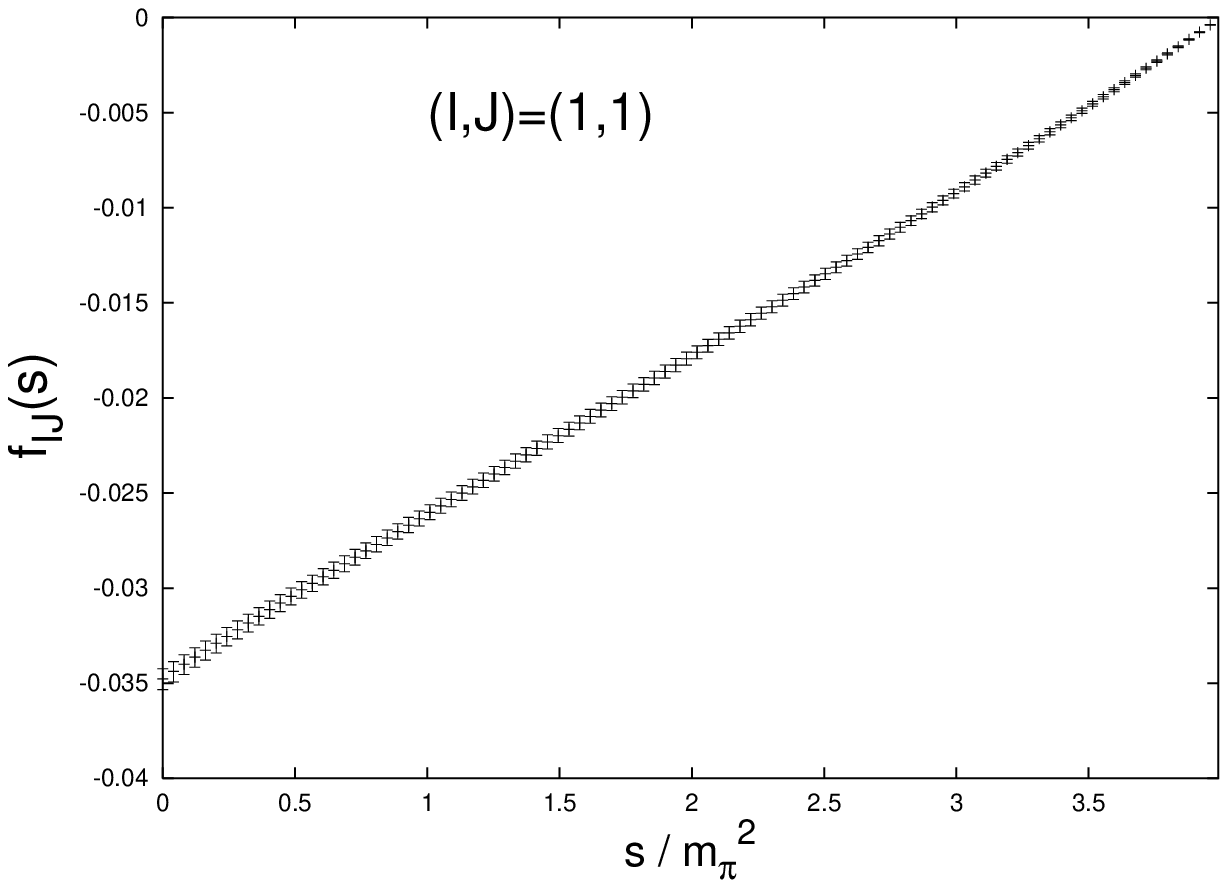,height=5.5cm,width=5.5cm}\epsfig{figure=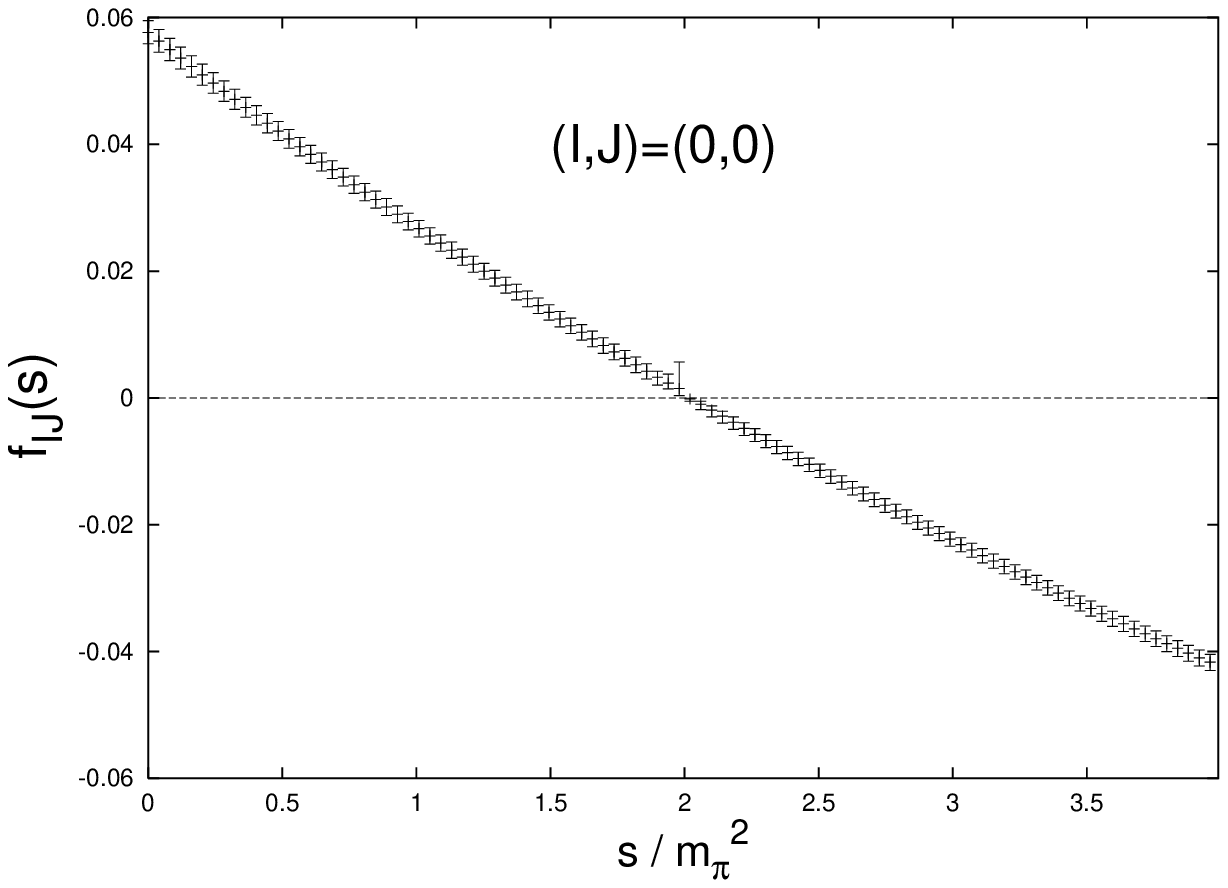,height=5.5cm,width=5.5cm}
\epsfig{figure=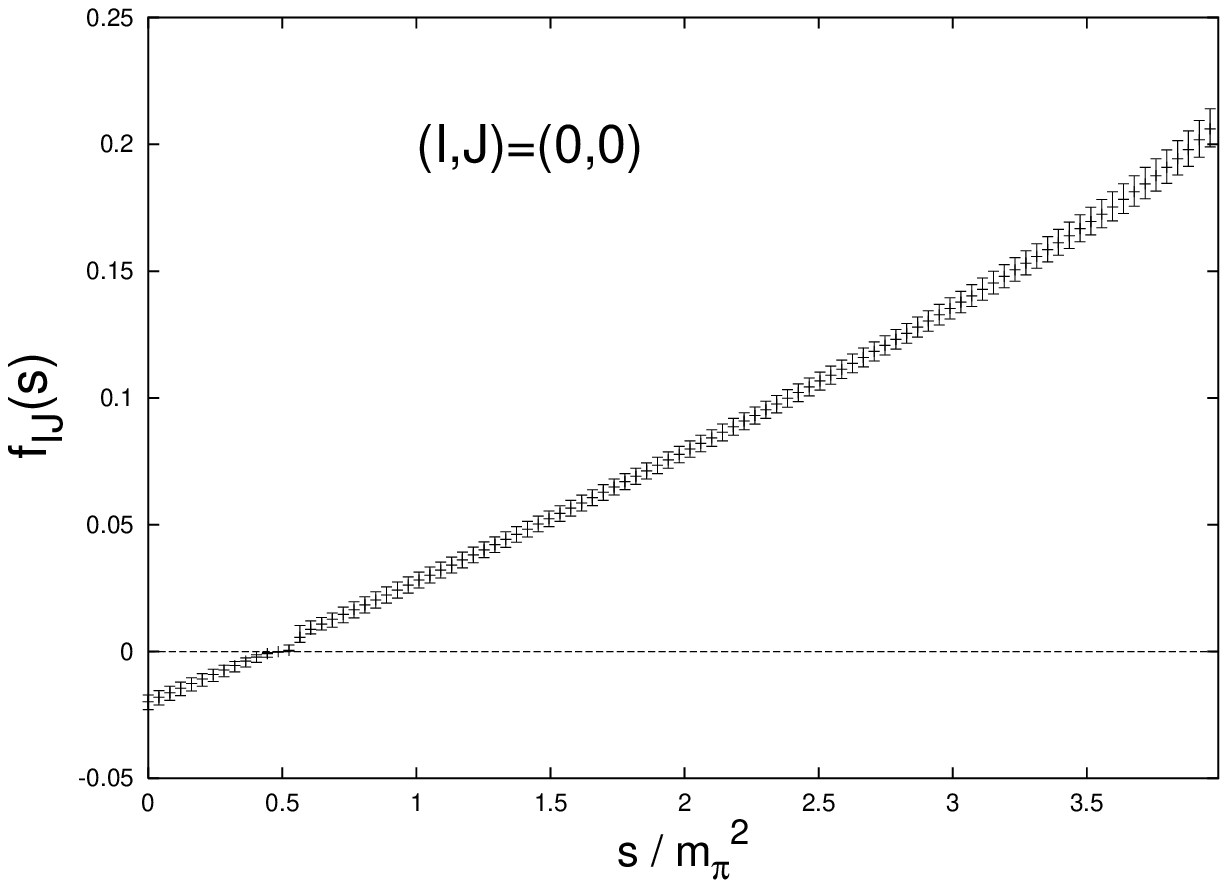,height=5.5cm,width=5.5cm}\epsfig{figure=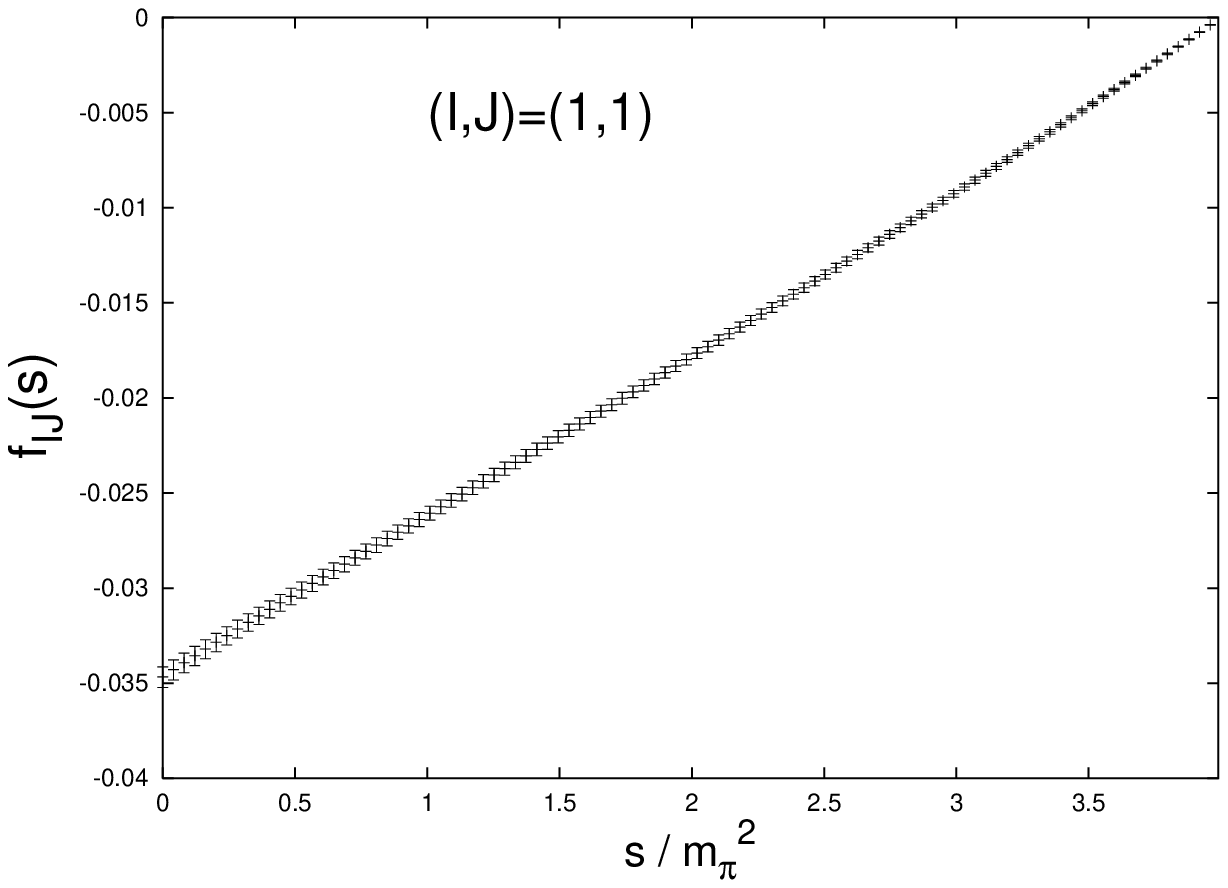,height=5.5cm,width=5.5cm}\epsfig{figure=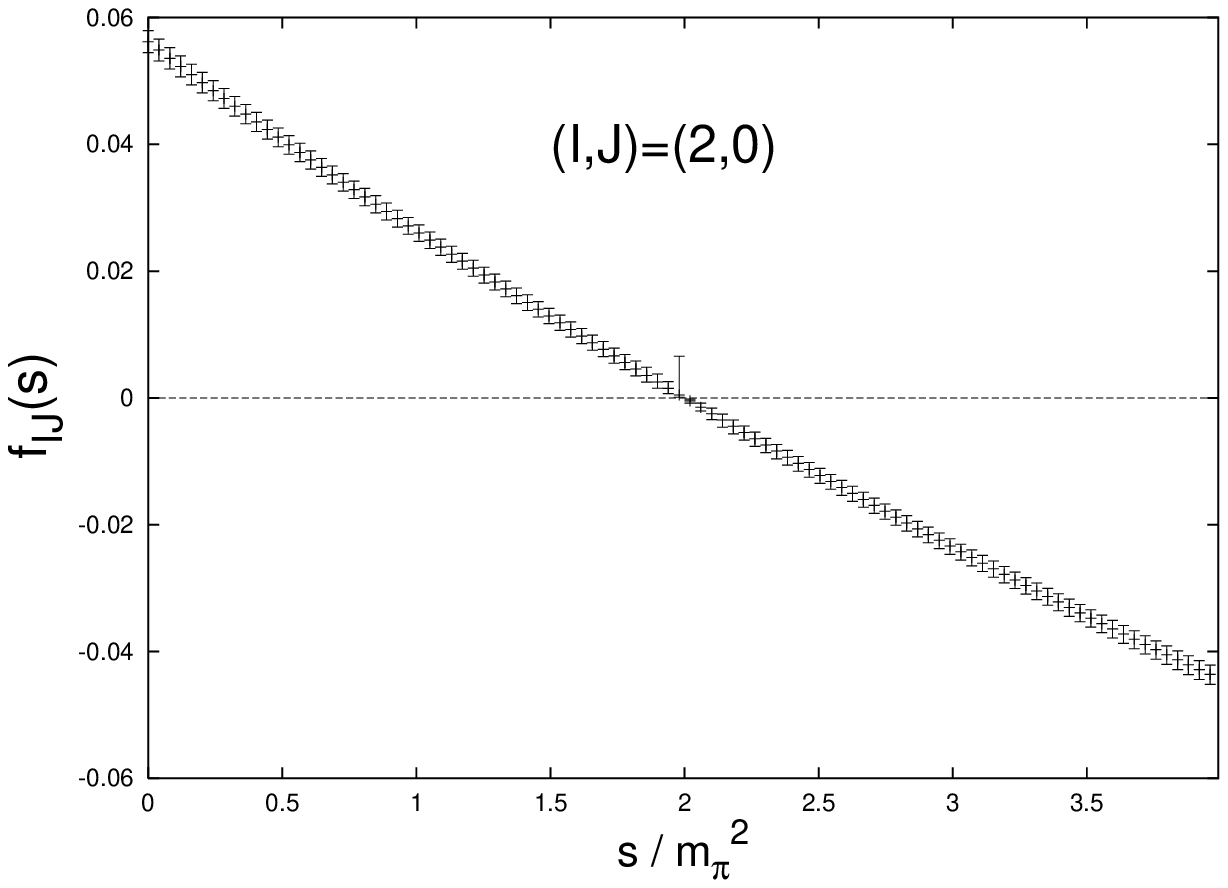,height=5.5cm,width=5.5cm}
\end{center} 
\caption{$S-$ and $P-$ partial wave amplitudes $ f_{IJ} (s) = \sqrt{s}
\, t_{IJ}(s) $ (in fm ) for $\pi\pi$ scattering in the IAM to two
loops in the unphysical region, $ 0 \le s \le 4 m_\pi^2$. Monte-Carlo
scheme. Upper panel: Set {\bf Ic} of Ref.~\cite{EJ00a}. Lower panel:
Set {\bf III} of Ref.~\cite{EJ00a}. Normalization is such that at
$s=4m_\pi^2$ one has for the $S-$ waves the scattering length. In the
scattering region ( $ s > 4 m_\pi^2 $ ) this figure corresponds to
Fig.~\ref{fig:separated}.}
\label{fig:unphys.iam}
\end{figure}

\subsubsection{Partial Fit scheme} 

The results of the two previous schemes suggest that some fine tuning
mechanism is needed, since the large difference between them is due to
either considering the $\bar b$ parameters as a whole or to explicitly
split them as different orders, according to $\bar b_i= \bar b^0_i +
\Delta \bar b_i $. This fact seem to indicate that their numerical
values may be constrained to a large extent by performing a
$\chi^2-$fit. As we have mentioned above this involves
10-parameters. Another possibility can be displayed by making
selective fits in some subsets of parameters propagating the errors in
the remaining ones.  After trying out several combinations, we have
found that, indeed, the low energy parameters of class A, i.e., those
not vanishing in the chiral limit, are enough for a satisfactory fit
to the data.  In practice, the procedure is as follows. For either Set
{\bf Ic} and Set {\bf III}, we generate a sufficiently large sample
($N=10^4$ proves large enough), of class B parameters, i.e., those
corresponding to chiral symmetry breaking terms in the $\pi\pi$
scattering amplitudes. For any member of the class B parameter
population a $\chi^2-$ fit of class A parameters is performed. This
procedure yields distributions for $\bar l_1$, $\bar l_2$, $r_5$ and
$r_6$ parameters whence straightforward error analysis may be
undertaken. The choice of these parameters rather than the $\bar b$'s
has the advantage that the separation $\bar b_i = \bar b^0_i + \Delta
\bar b_i $ may be explicitly done within the fitting procedure, and
hence a correct bookkeeping of chiral orders in the inverse amplitude
is implemented. The result of the fit is\footnote{We apply the
following energy cuts and scattering data. For the isoscalar $S-$wave
we cut at $\sqrt{s}=610 {\rm MeV}$ the data of
Refs.~\cite{pa73,hy73,em74,sr75,ro77,klr97}. For the isotensor $S-$
wave we cut at $970 {\rm MeV}$ the data of Refs.~\cite{ho77,lo74}. For
the isovector $P-$ wave we cut at ${\rm MeV} $.}
\begin{eqnarray}
\bar l_1=  -0.14\er{0.62}{0.74} \quad \bar l_2 = 4.4 \pm{0.1} \quad
10^4 r_5 = 1.07 \er{0.31}{0.35} \quad 10^4 r_6 = -0.35\er{0.13}{0.34}
\label{eq:fit-2loop}
\end{eqnarray} 
which produces $\chi^2 /{\rm d.o.f.} = 69.9 / (67-4)=1.11 $, a rather
satisfactory value as can also be clearly seen in
Fig.~\ref{fig:mc-fit.rl}.  Besides, we obtain the correlation
coefficient $r( \bar l_1 , \bar l_2)=0.22$, There are no Set {\bf Ic}
and Set {\bf III} labels because $\bar l_1$ and $\bar l_2$, which
provide this label, are determined from the fit. Both class A and
class B parameters contribute to the total errors. The errors
corresponding to fitted (class A) low energy parameters can be
determined by employing the standard procedure of changing the
$\chi^2$ from its minimal value by one unit.  We find that they are
rather small, so the quoted errors in Eq.~(\ref{eq:fit-2loop}) are
dominated by the uncertainties in the class B low energy parameters
and the scale at which resonance saturation is assumed, $\mu = 750 \pm
250 {\rm MeV}$.
 
As we see, the out-coming values of the fitted
parameters, particularly $\bar l_1$ and $\bar l_2$, are very much in
agreement with the standard ChPT estimates of Ref.~\cite{EJ00a}. It is
also interesting to note that the values of $r_5$ and $r_6$ are
consistent within error-bars with those assumed from resonance
saturation provided with a $100 \%$ uncertainty as suggested in
Ref.~\cite{gir97}. Transforming these values into $\bar b$ parameters
we get
\begin{eqnarray} 
\bar b_1 = -12.1 \er{1.9}{2.2} \qquad \bar b_2 &=& 11.5 \er{1.2}{1.0}
\quad \, \bar b_3 = -0.29 \er{0.20}{0.27} \nonumber \\ \bar b_4 = 0.75
\pm{0.02} \quad \, \bar b_5 &=& 3.14 \er{0.42}{0.56} \quad \bar b_6 =
0.55 \er{0.46}{0.64}
\end{eqnarray} 
These numbers have been constructed by adding the out-coming $\bar
b^0_i $ to the second order $\Delta \bar b_i$ contribution, although
both contributions enter the fit in a non-additive way. As we see, the
resulting values agree with those expected from standard ChPT analyses
within uncertainties, with the sole exception of $\bar b_6$ which
turns out to be inconsistent. The reason is that the corresponding
value for $r_6$ comes out exactly opposite sign as that expected in
resonance saturation.

Finally, we present in Fig.~\ref{fig:unphys.iam} the IAM unitarized
partial $S-$ and $P-$wave amplitudes in the unphysical region $0 \le s
\le 4 m_\pi^2$ using the Monte Carlo scheme for both Sets {\bf Ic} and
{\bf III}. Obviously, the kinematical zero of the $P-$wave at $s=4
m_\pi^2$ remains fixed. On the other hand, the non-kinematical Adler
zeros of the $S-$wave amplitudes do not move from their lowest order
locations given by Eq.~(\ref{eq:adler0}) but become higher order
zeros, as can be observed from the figures in the small bumps around
the zeros and analytically in Eq.~(\ref{eq:iam}). Although this effect is
undesirable, we see that from a direct comparison of the standard ChPT
amplitudes of Fig.~\ref{fig:unphys.chpt} with the IAM unitarized ones
of Fig.~\ref{fig:unphys.iam} one may conclude that the violation of
the order of the non-kinematical zero does not have quantitative
dramatic consequences.

\subsection{Crossing violations} 

As we have said, the IAM restores unitarity but violates crossing
symmetry. The interesting point is that although unitarization, which
has to do with the right-hand cut, may provide a satisfactory energy
dependence and, hence a description of the data in the scattering
region, it can still do so with low energy constants (LEC's) which
differ from those expected in standard ChPT. The precise numerical
values of the LEC's depend on how the left-hand cut is handled and
approximated before or after unitarization. On the other hand a proper
left-hand cut is a direct consequence of crossing symmetry in the
$s,t,u$ representation. It is therefore interesting to study these
crossing violations on a quantitative basis. A particular way of doing
this at the level of partial wave amplitudes is by using Roskies sum
rules \cite{Ro70,CP71}. Two ways at least have been introduced in the
literature to characterize crossing violations in a quantitative
way. Defining the quantities introduced in Ref.~\cite{BP97}
\begin{eqnarray}
C_1 &=&\int_0^{4 m_\pi^2} ds (4m_\pi^2 -s ) (3s- 4 m_\pi^2)
\left( t_{00} (s) + 2 t_{20}(s) \right) \nonumber \\
C_2 &=&\int_0^{4 m_\pi^2} ds (4m_\pi^2 -s ) \left(2 t_{00} (s) - 5
t_{20}(s) \right) \nonumber \\ 
C_3 &=&\int_0^{4 m_\pi^2} ds \left[ (4m_\pi^2 -s ) (3s-4 m_\pi^2) 
\left( 2 t_{00} (s) - 5 t_{20}(s) \right) + 9 
(4m_\pi^2 -s)^2  t_{11} (s) \right]  \\
C_4 &=&\int_0^{4 m_\pi^2} ds \left[ (4m_\pi^2 -s) s^2 \left( 2 t_{00}
(s) - 5 t_{20}(s) \right) + 3 (4m_\pi^2 -s)^3 t_{11} (s) \right]
\nonumber \\
C_5 &=&\int_0^{4 m_\pi^2} ds \left[ (4m_\pi^2 -s)^2 s^2 \left( 2 t_{00}
(s) - 5 t_{20}(s) \right) + 3 (4m_\pi^2 -s)^2 ( 8m_\pi^2 - 3 s) s \, 
t_{11} (s) \right] \nonumber
\end{eqnarray} 
These relations can be written in the general form 
\begin{eqnarray}
C_i = \int_0^{4 m_\pi^4} ds \sum_{IJ} \omega_{IJ,i} (s) t_{IJ} (s) 
\end{eqnarray} 
Crossing symmmetry implies $C_i=0$. We also consider the definitions
introduced in Ref.~\cite{CB01}
\begin{eqnarray}
A_1 &=& 2 \int_0^{4 m_\pi^2} ds  (s- 4m_\pi^2 ) t_{00} (s)\nonumber \\
B_1 &=& 5 \int_0^{4 m_\pi^2} ds  (s- 4m_\pi^2 ) t_{20} (s)\nonumber \\
A_2 &=& \int_0^{4 m_\pi^2} ds  (s- 4m_\pi^2 ) (3s-4 m_\pi^2)
t_{00} (s)\nonumber \\ B_2 &=& -2 \int_0^{4 m_\pi^2} ds  (s- 4m_\pi^2
)(3s-4 m_\pi^2) t_{20} (s) \nonumber \\ 
A_3 &=& \int_0^{4 m_\pi^2} ds 
(s- 4m_\pi^2 ) (3s-4 m_\pi^2) t_{00} (s)\nonumber \\ 
B_3 &=& 2 \int_0^{4 m_\pi^2} ds  (s- 4m_\pi^2 )^2 t_{11} (s)\nonumber \\
A_4 &=& \int_0^{4 m_\pi^2} ds  (s- 4m_\pi^2 ) (3s-4 m_\pi^2)
\left\{ 2 t_{00} (s) - 5 t_{20} (s) \right\}  \\
B_4 &=& 9 \int_0^{4 m_\pi^2} ds  (s- 4m_\pi^2 )^2 t_{11} (s)\nonumber \\ 
A_5 &=& \int_0^{4 m_\pi^2} ds  (s- 4m_\pi^2 ) (10 s^2 - 32 s
m_\pi^2 + 16 m_\pi^4 ) \left\{ 2 t_{00} (s) - 5 t_{20} (s) \right\} \nonumber \\
B_5 &=& -6 \int_0^{4 m_\pi^2} ds  (s- 4m_\pi^2 )^2 (5 s - 4
m_\pi^2 ) t_{11} (s)\nonumber \\ 
A_6 &=& \int_0^{4 m_\pi^2} ds  (s- 4m_\pi^2 ) (35 s^3 - 180
s^2 m_\pi^2 + 240 s m_\pi^4 - 64 m_\pi^6) 
\left\{ 2 t_{00} (s) - 5 t_{20} (s) \right\} \nonumber \\
B_6 &=& 15 \int_0^{4 m_\pi^2} ds  (s- 4m_\pi^2 )^2 (21 s^2 -
48 s m_\pi^2 + 16 m_\pi^4 ) t_{11} (s) \nonumber 
\end{eqnarray} 
Crossing symmetry implies in this case $ A_i - B_i =0 $ for $i$ from 1
to 6.  Formally, if the unitarized amplitudes embody ChPT to some
order, these sum rules will be identically verified to the same
order. In previous works, the numerical fulfillment of the sum rules
has been tested to one and two loops, but no error estimates have been
taken into account. Due to this, some {\it ad hoc} dimensionless
crossing violations have been defined \cite{BP97,CB01}. In
Refs.~\cite{BP97,Ha99} the ratio
\begin{equation}
R_i = 100 \times  \frac{ \int_0^{4 m_\pi^4} ds \sum_{IJ}
\omega_{IJ,i} (s) t_{IJ} (s)}{ \int_0^{4 m_\pi^4} ds \left| \sum_{IJ}
\omega_{IJ,i} (s) t_{IJ} (s) \right| }
\label{eq:viola2} 
\end{equation}   
is introduced\footnote{We use specifically the formula proposed in
Ref.~\cite{Ha99} since it is unambiguous. The formula of
Ref.~\cite{BP97} is misleading, probably due to a misprint.}  whereas
in Ref.~\cite{CB01} the violation
\begin{equation}
V_i = 100 \times \left | \frac {A_i-B_i}{A_i+B_i} \right| \, .
\label{eq:viola}
\end{equation}
is defined. The advantage of providing error bars to the crossing sum
rules $C_i=0$ or $A_i-B_i =0 $ is obvious; the dimensionless quantity
is naturally defined as the size of the uncertainty relative to the
mean value.  This allows to make a definite statement on crossing
violations in terms of statistical uncertainties. Nevertheless, we
quote in Tables~\ref{tab:v} and \ref{tab:r} the $V_i$ and $R_i
$ values of Ref.~\cite{CB01} and Refs.~\cite{BP97,Ha99} respectively
since they give an idea of how large are these violations in
percentage terms.
\begin{table}[t]
%\vspace{-0.3cm}
\begin{center}
\begin{tabular}{c|c|c|c|c|c|c}  
{\bf IAM} & $V_1$ & $V_2$ & $V_3$ & $V_4$ & $V_5$ & $V_6$ \\
\hline\tstrut Set {\bf Ic} Naive. Fig.~\ref{fig:naive} & $0.9
\er{0.6}{0.7}$ & $0.9 \er{0.7}{0.5}$ & $0.9 \er{0.8}{0.5} $ & $0.4
\er{0.4}{0.3}$ & $33 \er{62}{21}$ & $ 6 \er{13}{67} $ \\

Set {\bf III} Naive. Fig.~\ref{fig:naive}& $0.3 \er{0.2}{0.6} $  & $0.4 \er{0.6}{0.2}$ & $0.4 \er{0.6}{0.3} $ & $ 0.15\er{0.17}{0.30} $&  $ 14 \er{40}{13} $& $ 2 \er{22}{61}$  \\ 

Set {\bf Ic} Non-Naive. Fig.~\ref{fig:separated} & $0.7 \er{0.4}{0.6}$  & $0.7 \er{0.6}{0.5}$ & $0.9 \er{0.8}{0.6} $ & $0.5 \er{0.5}{0.4}$ & $22 \er{29}{19}$ & $ 4 \er{12}{14} $  \\

Set {\bf III} Non-Naive. Fig.~\ref{fig:separated} & $0.3 \er{0.5}{0.1}$  & $0.3 \er{0.5}{0.2}$ & $0.4 \er{0.6}{0.3} $ & $0.2 \er{0.4}{0.2}$ & $9 \er{23}{15}$ & $ 6 \er{10}{15} $  \\

Set {\bf III} Fit. Fig.~\ref{fig:mc-fit.rl}& $ 0.5\er{0.5}{0.3}$  & $ 0.5\er{0.5}{0.3}$ & $ 0.5\er{0.6}{0.4} $ & $ 0.3\er{0.3}{0.3}$ & $ 12\er{19}{25}$ & $  7\er{21}{7} $  \\ 
\end{tabular}
\end{center}
\caption{\footnotesize Roskies sum rules violations in percentage as
defined by Eq.~(\ref{eq:viola}) and introduced in Ref.~\cite{CB01} for
the IAM method and the parameter sets {\bf Ic} and {\bf III} of
Ref.~\cite{EJ00a}. We also indicate the phase-shift figures which
generate these violations}
\label{tab:v} 
\end{table}

\begin{table}[t]
%\vspace{-0.3cm}
\begin{center}
\begin{tabular}{c|c|c|c|c|c}  
{\bf IAM} & $R_1$ & $R_2$ & $R_3$ & $R_4$ & $R_5$  \\ 
\hline\tstrut 
Set {\bf Ic} Naive. Fig.~\ref{fig:naive} & $1.2 \er{0.9}{0.7}$  & $-0.9 \er{0.6}{0.8}$ & $0.9 \er{0.8}{0.6} $ & $0.03 \er{0.12}{0.08}$ & $-0.05 \er{0.09}{0,10}$     \\ 

Set {\bf III} Naive. Fig.~\ref{fig:naive}& $0.5 \er{0.8}{0.3} $ &
$-0.3 \er{0.2}{0.7}$ & $0.3 \er{0.6}{0.3} $ & $ -0.07 \er{0.18}{0.07} $&
$ -0.02 \er{0.13}{0.07} $  \\

Set {\bf Ic} Non-Naive. Fig.~\ref{fig:separated} & $0.9
\er{0.7}{0.6}$ & $-0.7 \er{0.5}{0.6}$ & $1.0 \er{1.0}{0.7} $ & $0.1
\er{0.2}{0.1}$ & $0.08 \er{0.2}{0.1}$  \\

Set {\bf III} Non-Naive. Fig.~\ref{fig:separated} & $0.4
\er{0.7}{0.3}$ & $-0.3 \er{0.1}{0.6}$ & $0.4 \er{0.8}{0.4} $ & $0.06
\er{0.2}{0.1}$ & $0.04 \er{0.2}{0.1}$  \\

Set {\bf III} Fit. Fig.~\ref{fig:mc-fit.rl}& $ 0.6\er{0.6}{0.4}$ & $
-0.5\er{0.3}{0.5}$ & $ 0.6 \er{0.7}{0.5} $ & $ 0.06\er{0.1}{0.1}$ & $
0.01\er{0.1}{0.2}$  \\
\end{tabular}
\end{center}
\caption{\footnotesize Sum rules violations in percentage as defined
by Eq.~(\ref{eq:viola2}) and introduced in Refs.~\cite{BP97,Ha99}for the
IAM method and the parameter sets {\bf Ic} and {\bf III} of
Ref.~\cite{EJ00a}. We also indicate the phase-shift figures which
generate these violations}
\label{tab:r} 
\end{table}

\begin{table}[t]
%\vspace{-0.3cm}
\begin{center}
\begin{tabular}{c|c|c|c|c|c|c}  
 & $a_{00} m_\pi $ & $b_{00} m_\pi^3 $ & $10 \cdot a_{11} m_\pi^3 $ &
$ 10 \cdot b_{11} m_\pi^5 $ & $10 \cdot a_{20} m_\pi $ & $ 10 \cdot
b_{20} m_\pi^3 $ \\ 

\hline\tstrut 

ChPT-{\bf Ic \,} Fig.~\ref{fig:chpt} & $0.214(5)$ & $0.27(1)$ &
$0.37(1) $ & $ 0.06(1) $ & $-0.42(1)$ & $ -0.76(2) $ \\

ChPT-{\bf III} Fig.~\ref{fig:chpt} & $0.208(6) $ & $0.25(1) $ &
$0.374(8) $ & $0.053(7) $ & $-0.44(1)$ & $ -0.80(2) $ \\

\hline\tstrut 

ChPT'-{\bf Ic \,} Fig.~\ref{fig:naive-chpt} & $0.214(5)$ & $0.27(1)$ &
$0.37(1) $ & $ 0.06(1) $ & $-0.42(1)$ & $ -0.76(2) $ \\

ChPT'-{\bf III} Fig.~\ref{fig:naive-chpt} & $0.208(6) $ & $0.25(1) $ &
$0.374(8) $ & $0.053(7) $ & $-0.44(1)$ & $ -0.80(2) $ \\

\hline\tstrut 

IAM-{\bf Ic \,} Fig.~\ref{fig:naive} & $0.220(7)$ & $0.30(2)$ & 
$0.37(8) $ & $0.048(6) $ & $-0.42(1)$ & $ -0.76(2) $ \\

IAM-{\bf III} Fig.~\ref{fig:naive} & $0.211(8)$ & $0.27(2)$ & 
$0.37(6) $ & $0.046(5) $ & $-0.44(1)$ & $ -0.80(2) $ \\

\hline\tstrut 

IAM'-{\bf Ic \,} Fig.~\ref{fig:separated} & $0.221(8) $  & $0.29(2) $ &
 $0.38(1)$ & $0.072 (1)$ & $-0.42(1)$ & $ -0.76(2)$  \\

IAM'-{\bf III} Fig.~\ref{fig:separated} & $0.213(8) $  & $0.27(2) $ &
 $0.381(9)$ & $0.064 (1)$ & $-0.44(1)$ & $ -0.80(2)$  \\

\hline\tstrut 

FIT-{\bf III} Fig.~\ref{fig:mc-fit.rl} & $ 0.216(5) $ & $ 0.280(7) $ & $
0.376(6) $ & $ 0.058(5)$ & $ -0.43 (1) $ & $ -0.79(1) $
\\
\end{tabular}
\end{center}
\caption{\footnotesize Scattering lengths, $a_{IJ}$ and slopes
$b_{IJ}$ defined by Eq.~(\ref{eq:threshold}) for the IAM method and
the parameter Sets {\bf Ic} and {\bf III} of Ref.~\cite{EJ00a}. We
also indicate the phase-shift figures which correspond to these
threshold parameters.}
\label{tab:threshold} 
\end{table}

As can be inferred from Table~\ref{tab:v}, crossing violations as
introduced in Ref.~\cite{CB01} do not seem to be dramatically large,
although this depends on their particular definition. The most serious
violations appear in the $V_5$ rule, which combines both isospin
$S-$wave channels and the $P-$ wave. The computed uncertainties
provide a less pessimistic impression, since in some cases these
violations are compatible with zero. Moreover, if the crossing
violation definition of Refs.~\cite{BP97,Ha99} is evaluated we see
from Table~\ref{tab:r} that these sum rules are better fulfilled. The
effect of uncertainties on the violations has been overlooked in
previous works \cite{BP97,Ha97}. Nevertheless, we point out that
generally speaking there are systematic, though small, crossing
violations. In the partial fit scheme, corresponding to
Fig.~\ref{fig:mc-fit.rl} the crossing violation defined in
Refs.~\cite{BP97,Ha99} are compatible with being smaller than $0.1 \%
$.

\subsection{Threshold parameters} 

The chiral expansion is expected to work best at low energies. But
even so, threshold parameters such as scattering lengths $a_{IJ} $ and
effective ranges $b_{IJ}$ defined by Eq.~(\ref{eq:threshold}) turn out
to get corrections at each order of the expansion. The IAM method is
constructed to reproduce ChPT at all energies but in the lowest orders
of the $1 / f_\pi^2 $ expansion. Thus, if we go to the threshold
region we do not exactly reproduce the standard ChPT behaviour.
Nevertheless, as can be seen from the figures the difference of the
standard ChPT amplitudes and the unitarized ones is actually very
small. Our results for the two loop IAM threshold parameters are
presented in Table~\ref{tab:threshold}. The fact that ChPT and ChPT'
entries of the table are the same within errors is not accidental;
it merely reflects the additive combination $\bar b_i= \bar b^0_i +
\Delta \bar b_i$ and the smallness of $\Delta b_i$. A more detailed
table containing the explicit separation in tree-level, one-loop and
two loops contributions as well as 68 \% ellipses of the $S-$wave
scattering lengths can be found in Ref.~\cite{EJ00a}. In general we
see that the IAM threshold parameters are compatible within errors
with the ChPT ones. The only exception is the slope $b_{11}$ in the
$\rho$ channel for the two Monte-Carlo schemes. The partial fit scheme
provides compatible $a_{11}$ and $b_{11}$ values with slightly better
accuracy.

\begin{table}[t]
%\vspace{-0.3cm}
\begin{center}
\begin{tabular}{c|c|c|c|c|c|c}  
{\bf Generalized IAM} & $V_1$ & $V_2$ & $V_3$ & $V_4$ & $V_5$ & $V_6$

\\ \hline\tstrut 

Set {\bf Ic} Naive & $-0.06
\er{0.12}{0.11}$ & $0.6 \er{0.02}{0.09}$ & $0.01 \er{0.04}{0.04} $ & $-0.02
\er{0.04}{0.05}$ & $14 \er{23}{9}$ & $ 6 \er{4}{5} $ \\

Set {\bf III} Naive & $0.02 \er{0.01}{0.01}$ & $0.06 \er{0.02}{0.02}$
& $0.01 \er{0.03}{0.04} $ & $ -0.02 \er{0.04}{0.04} $& $ 8
\er{10}{4} $& $ 2 \er{4}{6}$ \\

Set {\bf Ic} Non-Naive & $-0.3 \er{0.1}{0.1}$ & $-0.09
\er{0.03}{0.04}$ & $-0.04 \er{0.04}{0.04} $ & $0.02 \er{0.05}{0.04}$ &
$1 \er{5}{2}$ & $ 2 \er{4}{5} $ \\

Set {\bf III} Non-Naive & $-0.3 \er{0.1}{0.1}$ & $-0.06
\er{0.03}{0.03}$ & $-0.01 \er{0.04}{0.03} $ & $0.02 \er{0.05}{0.04}$ &
$1 \er{2}{1}$ & $ 2 \er{4}{5} $ \\

\end{tabular}
\end{center}
\caption{\footnotesize Roskies sum rules violations in percentage as
defined by Eq.~(\ref{eq:viola}) and introduced in Ref.~\cite{CB01} for
the generalized IAM method of Ref.~\cite{Ha99} and the parameter sets
{\bf Ic} and {\bf III} of Ref.~\cite{EJ00a}.}
\label{tab:v-generalized} 
\end{table}

\begin{table}[t]
%\vspace{-0.3cm}
\begin{center}
\begin{tabular}{c|c|c|c|c|c}  
{\bf Generalized IAM} & $R_1$ & $R_2$ & $R_3$ & $R_4$ & $R_5$ \\

\hline\tstrut 

Set {\bf Ic} Naive & $0.08 \er{0.03}{0.02}$ & $-0.06 \er{0.1}{0.1}$ &
$-0.04 \er{0.09}{0.10} $ & $-0.06 \er{0.03}{0.03}$ & $-0.05
\er{0.05}{0.04}$ \\

Set {\bf III} Naive & $0.08 \er{0.02}{0.03}$ & $0.02 \er{0.1}{0.1}$ &
$-0.05 \er{0.08}{0.09} $ & $-0.03 \er{0.03}{0.03}$ & $-0.004
\er{0.06}{0.05}$ \\

Set {\bf Ic} Non-Naive & $-0.11 \er{0.04}{0.05} $ & $-0.3
\er{0.1}{0.1}$ & $0.03 \er{0.10}{0.09} $ & $ -0.12 \er{0.03}{0.04} $&
$ -0.2 \er{0.05}{0.06} $ \\

Set {\bf III} Non-Naive & $-0.08 \er{0.05}{0.04}$ & $-0.3 \er{0.1}{0.1}$
& $0.04 \er{0.10}{0.08} $ & $-0.10 \er{0.03}{0.04}$ & $-0.2 \er{0.2}{0.1}$
\\

\end{tabular}
\end{center}
\caption{\footnotesize Sum rules violations in percentage as defined
by Eq.~(\ref{eq:viola2}) and introduced in Refs.~\cite{BP97,Ha99}for
the generalized IAM method of Ref.~\cite{Ha99} and the parameter sets
{\bf Ic} and {\bf III} of Ref.~\cite{EJ00a}.}
\label{tab:r-generalized} 
\end{table}

\subsection{Generalized IAM} \label{sec:giam}

Roskies sum rules provide a set of necessary conditions for a crossing
symmetric $\pi\pi$ scattering amplitude. It is has been noticed that
the IAM method transforms the non-kinematical single zeros of the
partial wave amplitudes into $N+1$-order zeros of the IAM unitarized
amplitudes, being $N$ the order of the chiral expansion (see the
denominators $[t_{IJ}^{(2)}(s)]^{N+1}$ of Eq.~(\ref{eq:iam})). Since
the integrals involve the interval $ 0 \le s \le 4 m_\pi^2 $ between
the right- and left-hand cuts, this higher order zeros clearly
influence the fulfillment of the crossing sum rules. However, there is
no unique way how to modify the chiral zeros behaviour in order to
achieve a better fulfilment of crossing. To overcome this difficulty
several interesting methods have been proposed effectively rectifying
the amplitude behaviour in the unphysical region, though one should
say that none of them is entirely satisfactory from the point of view
of the mathematical properties that one wants to impose {\it a priori}
on the amplitude. In Ref.~\cite{BP97} it has been suggested ( {\it
scheme II} of that work) to use a dispersion relation for the inverse
amplitude, $t_{IJ}(s)^{-1}$. In such a way, not only the unitarity cut
but also the position of the single chiral zero, which becomes a
single pole for the inverse amplitude may be enforced from the
beginning. The left-hand cut is assumed to be that one of ChPT up to a
certain negative $s=-\Lambda^2 $ value ( $\Lambda^2 =0.5-0.6 {\rm
GeV}^2 $) and a constant up to $s=-\infty$. This procedure has the
disadvantage of introducing a new variable (the cut-off $\Lambda$)
into the problem not present in the ChPT amplitude. In addition, it
does not take the shift in the non-kinematical chiral zeros into
account, and imposes the tree level ones. More recently, in
Ref.~\cite{Ha99} a generalized IAM to two loops has been proposed. If
$s_0$ is the chiral Adler zero to lowest order $ t_{IJ}^{(2)}(s_0)=0$
then the following expression for the inverse amplitude is suggested
\cite{Ha99}
\begin{eqnarray}
t_{IJ}^{-1}(s)&=&\frac{t_{IJ}^{(2)}(s)-t_{IJ}^{(4)}(s)+
\frac{t_{IJ}^{(4)}(s)^2}{t_{IJ}^{(2)}(s)}-t_{IJ}^{(6)}(s) + 2 t_{IJ}^{(4)}(s_0) \left(
1-\frac{t_{IJ}^{(4)}(s)}{t_{IJ}^{(2)}(s)} \right) + 2 t_{IJ}^{(6)}(s_0) +
\frac{t_{IJ}^{(4)}(s_0)^2}{t_{IJ}^{(2)}(s)}}{\left[t_{IJ}^{(2)}(s)+t_{IJ}^{(4)}(s_0
)+t_{IJ}^{(6)}(s_0)\right]^2} \nonumber \\ 
\label{eq:iam-gen} 
\end{eqnarray} 
This expression violates exact unitarity since  
\begin{eqnarray}
{\rm Im} t_{IJ}^{-1} (s) = -\sigma(s) \frac{t_{IJ}^{(2)}(s)^2 + 2 t_{IJ}^{(4)}(s_0)
t_{IJ}^{(2)}(s)}{\left[t_{IJ}^{(2)}(s)+t_{IJ}^{(4)}( s_0 )+t_{IJ}^{(6)}(s_0)\right]^2}
\end{eqnarray} 
and, in addition, has a single zero at the lowest order approximation
of the chiral zero. The slope coincides with the one obtained in ChPT
as can be seen from the formula
\begin{eqnarray}
t_{IJ}(s) = \left[ t_{IJ}^{(2)} (s_0 )'+t_{IJ}^{(4)} ( s_0 )' +
t_{IJ}^{(6)} (s_0)' \right] (s-s_0) + {\cal O} \left[ (s-s_0)^2
\right]
\end{eqnarray} 
In the limit $ t_{IJ}^{(4)}( s_0)+ t_{IJ}^{(6)}(s_0) \to 0 $ in
Eq.~(\ref{eq:iam-gen}) the generalized IAM of Ref.~\cite{Ha99} reduces
to the standard IAM method of Eq.~(\ref{eq:iam}) and also unitarity is
exactly fulfilled. In practice, both unitarity violations and the
absence of a shift for non-kinematical zeros are numerically small.
It has been shown in Ref.~\cite{Ha99} that the generalized IAM
improves the fulfillment of the Roskies sum rules, but no
uncertainties estimates were considered. Using the two definitions of
crossing violations given by Eqs.~(\ref{eq:viola2}) and
(\ref{eq:viola}) suggested in Refs.~\cite{BP97,Ha99} and \cite{CB01}
we show in Tables~\ref{tab:v-generalized} and \ref{tab:r-generalized}
respectively that this is indeed the case, provided uncertainties in
the parameters are taken into account.

\begin{figure}[t]
\begin{center}
\epsfig{figure=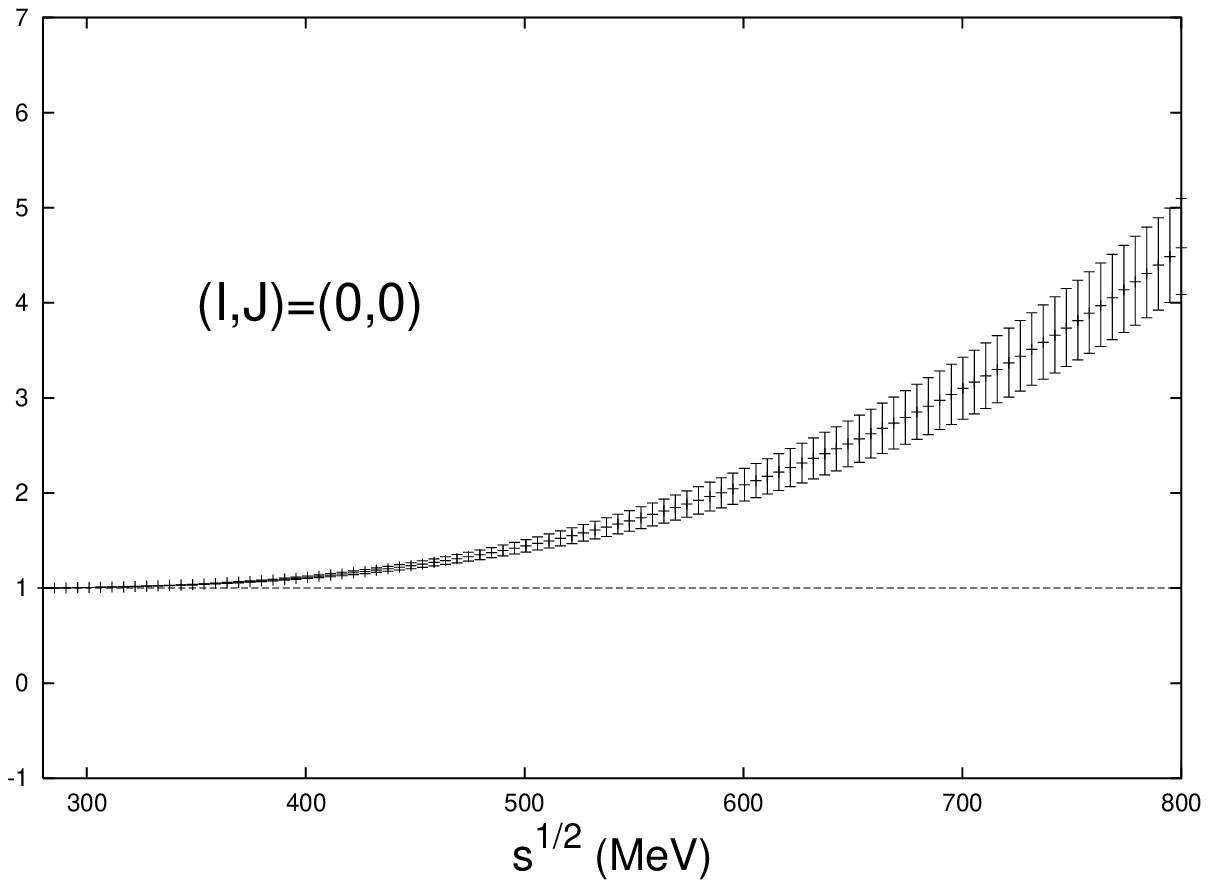,height=5.5cm,width=5.5cm}\epsfig{figure=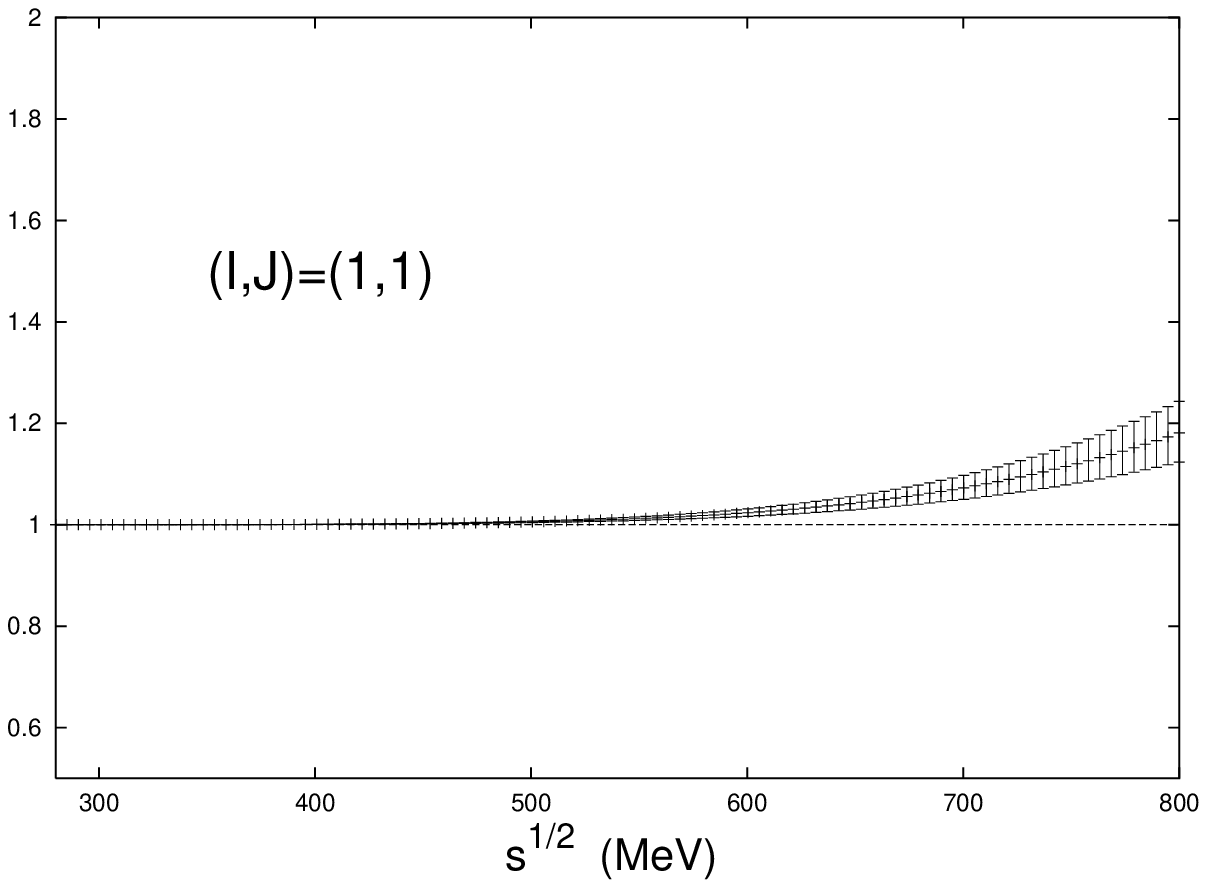,height=5.5cm,width=5.5cm}\epsfig{figure=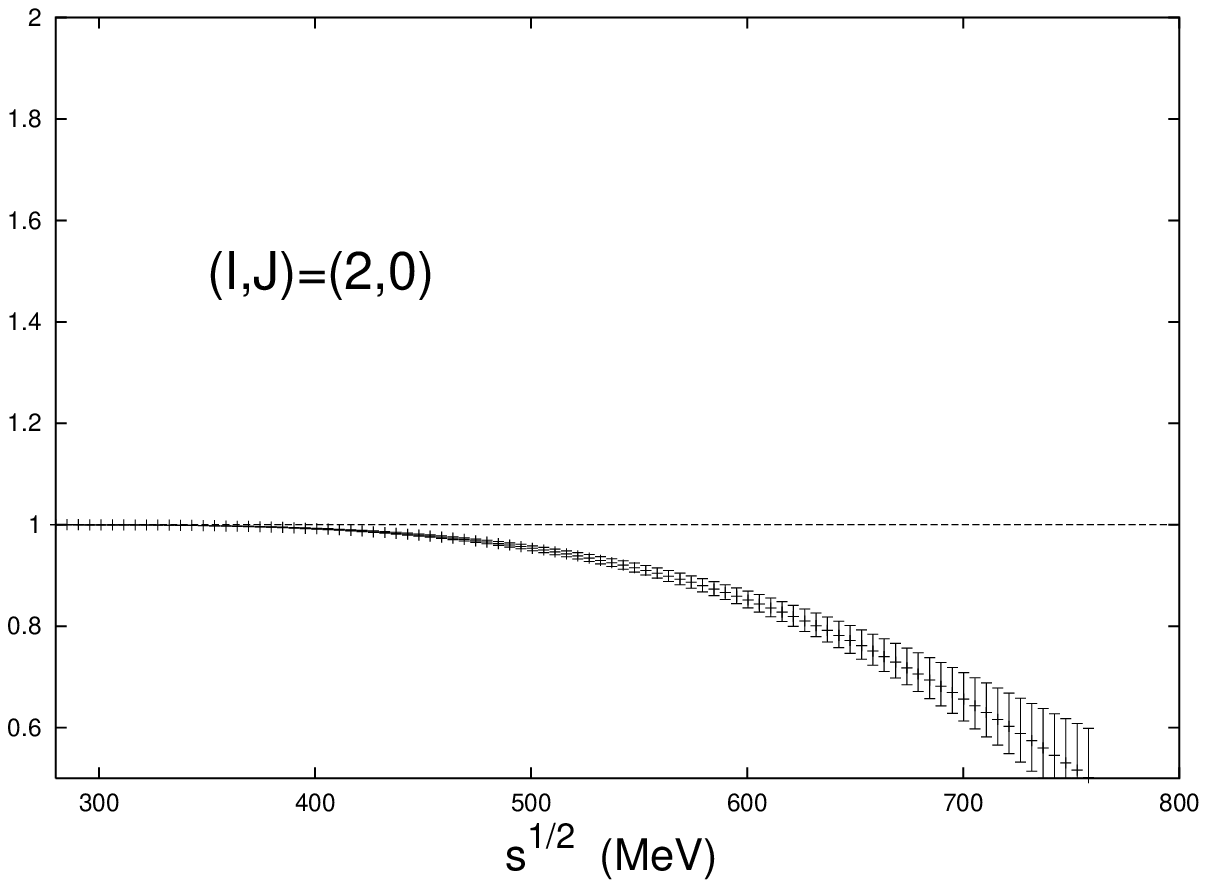,height=5.5cm,width=5.5cm}
\epsfig{figure=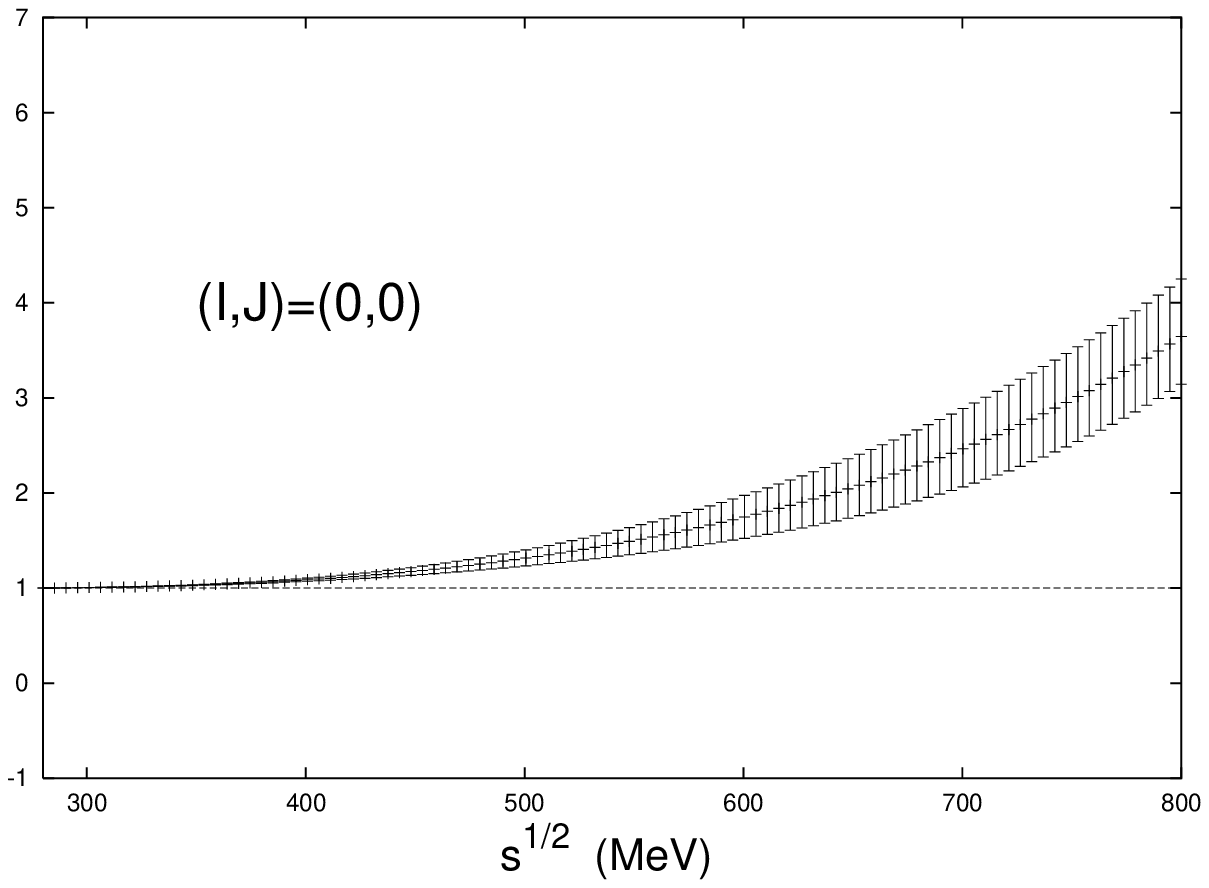,height=5.5cm,width=5.5cm}\epsfig{figure=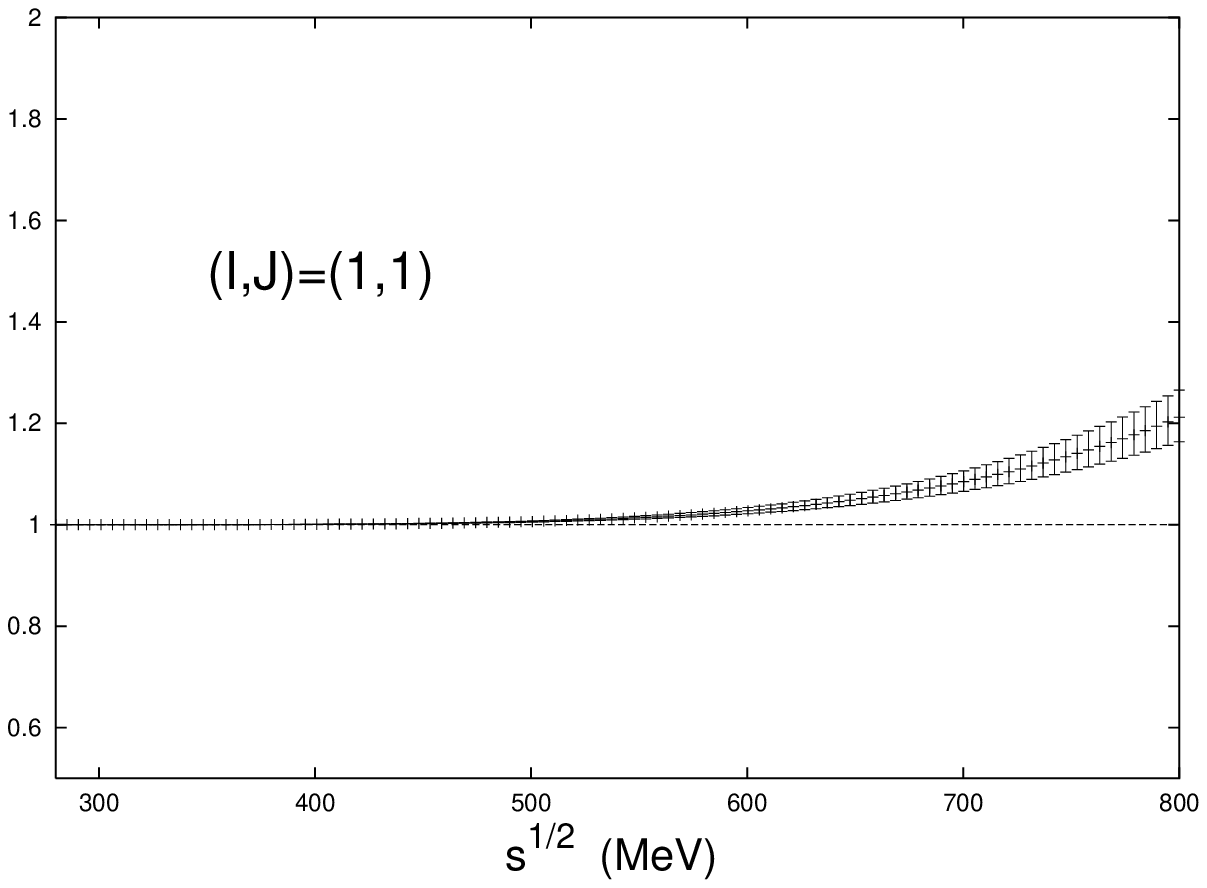,height=5.5cm,width=5.5cm}\epsfig{figure=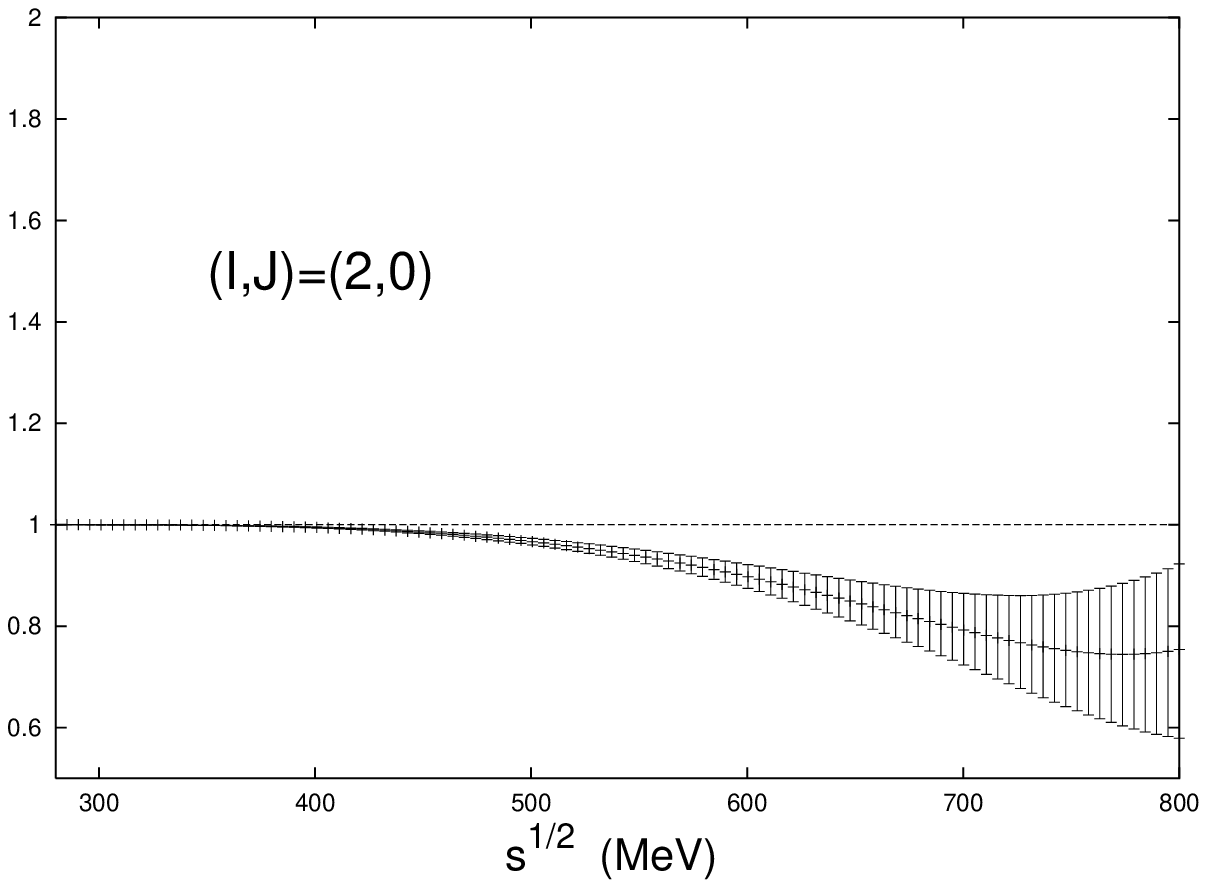,height=5.5cm,width=5.5cm}
\end{center} 
\caption{Unitarity condition for standard NLO-ChPT amplitudes in
$\pi\pi$ scattering for $S-$ and $P-$ waves defined by $U_{IJ} (s) = |
1 + 2 i \sigma(s) t_{IJ} (s) |$ . Upper panel: Set {\bf Ic} of
Ref.~\cite{EJ00a}. Lower panel: Set {\bf III} of Ref.~\cite{EJ00a} The
unitarity condition requires $U_{IJ} (s)=1$.}
\label{fig:unitarity.NLO}
\end{figure}

\begin{figure}[t]
\begin{center}
\epsfig{figure=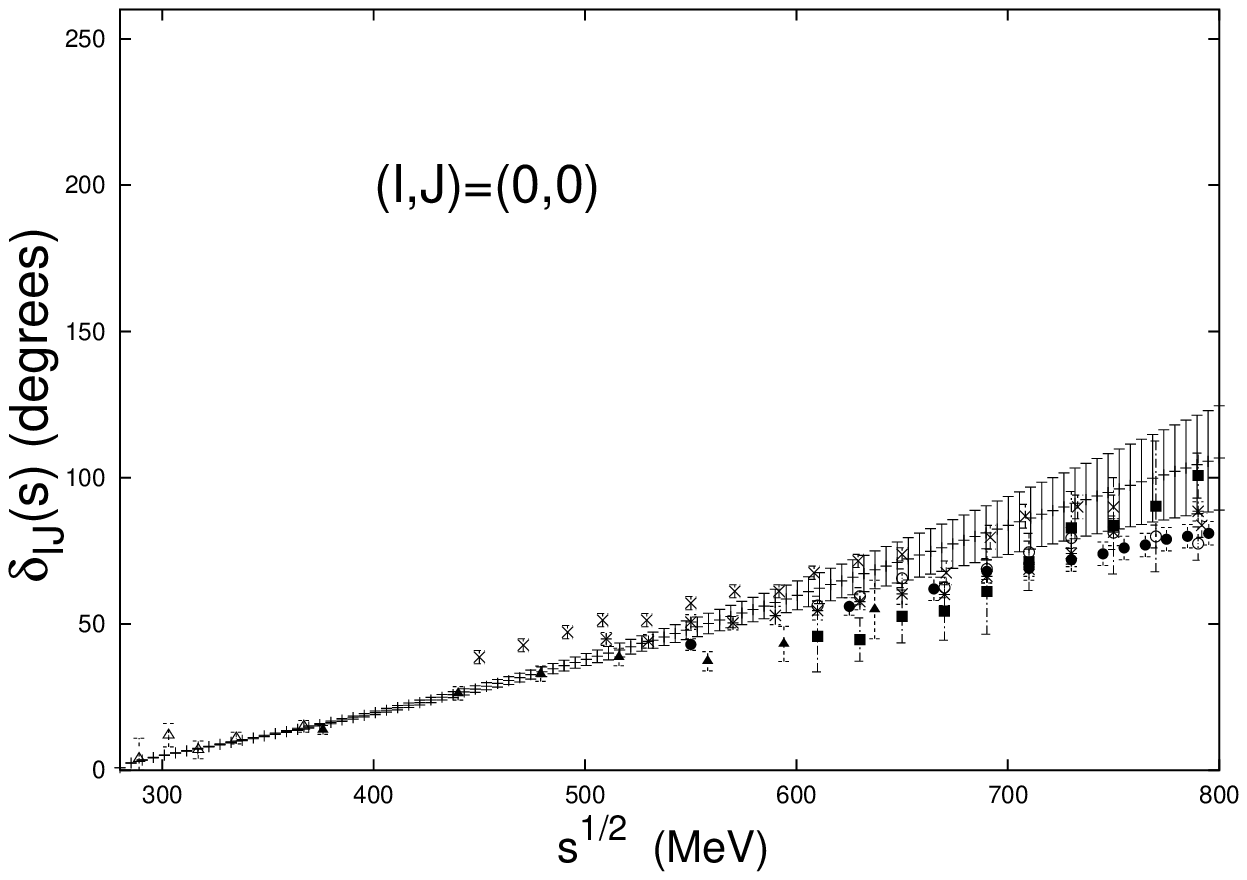,height=5.5cm,width=5.5cm}\epsfig{figure=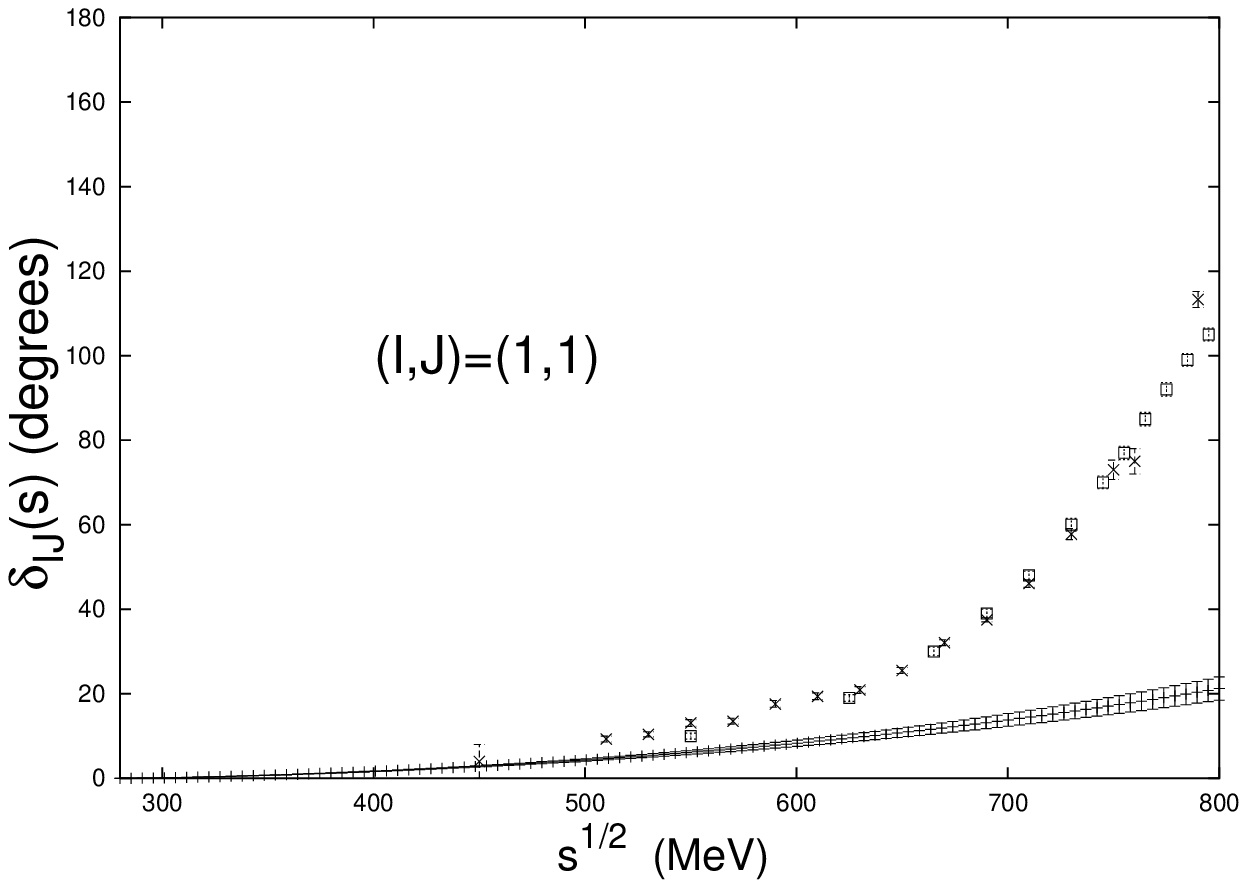,height=5.5cm,width=5.5cm}\epsfig{figure=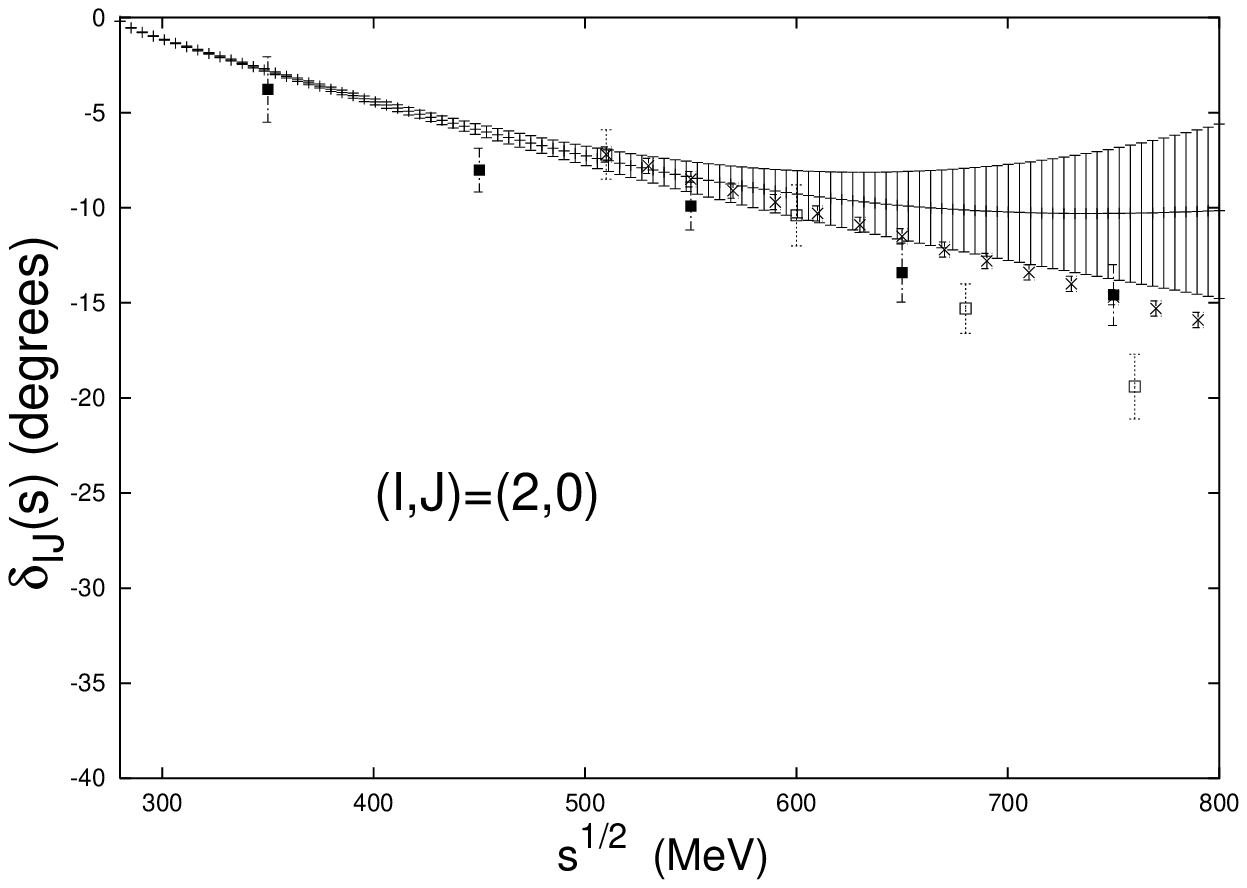,height=5.5cm,width=5.5cm}
\epsfig{figure=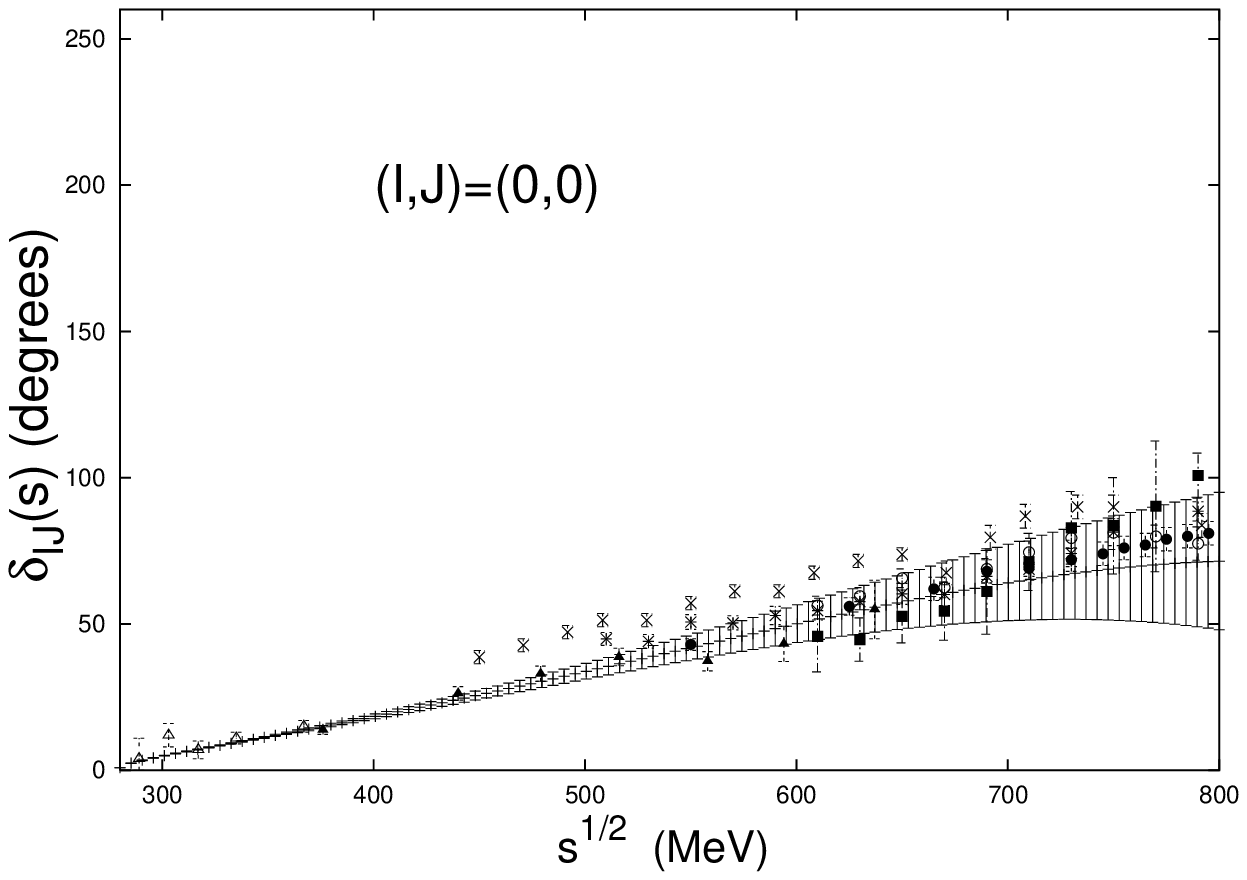,height=5.5cm,width=5.5cm}\epsfig{figure=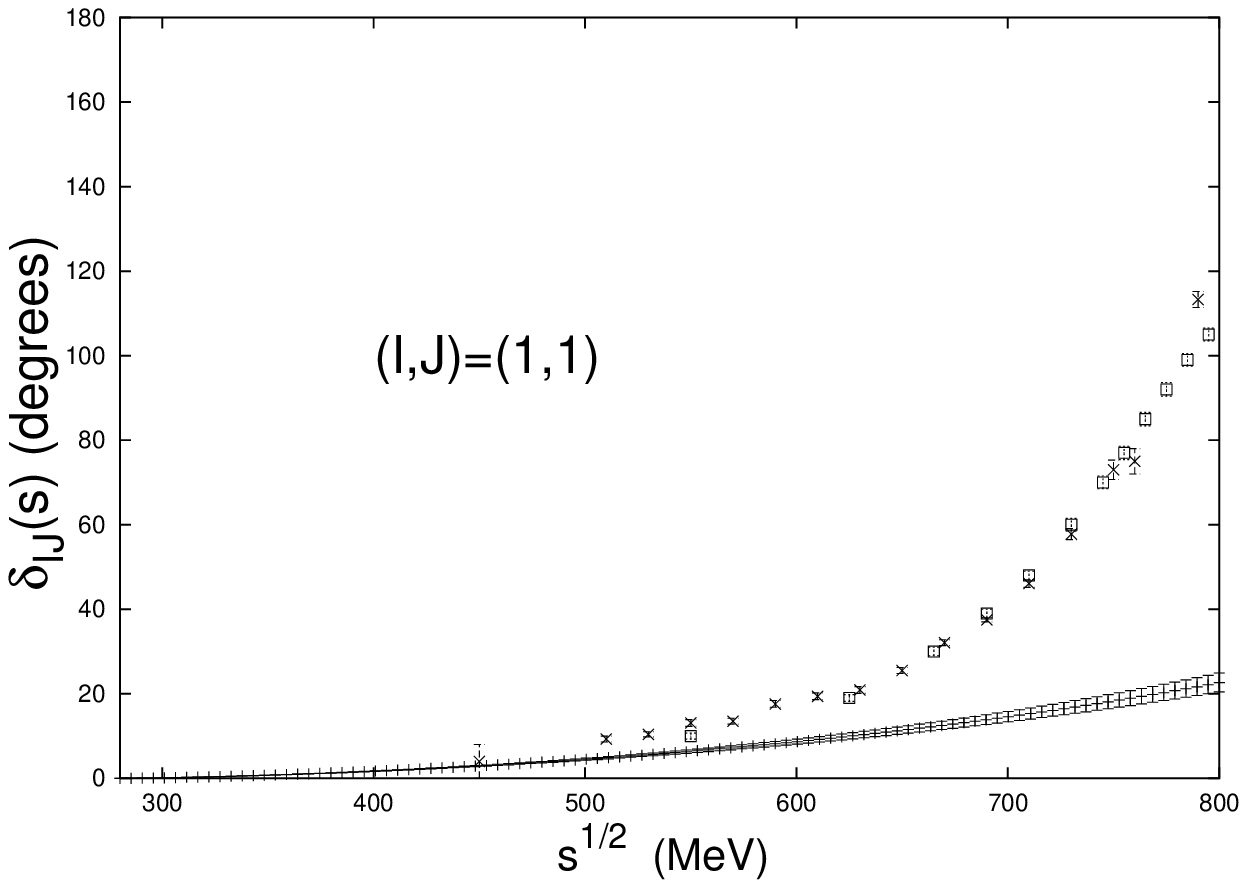,height=5.5cm,width=5.5cm}\epsfig{figure=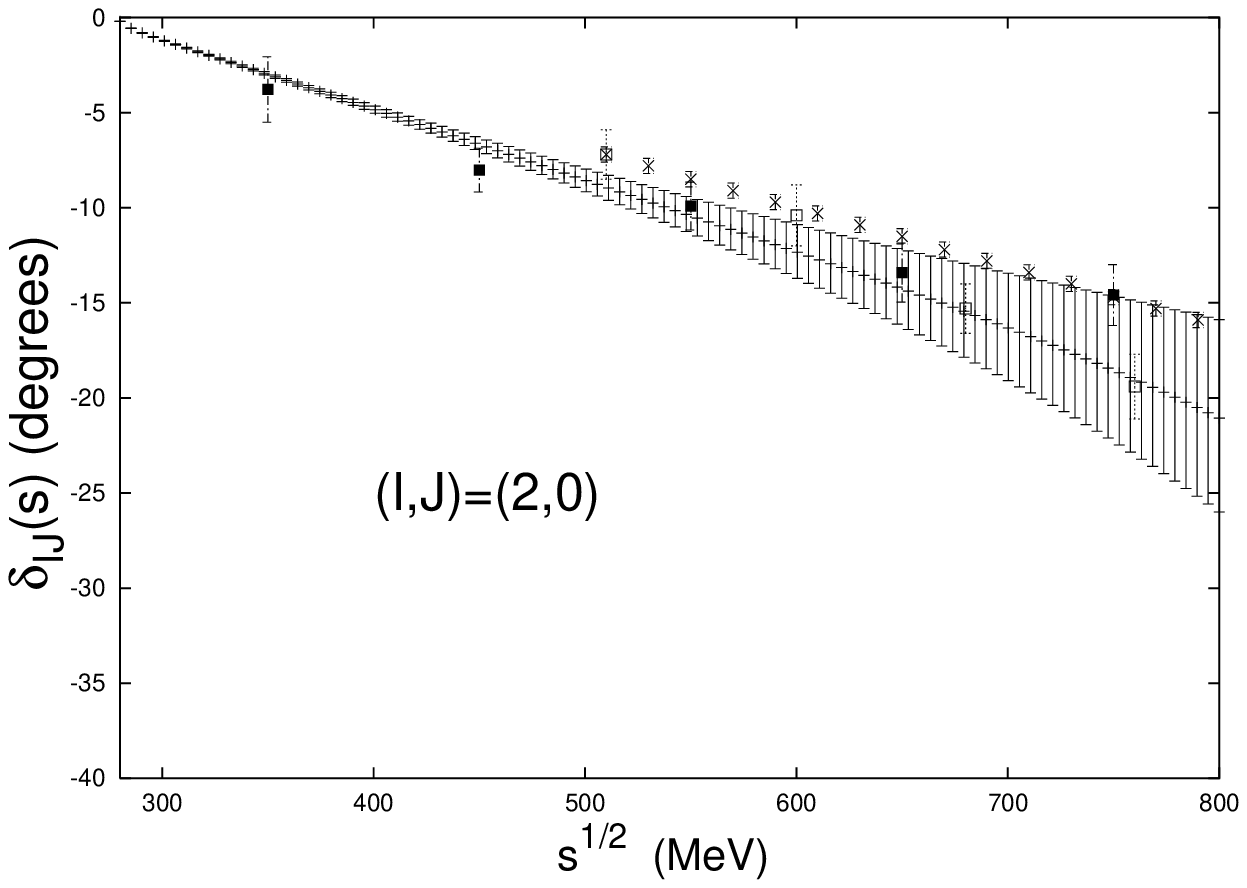,height=5.5cm,width=5.5cm}
\end{center} 
\caption{Standard NLO-ChPT phase shifts (in degrees) for $\pi\pi$
scattering for $S-$ and $P-$ waves after
Eq.~(\ref{eq:delta_chpt}). Upper panel: Set {\bf Ic} of
Ref.~\cite{EJ00a}. Lower panel: Set {\bf III} of
Ref.~\cite{EJ00a}.Combined data from Refs.~\cite{pa73}-\cite{fp77}.}
\label{fig:chpt.NLO}
\end{figure}

\begin{figure}[t]
\begin{center}
\epsfig{figure=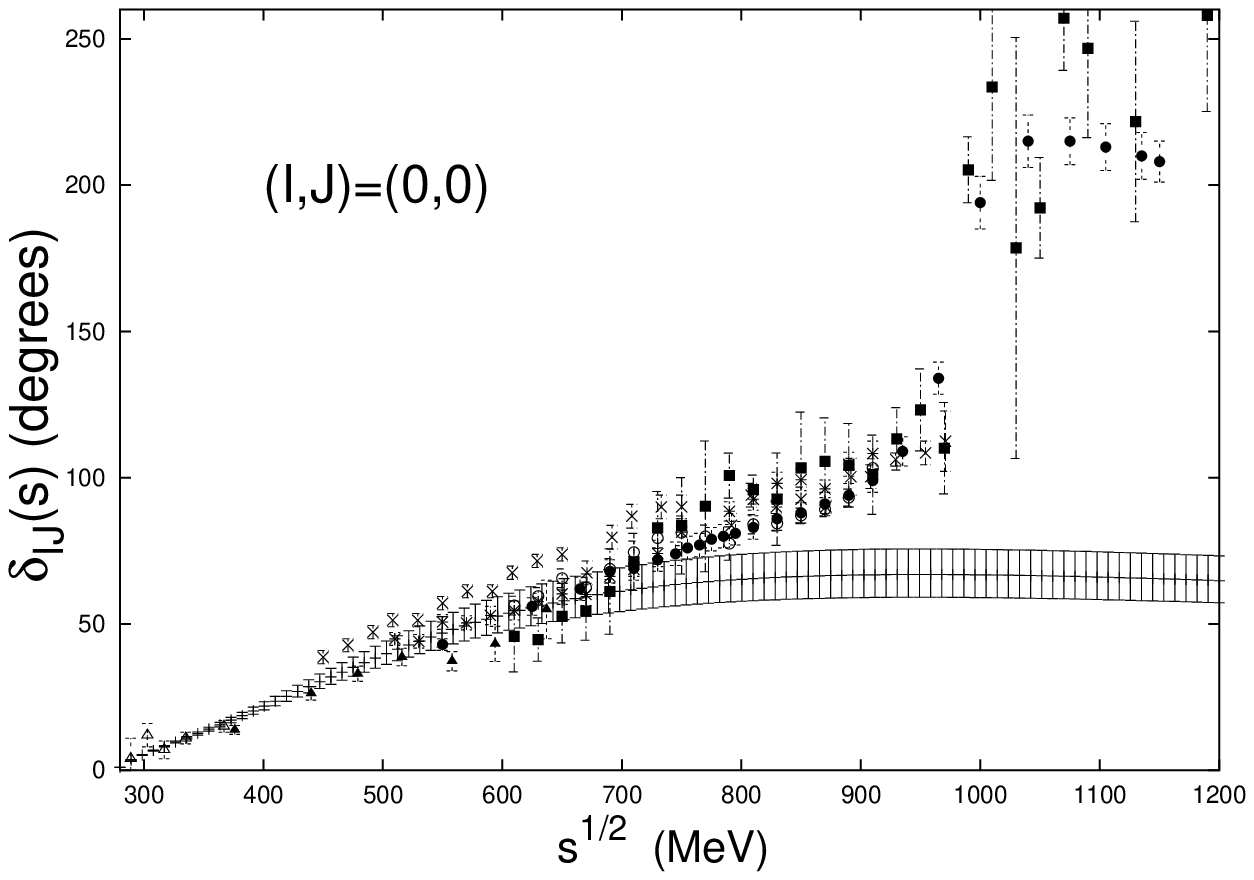,height=5.5cm,width=5.5cm}\epsfig{figure=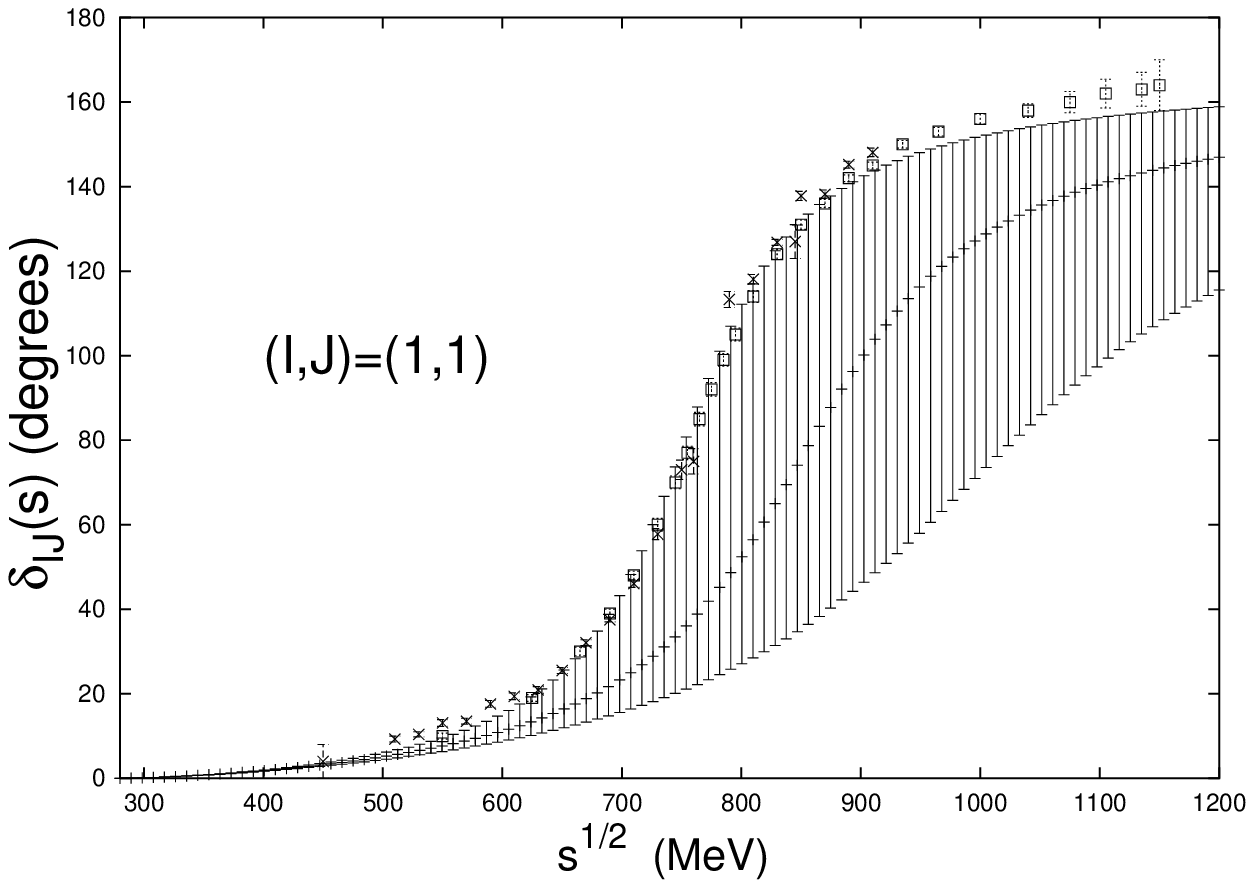,height=5.5cm,width=5.5cm}\epsfig{figure=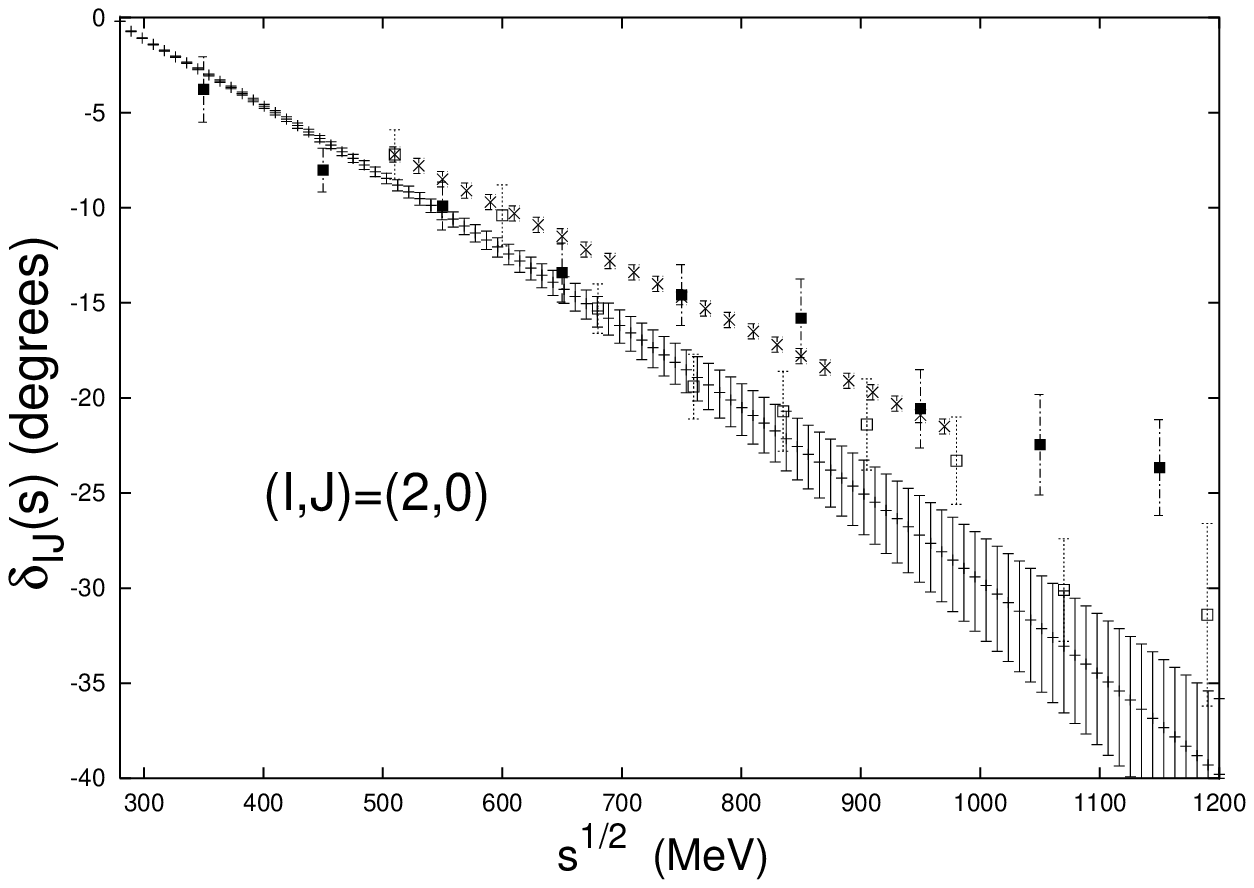,height=5.5cm,width=5.5cm}
\epsfig{figure=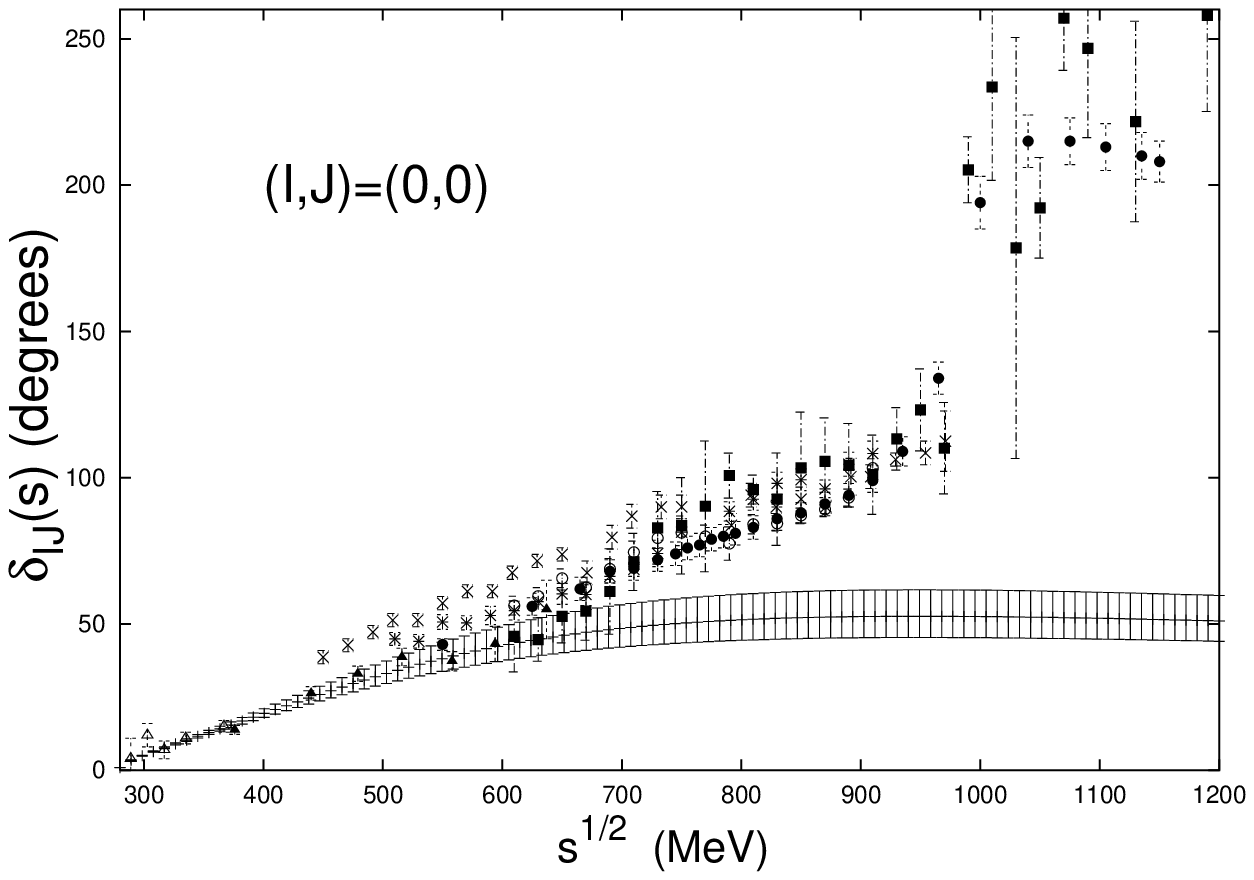,height=5.5cm,width=5.5cm}\epsfig{figure=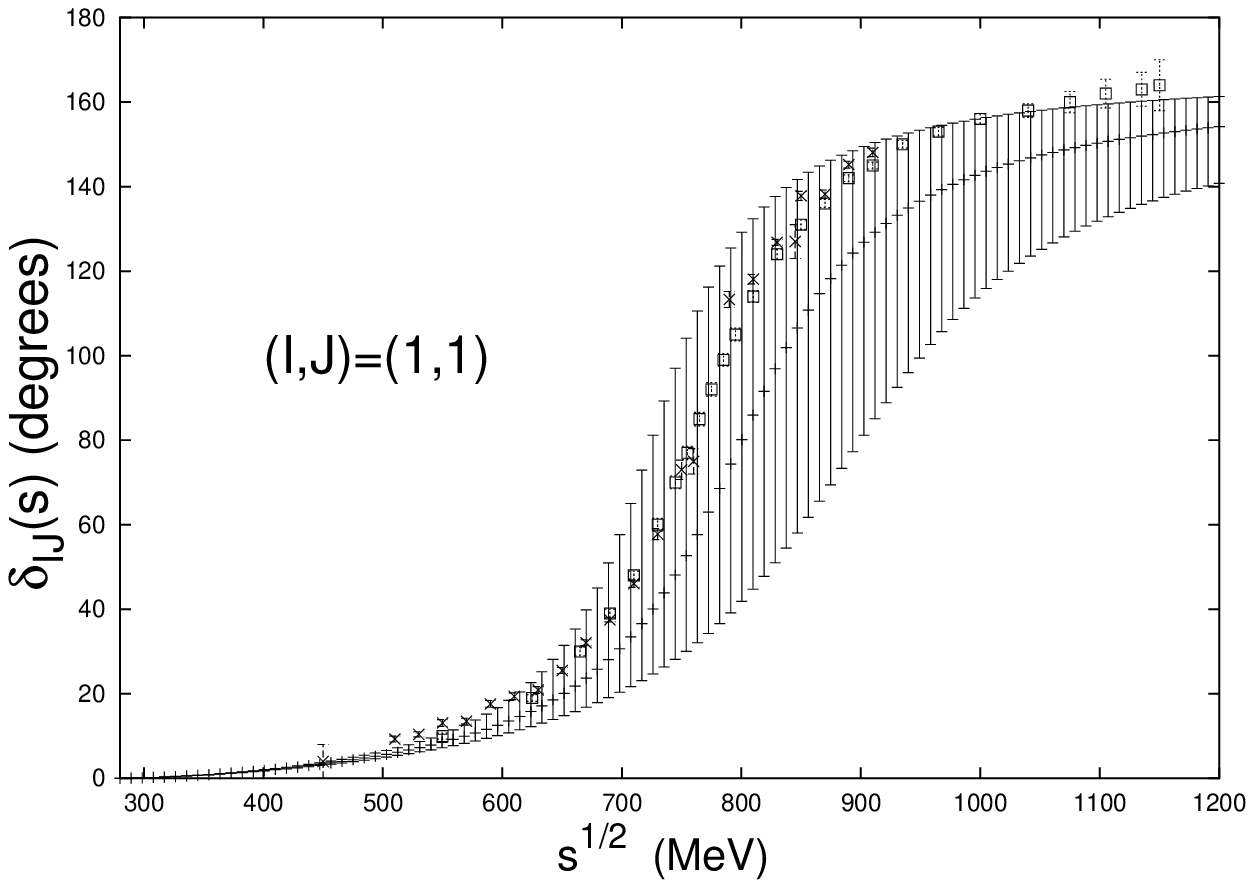,height=5.5cm,width=5.5cm}\epsfig{figure=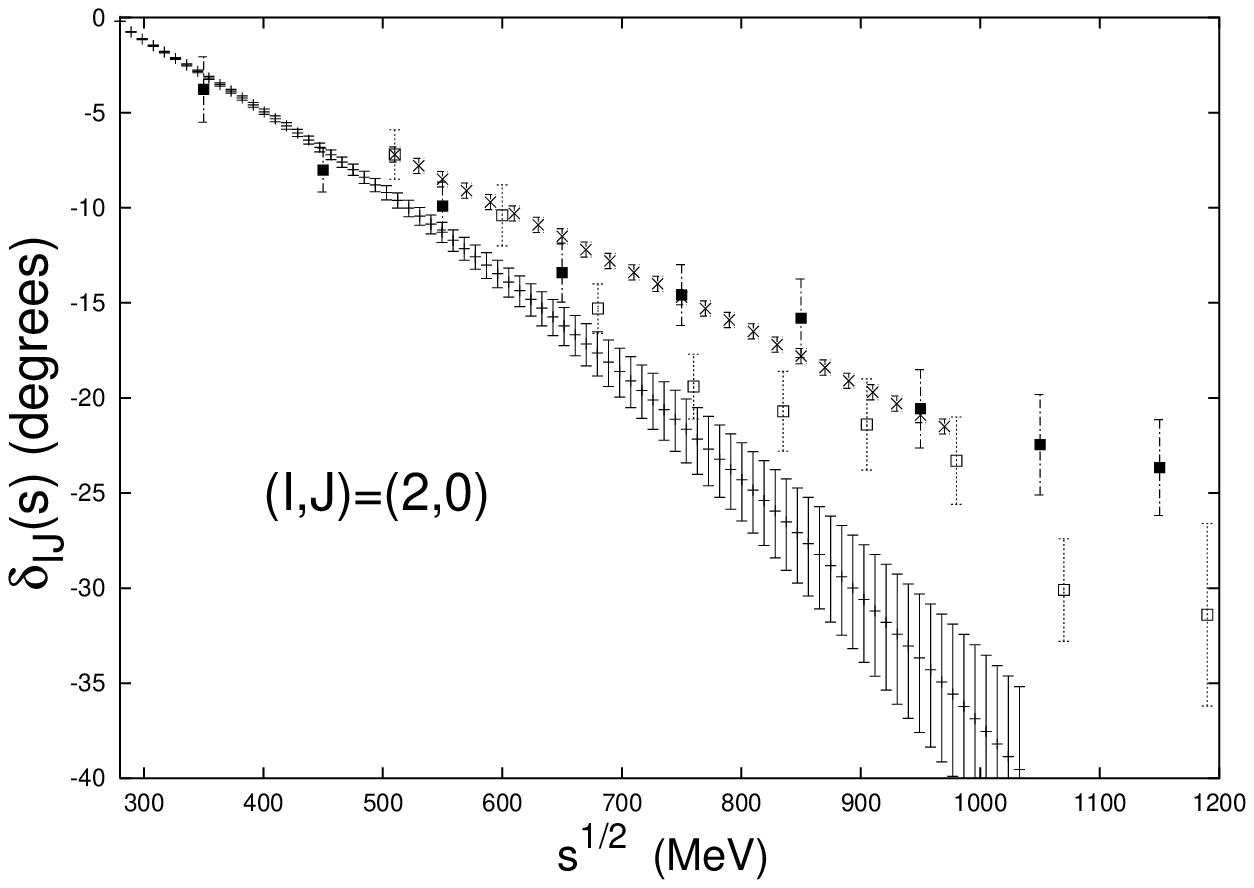,height=5.5cm,width=5.5cm}
\end{center} 
\caption{NLO-IAM Unitarized phase shifts (in degrees) for $\pi\pi$
scattering for $S-$ and $P-$ waves. Monte Carlo scheme (see main
text). Upper panel: Set {\bf Ic} of Ref.~\cite{EJ00a}. Lower panel:
Set {\bf III} of Ref.~\cite{EJ00a}. Combined data from
Refs.~\cite{pa73}-\cite{fp77}. }
\label{fig:iam.NLO}
\end{figure}

\begin{figure}[t]
\begin{center}
\epsfig{figure=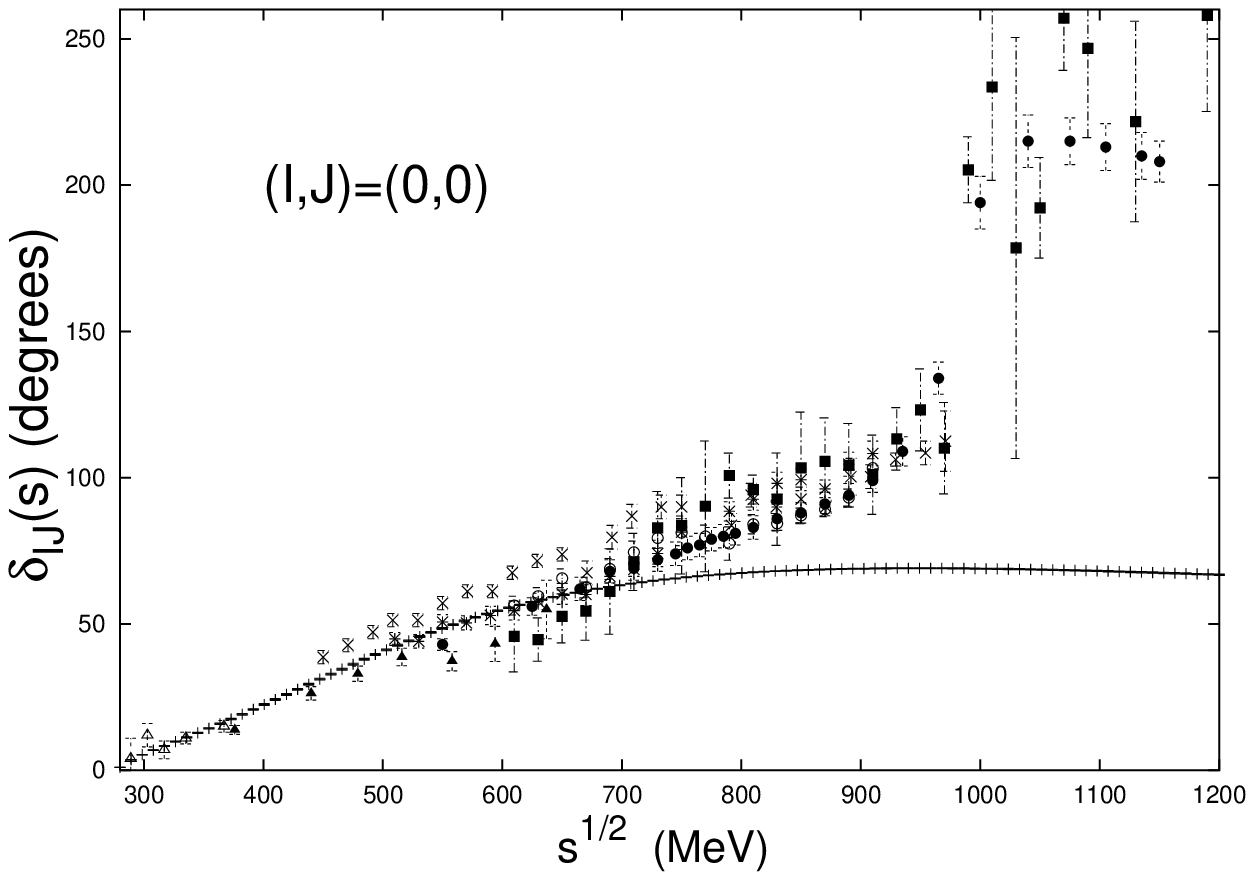,height=5.5cm,width=5.5cm}\epsfig{figure=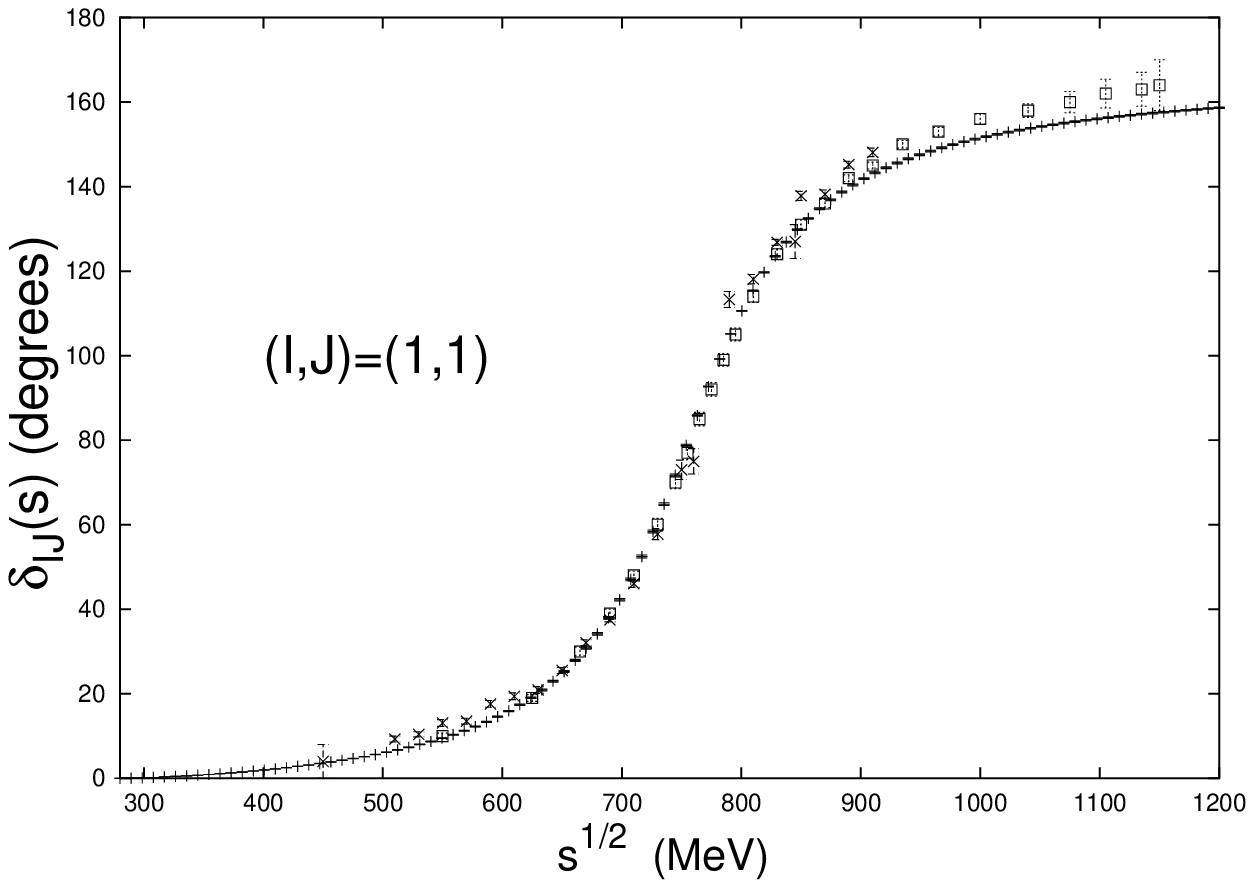,height=5.5cm,width=5.5cm}\epsfig{figure=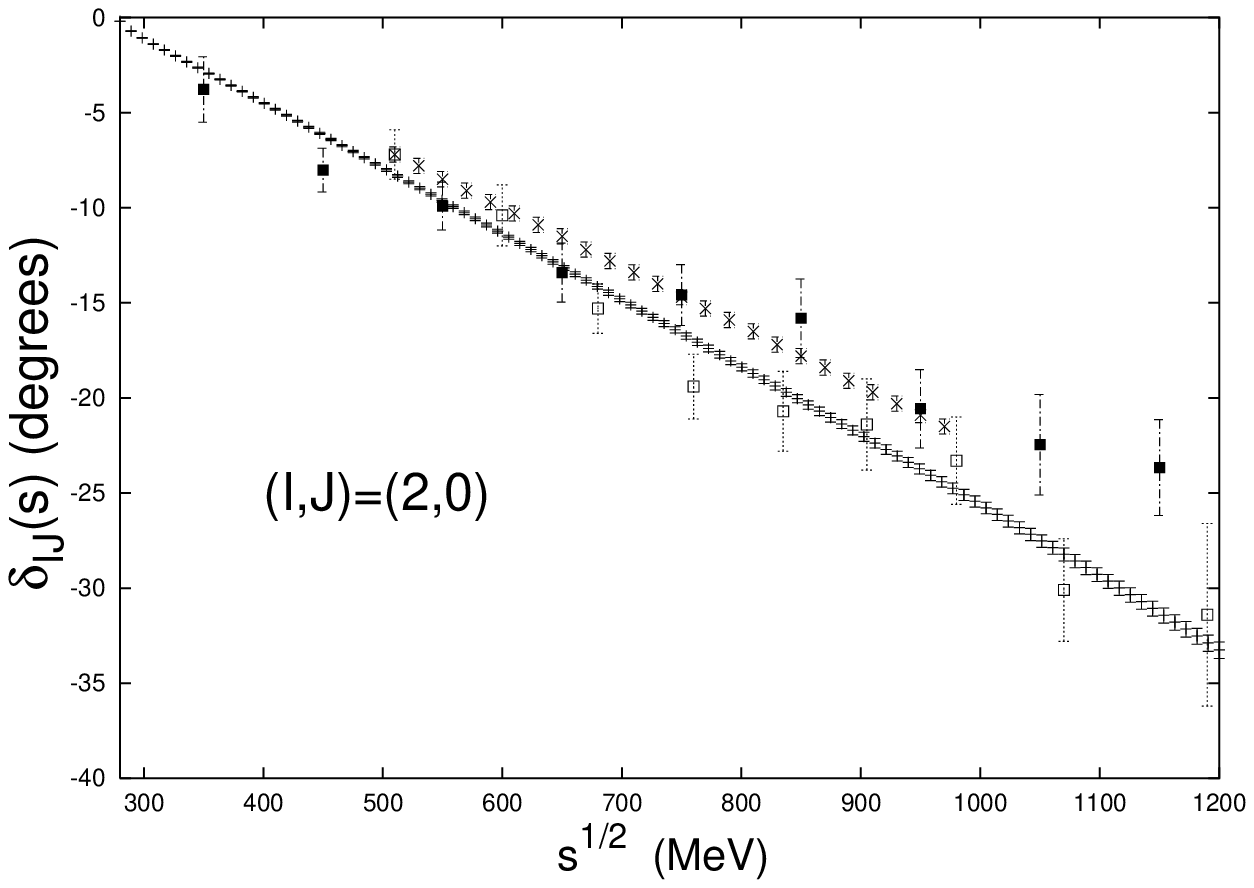,height=5.5cm,width=5.5cm}
\end{center} 
\caption{NLO-IAM Unitarized phase shifts (in degrees) for $\pi\pi$
scattering for $S-$ and $P-$ waves. Partial fit scheme (see main
text). The errors in the curves are due to uncertainties in $\bar l_3$
and $\bar l_4$. Combined data from Refs.~\cite{pa73}-\cite{fp77}. }
\label{fig:iam.fit.NLO}
\end{figure}

\section{On the convergence of the IAM method} \label{sec:conv} 

The IAM method can be systematically implemented to any order in the
chiral expansion with no additional LEC's than those required by
standard ChPT. There arises the question to what extent is this method
convergent. To answer this question in practice we can only compare
one-loop and two loop predictions for the unitarized phase
shifts. Such a comparison makes sense only if errors in the LECS are
also transported, as we have repeatedly done along this
work. Actually, in Ref.~\cite{DP97a} the one loop error analysis of
the IAM phase shifts was estimated by varying the low energy
constants. In this section we reanalyze this question by using the
updated values of the one loop coefficients given by Set {\bf Ic} and
Set {\bf III} taking into account by means of a Monte-Carlo simulation
the important anticorrelations between $\bar l_1 $ and $\bar l_2$
determined in Ref.~\cite{EJ00a}. Following the same systematics of the
two loop calculation, we show in Fig.~\ref{fig:unitarity.NLO} the
unitarity condition of Eq.~(\ref{eq:uni_viol}). As one would expect,
unitarity violations of one-loop ChPT occurr at lower energies. The
NLO ChPT phase shifts, defined through Eq.~(\ref{eq:delta_chpt}) are
depicted in Fig.~\ref{fig:chpt.NLO}. The general trend follows a
similar pattern to the two loop calculation, although some important
differences emerge. Firstly, the uncertainties in the phase shifts are
smaller at NNLO than at NLO in the threshold region, as one would
expect from the fact that threshold parameters are more accurately
determined at NNLO than at NLO \cite{EJ00a}\footnote{This circumstance
is not trivial and it only happens for Sets{\bf Ic} and {\bf III}. The
Set {\bf II} of Ref.~\cite{EJ00a} shows cases where predictive in the
threshold parameters is lost, and the NNLO result is no more accurate
than the NLO.}. In the region above threshold the situation is exactly
the opposite, the two loop calculation produces larger uncertainties
than the one loop one. In addition, by comparison of
Fig.~\ref{fig:chpt.NLO} and Figs.~\ref{fig:chpt} and
(\ref{fig:naive-chpt} in the $\rho$ channel) one sees that the
discrepances in the region above threshold are larger than the estimated
uncertainties, with an overall trend to improvement in the two loop
calculation. A similar trend, although in a less clear manner, is
observed in the two $S-$waves. The unitarity condition,
Eq.~(\ref{eq:uni_viol}), gives us a good idea on the applicability of
standard ChPT to one and two loop approximations. Nevertheless, the
agreement of the perturbative phase shifts, Eq.~(\ref{eq:delta_chpt}),
with experiment seems to extend up to a region where the unitarity
violation may be as large as $10-20 \%$.

The one loop IAM phase shifts are depicted in
Fig.~\ref{fig:iam.NLO}. Apparently, the general picture provided by
NLO-ChPT looks better than what one obtains by comparing any of the
two Monte-Carlo schemes studied in Sect.~\ref{sec:iam} based in
NNLO-ChPT. By looking at any of these two-loop schemes,
Fig.~\ref{fig:naive} and Fig.~\ref{fig:separated}, we realize that
there is a clear loss of predictive power; the errors in the two loop
phase shifts are larger than the discrepancy between their mean value
and the one loop mean value. Finally, in Fig.~\ref{fig:iam.fit.NLO}
a partial one-loop fit procedure in $\bar l_1$ and $\bar l_2$
parameters to the data is presented, where variations on the $\bar
l_3$ and $\bar l_4$ are taken into account. The result of the fit is 
\begin{eqnarray}
\bar l_1 = -0.44 \pm 0.02 \qquad \bar l_2 = 5.51 \pm 0.04 \qquad
r(\bar l_1 , \bar l_2) = -0.81 
\end{eqnarray}
where the errors reflect the uncertainites in $\bar l_3$ and $\bar
l_4$. Here $\chi^2 / {\rm d.o.f.} = 191 /(67-2) = 2.94 $ almost three
times larger than in the two loops case (1.11), and too large to be
considered a satisfactory drescription of the scattering data. Such a
large $\chi^2$ value makes the determination of the uncertainties of
$\bar l_1$ and $\bar l_2$ due to the error bars in the fitted data
meaningless. As we see, the obtained values for the fitted parameters
are compatible with the corresponding two loop partial fit procedure,
Eq.~(\ref{eq:fit-2loop}), although the errors in the one-loop case due
to uncertainties in $\bar l_3$ and $\bar l_4$ are much smaller than in
the two-loop case. This may be an indirect consequence of the large
$\chi^2 $ value. 

\section{Conclusions} \label{sec:concl} 

In the present work we have presented a thorough study of the Inverse
Amplitude Method to unitarize the NLO (one-loop) and NNLO (two-loop)
ChPT $\pi\pi$ scattering amplitudes below the $K \bar K$ threshold. To
this end, we have considered several one-loop $\bar l_{1,2,3,4}$ and
two loop $\bar b_{1,2,3,4,5,6}$ parameter sets along the lines
discussed in our previous work \cite{EJ00a}. Particularly interesting
in this work is the role played by the uncertainties in these
parameters. To complement the analysis and provide some quantitative
motivation we have determined unitarity violations within standard
ChPT, with error estimates. They take place at much lower energies
than the unitarity limit suggests. Moreover, we have also shown the
systematic discrepancy with the data in the region above threshold if
phase shifts are defined perturbatively. The discussion is complicated
by the fact that the two loop parameters $\bar b_{1,2,3,4,5,6} $ may
by splitted in a zeroth order contribution $\bar b_{1,2,3,4,5,6}^0 $
and a higher order contribution $\Delta \bar b_{1,2,3,4,5,6}$ which
slightly spoils the chiral counting. The distinction caused in the
phase shifts by including or not the higher contribution is small
within ChPT. Motivated by this we have unitarized the two loop
amplitude, and devised several schemes to predict the phase-shifts
from threshold up to the resonance region. The effect of consistently
treating or not $\Delta \bar b_{1,2,3,4,5,6}$ as higher order is much
stronger for the IAM unitarized phase shifts. Typically, a factor of
two difference or larger in the uncertainties is encountered. In any
case, they are rather large, although seem consistent with the
scattering data. This indicates a kind of fine tuning going on and
suggests a fit to the data to determine the low energy parameters
which remain in the chiral limit, keeping the remaining low energy
parameters within their error bars. The result of the fit is
satisfactory, although there appears a discrepancy in the $\bar b_6$
coefficient. Nevertheless, the predicted partially fitted phase shifts
vary within very small uncertainties, not far from the recent ChPT
analysis of Roy equations carried out in Ref.~\cite{CGL01}. Despite of
these features, the IAM produces crossing violations which have been
quantified in terms of Roskies sume rules. Generally speaking, they
are not very large in percentage terms, although in some cases the
uncertainties are so large that no conclusion can be drawn. We have
also studied some proposals to generalize the IAM in order to achieve
a better fulfillment of crossing properties. Finally, we have adressed
the convergence of an expansion based on the IAM and increasing order
of approximation in ChPT. By comparing NLO (one-loop) and NNLO
(two-loop) IAM predicted phase shifts we see at the present stage a
lack of predictive power; the errors in the two loop phase-shift are
larger than the difference between the central one-loop and two-loop
values. This is a direct consequence of the low accuracy in the two
loop parameters.

\appendix

\section{Correlations among NNLO low energy constants and threshold parameters in standard ChPT} \label{sec:app1}

The correlation matrix, defined as usual,
\begin{eqnarray}
r_{ij} &=& \langle  x_i x_j \rangle \nonumber\\
&&\nonumber\\
x_i &=& \frac{c_i -\langle c_i \rangle }{\sqrt{ \langle c_i^2\rangle -
\langle c_i \rangle^2}}\nonumber\\
&&\nonumber\\
\langle f(c_1,\dots,c_n) \rangle & = & 
\frac{1}{N} \sum_{\alpha=1}^N f(c_{1,\alpha},\dots,c_{n,\alpha})
\end{eqnarray}
being $c_i$ any of the low energy constants or threshold parameters is
provided below as obtained from our $N=10^4$ finite size samples both
for Set {\bf Ic} and Set {\bf III}. Taking into account the central
values and their errors given in Table \ref{tab:bbar} and Table
\ref{tab:threshold} and ignoring the slight error asymmetries the
parameter Sets are fully portable by going through diagonalization to
principal axis in parameter space and making a Monte-Carlo Gaussian
simulation in each principal direction.

\underline{\bf Set Ic}

\begin{eqnarray}
r(\bar b_i, \bar b_j) &=&\left(\matrix{ +1.00 & & & & & \cr -0.74 &
+1.00 & & & & \cr +0.53 & -0.77 &+1.00 & & & \cr -0.49 & +0.64 &-0.49
& +1.00 & & \cr +0.09 & +0.04 &-0.41 & +0.10 &+1.00 & \cr -0.08 &
+0.23 &-0.54 & +0.23 &+0.57 & +1.00} \right) \\ \nonumber \\ &&
\matrix{ \qquad a_{00} & \quad a_{11} & \quad a_{20} & \quad b_{00} &
\quad b_{11} & \quad b_{20} }
\nonumber \\ r_{ij}^{\rm Threshold} &=& \left(\matrix{ +1.00& & & & &
\cr +0.13& +1.00& & & & \cr +0.58& -0.21& +1.00& & & \cr +0.75& -0.13&
+ 0.37& +1.00& & \cr +0.02& +0.89& -0.07& -0.10& +1.00& \cr +0.20&
+0.06& +0.57& +0.38& +0.11& +1.00 } \right)
\end{eqnarray} 

\underline{\bf Set III}

\begin{eqnarray}
r(\bar b_i, \bar b_j) &=&\left(\matrix{
+1.00 &      &       &       &       &       \cr 
-0.73 &+1.00 &       &       &       &       \cr 
+0.53 &-0.76 & +1.00 &       &       &       \cr 
-0.12 &+0.20 & -0.14 & +1.00 &       &       \cr 
+0.22 &-0.16 & -0.19 & +0.21 & +1.00 &       \cr 
+0.06 &+0.02 & -0.29 & +0.19 & +0.47 & +1.00
} \right) \\ \nonumber \\ &&
\matrix{ \qquad a_{00} & \quad a_{11} & \quad a_{20} & \quad b_{00} &
\quad b_{11} & \quad b_{20} }
\nonumber \\ r_{ij}^{\rm Threshold} &=& \left(\matrix{
+1.00 &       &       &       &       &       \cr 
-0.08 &	+1.00 &       &       &       &       \cr 
+0.67 & -0.46 & +1.00 &       &       &       \cr 
+0.81 &	-0.32 & +0.55 &	+1.00 &       &       \cr
-0.11 & +0.82 &	-0.23 & -0.18 & +1.00 &       \cr 
+0.37 & -0.45 & +0.71 & +0.59 &	-0.21 & +1.00 \cr 
} \right)
\end{eqnarray}

%\newpage 

\section*{Acknowledgements} 

This work has been partially supported by DGES PB98-1367 and by the
Junta de Andaluc\'{\i}a FQM0225. The work of M.P.V. has been done in
part with a grant under the auspices of the Junta de Andaluc\'{\i}a.

\end{document}